{}

\documentclass[11pt,a4paper]{article}

\newif\ifpublic\publictrue

\ifpublic\else\usepackage{showkeys}\fi
\usepackage[nosort]{cite}
\usepackage[parsep]{collref}
\usepackage[english]{babel}
\usepackage[T1]{fontenc}
\usepackage[utf8]{inputenc}
\usepackage{lmodern}
\usepackage{amsfonts,amsbsy,dsfont,euscript,mathrsfs,fixmath}
\usepackage{amsmath,amssymb,mathtools}
\usepackage{ulem}
\usepackage{scalerel}
\usepackage{enumerate}
\usepackage[x11names]{xcolor}
\usepackage{graphicx}
\usepackage{adjustbox}
\usepackage{multirow}
\usepackage{color, colortbl}
\usepackage{empheq}
\usepackage{xfrac}

\usepackage{tikz,pgfplots}
\usetikzlibrary{arrows,calc,chains,decorations.pathmorphing,decorations.pathreplacing,fadings,matrix,positioning,shapes,shapes.geometric,shapes.symbols,trees}
\pgfplotsset{compat=1.9}
\usetikzlibrary{decorations.markings}
\usepackage{upgreek}
\renewcommand{\Delta}{\upDelta}
\renewcommand{\Omega}{\upOmega}
\let\originalTheta\Theta
\renewcommand{\Theta}{\mathrm{\originalTheta}}
\let\originalPhi\Phi
\renewcommand{\Phi}{\mathrm{\originalPhi}}
\let\originalPsi\Psi
\renewcommand{\Psi}{\mathrm{\originalPsi}}

\usepackage{cancel}

\def\showkeysrefformat#1{{\normalfont\tiny\ttfamily#1}}
\makeatletter
\def\SK@@ref#1>#2\SK@{{\@inlabelfalse\leavevmode\vbox to\z@{\vss\SK@refcolor\rlap{\vrule\raise .75em \hbox{\showkeysrefformat{#2}}}}}}
\makeatother

\usepackage[a4paper,text={160mm,247mm},centering]{geometry}
\allowdisplaybreaks[3]
\numberwithin{equation}{section}
\def\[{\begin{equation}\begin{aligned}}
\def\]{\end{aligned}\end{equation}}

\usepackage[font=small,labelfont=bf,width=0.85\textwidth]{caption}

\expandafter\def\expandafter\bfseries\expandafter{\bfseries\ifmmode\else\boldmath\fi}
\expandafter\def\expandafter\mdseries\expandafter{\mdseries\ifmmode\else\unboldmath\fi}
\expandafter\def\expandafter\normalfont\expandafter{\normalfont\ifmmode\else\unboldmath\fi}

\RequirePackage{verbatim}

\makeatletter
\newwrite\bibinl@out
\newenvironment{bibtex}[1][\jobname]{%
\immediate\openout\bibinl@out #1.bib%
\immediate\write\bibinl@out{\@percentchar generated from `\jobname' starting line \the\inputlineno^^J}%
\def\verbatim@processline{\immediate\write\bibinl@out{\the\verbatim@line}}%
\@bsphack\let\do\@makeother\dospecials\catcode`\^^M\active\verbatim@start%
}
{\immediate\closeout\bibinl@out\@esphack}
\makeatother

\let\barefrac=\frac
\renewcommand{\frac}[2]{\mathinner{\barefrac{#1}{#2}}}

\let\baresqrt=\sqrt
\makeatletter
\renewcommand{\sqrt}{\@ifnextchar[\@sqrt@space@a\@sqrt@space@b}
\def\@sqrt@space@a[#1]#2{\mathinner{\mathchoice{\mkern-3mu}{\mkern-3mu}{}{}\baresqrt[#1]{#2}}}
\def\@sqrt@space@b#1{\mathinner{\mathchoice{\mkern-3mu}{\mkern-3mu}{}{}\baresqrt{#1}}}
\makeatother

\makeatletter
\let\per@dot@old=\.
\def\.{\ifmmode\def\per@dot@sel{\mkern3mu}\else\def\per@dot@sel{\per@dot@old}\fi\per@dot@sel}
\makeatother

\let\barefootnote=\footnote
\renewcommand{\footnote}[1]{\barefootnote{#1\vspace{3pt}}}

\newcommand{\vfrac}[2]{\ifmmode\mathinner{\textstyle^{#1}\!/\!_{#2}}\else$^{#1}\!/\!_{#2}$\fi}

\DeclareMathOperator{\Tr}{Tr}

\DeclareMathOperator{\STr}{STr}

\makeatletter
\newcommand*\bigcdot{\mathpalette\bigcdot@{.5}}
\newcommand*\bigcdot@[2]{\mathbin{\vcenter{\hbox{\scalebox{#2}{$\m@th#1\bullet$}}}}}
\makeatother

\newcommand{\alg}[1]{\mathfrak{#1}}

\DeclareMathOperator{\ad}{ad}
\DeclareMathOperator{\Ad}{Ad}

\def\<{\big\langle}
\def\>{\big\rangle}

\DeclareFontEncoding{LS1}{}{}
\DeclareFontSubstitution{LS1}{stix}{m}{n}
\DeclareSymbolFont{stixsymbols}{LS1}{stixscr}{m}{n}
\SetSymbolFont{stixsymbols}{bold}{LS1}{stixscr}{b}{n}
\DeclareMathSymbol{\kay}{\mathalpha}{stixsymbols}{"6B}
\DeclareMathSymbol{\hay}{\mathalpha}{stixsymbols}{"68}
\DeclareMathAlphabet{\mathdsl}{U}{bbm}{m}{sl}

\newcommand{\ket}[1]{\big| #1 \big \rangle }
\newcommand{\bra}[1]{ \big \langle #1 \big|}
\newcommand{\braket}[2]{\big\langle #1 \big| #2 \big\rangle}

\newcommand{\R}{\mathcal{R}}
\newcommand{\F}{\mathcal{F}}

\newcommand{\h}{\mathsf{h}}
\newcommand{\e}{\mathsf{e}}
\newcommand{\f}{\mathsf{f}}
\newcommand{\Q}{\mathsf{Q}}
\newcommand{\sfi}{\mathsf{i}}
\newcommand{\sfj}{\mathsf{j}}
\newcommand{\sfa}{\mathsf{a}}

\newcommand{\thtsc}{ {\color{black}\frac{\sqrt{\lambda}}{2\pi}}}

\newcommand{\tithtsce}{ {\color{black}\tfrac{2\pi \eta}{\sqrt{\lambda}}}}
\newcommand{\ithtsce}{ {\color{black}\frac{2\pi \eta}{\sqrt{\lambda}}}}

\providecommand{\href}[2]{#2}

\makeatletter
\def\mr@ignsp#1 {\ifx\:#1\@empty\else #1\expandafter\mr@ignsp\fi}
\newcommand{\multiref}[1]{\begingroup%
\xdef\mr@no@sparg{\expandafter\mr@ignsp#1 \: }%
\def\mr@comma{}\def\mr@dash{-}%
\@for\mr@refs:=\mr@no@sparg\do{%
\ifx\mr@refs\mr@dash\def\mr@comma{}--\else%
\mr@comma\def\mr@comma{,}\ref{\mr@refs}\fi}%
\endgroup}
\renewcommand{\eqref}[1]{(\multiref{#1})}
\makeatother

\makeatletter
\newcommand{\namedref}[2]{\hyperref[#2]{#1~\ref*{#2}}}
\newcommand{\secref}{\@ifstar{\namedref{Section}}{\namedref{section}}}
\newcommand{\appref}{\@ifstar{\namedref{Appendix}}{\namedref{appendix}}}
\newcommand{\tabref}{\@ifstar{\namedref{Table}}{\namedref{table}}}
\newcommand{\figref}{\@ifstar{\namedref{Figure}}{\namedref{figure}}}
\newcommand{\fref}{\@ifstar{\namedref{Footnote}}{\namedref{footnote}}}
\makeatother

\newcommand{\tts}[1]{\text{\tiny #1}}

\makeatletter
\let\@keywords\@empty
\let\@subject\@empty
\providecommand{\keywords}[1]{\gdef\@keywords{#1}}
\providecommand{\subject}[1]{\gdef\@subject{#1}}
\def\thetitle{\@title}
\def\theauthor{\@author}
\def\thesubject{\@subject}
\def\thedate{\@date}
\def\thekeywords{\@keywords}
\makeatother
\AtBeginDocument{
\hypersetup{pdftitle={\thetitle}}
\hypersetup{pdfauthor={\theauthor}}
\hypersetup{pdfsubject={\thesubject}}
\hypersetup{pdfkeywords={\thekeywords}}}

\definecolor{Gray}{gray}{0.9}

\newif\ifshownote
\shownotetrue

\ifpublic\shownotefalse\fi

\ifshownote

\ifpdf\else\RequirePackage[active]{srcltx}\fi
\RequirePackage{xcolor}
\RequirePackage{ulem}

\newcommand{\remark}[2][]{{\normalfont\sffamily\hspace{1ex}
\def\emph{\textsl}
\def\textbullet{$\bullet$}
\def\tmparga{#1}
\def\tmpargb{BH}\ifx\tmparga\tmpargb\color[rgb]{0.7,0,0}\fi
\def\tmpargb{}\ifx\tmparga\tmpargb\normalfont\color{red}\fi
\def\tmpargb{}\ifx\tmparga\tmpargb\else \textbf{#1:}\fi
#2\hspace{1ex}}}

\else

\newcommand{\remark}[2][]{\ignorespaces}

\fi

\usepackage[
  hidelinks,
  bookmarks=true,
  bookmarksnumbered=true,
  bookmarksopen=true,
  unicode
]{hyperref}
\usepackage{bookmark}

\hypersetup{
  pdfpagemode=UseOutlines,
  plainpages=false,
  pdfstartview=FitH,
  colorlinks=false,
  citebordercolor={1 1 1},
  urlbordercolor={1 1 1},
  linkbordercolor={1 1 1}
}

\hypersetup{
  pdfpagemode=UseOutlines,  
  plainpages=false,
  pdfstartview=FitH,
}

\let\oldbib=\thebibliography
\def\thebibliography{%
  \phantomsection
  \addcontentsline{toc}{section}{\refname}%
  \oldbib}

\title{\Large Jordanian  spin chains for twisted strings in $AdS_5\times S^5$}
\author{\large Sibylle Driezen~~and~~Adrien Molines}

\begin{document}

\thispagestyle{empty}


\vspace*{2cm}
\begin{center}
\begingroup\Large\bfseries\thetitle\par\endgroup
\vspace{1cm}

\begingroup\theauthor\par\endgroup
\vspace{1cm}

\textit{Institut f\"ur Theoretische Physik, \\
Eidgen\"ossische Technische Hochschule Z\"urich,\\
Wolfgang-Pauli-Stra\ss e 27, 8093 Z\"urich, Switzerland}
\vspace{5mm}


\vspace{2cm}

\textbf{Abstract}\vspace{5mm}

\begin{minipage}{14cm}
We study the proposed integrable spin chain formulation of  Jordanian  
deformations of the $AdS_5\times S^5$ superstring,  realised via Drinfel'd twists. Among these models, we first identify a unique supergravity deformation   confined to an $SL(2,\mathbb{R})$ sector and with constant dilaton. We then develop a general framework for closed Drinfel'd twisted spin chains and construct an explicit map to undeformed models with twisted-boundary conditions. Applied to the non-compact $\mathrm{XXX}_{-1/2}$ spin chain, the Jordanian twist breaks the Cartan generator labelling magnon excitations, obstructing the standard Bethe methods.
 Instead, using the  twisted-boundary formulation, we initiate the spectral problem  based on a residual root generator both in the continuum limit and for short chains.
We find that the ground state is non-trivially deformed, and is in agreement with the classical string result, while our analysis does not capture higher-spin excited states.  
We also  study  the asymptotics of the associated $Q$-system, which is well-behaved and compatible with the boundary-twist. 
While a full  understanding of the spectrum remains open, our work provides concrete steps toward a spectral description of non-abelian twisted integrable models.

\end{minipage}

\end{center}

\vfill
\noindent\rule{0.36\linewidth}{0.4pt}\\[0.5ex]
\begingroup\footnotesize\ttfamily
 ~~\texttt{sdriezen@phys.ethz.ch} \\ 
 ~~\texttt{amolines@phys.ethz.ch}
\endgroup


 \cleardoublepage
  \phantomsection
  \bookmark[dest=toc,level=0]{\contentsname}
  \tableofcontents

\section{Introduction}\label{sec:intro}

Integrable deformations of   superstring theories provide a rich setting for probing the structure of the holographic correspondence beyond the well-established case of type IIB strings on $AdS_5\times S^5$. The planar dilatation operator of the  ${\cal N}=4$ Super Yang-Mills (SYM) dual acts as the Hamiltonian of a quantum integrable spin chain, whose excitations encode the spectrum of anomalous dimensions. The same spectral problem appears on the string side, where  the worldsheet theory is described by an integrable sigma-model. These two  formulations overlap in the limit of long (continuous) spin chains and semi-classical strings, also called the Landau-Lifshitz regime, while their integrability is underpinned by the same quantum group structure and associated $R$-matrices.  
See \cite{Beisert:2010jr,Bombardelli:2016rwb} for overviews. 


Within this framework, Homogeneous Yang–Baxter (HYB) deformations \cite{Klimcik:2008eq,Delduc:2013fga,Delduc:2013qra} offer a controlled way of deforming both the string geometry and the symmetry algebra while preserving integrability. 
The resulting deformed sigma-model is defined by a classical $R$-matrix, which up to equivalence relations corresponds one-to-one with Drinfel’d twists \cite{drinfeld1983constant,Giaquinto:1994jx}, inducing a twisted classical  Yangian symmetry structure \cite{vanTongeren:2015uha}.
When the $R$-matrix satisfies the additional unimodularity condition, the deformed background continues to solve the type IIB supergravity equations of motion \cite{Borsato:2016ose}.  HYB deformations can also be understood as non-local canonical transformations on the worldsheet \cite{Vicedo:2015pna}, and therefore admit an  equivalent formulation (at least on-shell) to an undeformed string sigma-model subject to twisted-boundary conditions  \cite{Frolov:2005dj,Matsumoto:2015jja,Vicedo:2015pna,vanTongeren:2018wek,Borsato:2021fuy}. For explicit local expressions implementing the twist, see \cite{Frolov:2005dj,Borsato:2021fuy,Borsato:2022drc}. 
Together, these features suggest the existence of a 
 holographic interpretation  that is governed by a Drinfel'd twisted integrable structure. On the gauge theory side, the dual deformation is  expected to manifest via a (typically non-commutative) star product \cite{Seiberg:1999vs} induced by the Drinfel’d twist \cite{vanTongeren:2015uha,vanTongeren:2016eeb,Araujo:2017jkb}, while the same twist acts on the underlying symmetry algebra of the planar dilatation operator, which then should yield a Drinfel’d-twisted spin chain description. 
This correspondence has also been illustrated  through brane constructions involving successive T- and S-duality transformations \cite{vanTongeren:2015uha}.

The integrable structure of \textit{abelian} HYB deformations, which correspond to T-duality-shift-T-duality (TsT) transformations \cite{Osten:2016dvf}, is relatively well understood in the holographic context, see e.g.~\cite{Beisert:2005if,McLoughlin:2006cg,deLeeuw:2012hp,Kazakov:2018ugh,Guica:2017mtd,Meier:2023kzt,Meier:2023lku,Meier:2025tjq} among many others. 
Notable examples include the  $\beta$-deformation \cite{Leigh:1995ep,Lunin:2005jy}   and the dipole deformation \cite{Bergman:2000cw,Alishahiha:2003ru}, which break internal and spacetime symmetries respectively. 
The twist in this case is of Drinfel'd-Reshetikhin type \cite{Reshetikhin:1990ep},  constructed from commuting symmetry generators of $\mathfrak{psu}(2,2|4)$, which affects the physics of the model only in their non-zero charge sectors. Inclusion of such twists on the canonical $\mathrm{XXX}$ spin chain has elucidated 
the completeness problem of the Bethe root structure and the construction of $Q$-operators, see e.g.~\cite{Bazhanov:2010ts},  which are foundational to advanced integrable methods such as the Quantum Spectral Curve.

By contrast, \textit{non-abelian} HYB deformations,  particularly those of Jordanian type, remain much less explored. 
These deformations necessarily act on a non-compact $\mathfrak{sl}(2,\mathbb{R})$ subalgebra of $\mathfrak{psu}(2,2|4)$  and are thus natural candidates for studying symmetry-breaking effects in this sector. In particular, they lead to  new integrable systems which are relevant for non-AdS holography. At present, however, little is known about their explicit  gauge theory dual or spin chain formulation. 
An exception is the recent work \cite{Borsato:2025smn}, which studied the spectrum of the Jordanian Drinfel'd twisted \cite{Gerstenhaber:1992,Ogievetsky:1992ph,Kulish2009}   $\mathrm{XXX}_{-1/2}$ spin chain relevant to the  $\mathfrak{sl}(2,\mathbb{R})$ sector  at one-loop \cite{Beisert:2003jj,Beisert:2003yb}. In \cite{Borsato:2025smn}, using an organisation  in terms of $\mathfrak{sl}(2,\mathbb{R})$ Cartan states,   the spectrum was found to be  undeformed---a result which appears to conflict with the classical string energy and its corrections computed for one of the simplest deformed sigma-model solutions  \cite{Borsato:2022drc}. Furthermore, on the string theory side, also the integrable S-matrix framework remains elusive: recent results point  to particle production processes involving  massive asymptotic states \cite{Borsato:2024sru}, which appear incompatible with   Bethe Ansatz approaches based on factorised S-matrices.\footnote{See also \cite{deLeeuw:2025sfs} for a related model with soft production} These features indicate substantial challenges for the use of Drinfel'd twisted integrability  in non-AdS holography. 

In this work,  we revisit the spin chain  description aligned with  Jordanian sigma-models, aiming to clarify their physical properties and integrable structure within the context of AdS/CFT. 
Motivated by the conjectured Drinfel'd twisted extension of the correspondence, we still expect that the continuum  (Landau-Lifshitz) limit of the twisted spin chain reproduces the deformed semi-classical string spectrum, in parallel with the undeformed case.  
However, as the precise formulation of the gauge-theory dual remains unclear, we do not attempt a direct identification of the corresponding spin chain from first principles here. 
Instead, to make progress, we  start directly from the
  Jordanian-twisted chain  and treat it as an integrable ``avatar'' of the twisted dilatation operator, that we will use to test and organise the spectral structure for holographic Jordanian models. As we will show, this identification is indeed supported in the Landau-Lifshitz regime.

We begin with a review of the sigma-model side, and perform a systematic analysis of the dilaton profiles of all unimodular Jordanian deformations of $AdS_5\times S^5$ which have been classified in \cite{Borsato:2022ubq}.  Among these,  we identify a unique  Jordanian twist with constant dilaton and which is non-trivial only in an $\mathfrak{sl}(2,\mathbb{R})$ subalgebra through a single deformation parameter. In  the context of holography, this model is therefore the simplest Jordanian example with a valid supergravity approximation  at large 't Hooft coupling.   As it turns out, this ``minimal'' example was already studied from various angles: it can be obtained from combinations of TsT and S-dualities \cite{Matsumoto:2014ubv}, allowing a brane construction  \cite{vanTongeren:2015uha}, its semiclassical energy corrections were derived from the spectral curve in \cite{Borsato:2022drc}, and its scattering processes in light-cone gauge was analysed in \cite{Borsato:2024sru}. It shares also several important features with the (abelian) dipole deformation studied in \cite{Guica:2017mtd,Ouyang:2017yko}, and we will therefore compare to this case where appropriate throughout this paper.  
We will discuss its residual (Schr\"odinger-type) symmetry structure, and discuss its simplest classical solution (a BMN-like string) both in the deformed and twisted-boundary formulations.

In the rest of this paper, we  consider the twisted spin chain. Before specifying to the Jordanian case, we construct a systematic framework for integrability studies of closed Drinfel'd twisted spin chains without specifying Hamiltonian nor $R$-matrix. Building on the Hopf-algebraic formulation of Maillet and de Santos \cite{Maillet:1996yy}, we derive universal expressions of the twisted monodromy and transfer matrices, and propose a simple ``gluing'' prescription that maps spin chains with deformed bulk interactions under periodic boundary conditions to explicit boundary-twist operators on an undeformed chain.  This parallels the on-shell equivalence of sigma-models with Drinfel’d twists and twisted boundary conditions \cite{Frolov:2005dj,Borsato:2021fuy}. 

We then apply our construction to  the Jordanian Drinfel'd twist of the $\mathfrak{sl}(2,\mathbb{R})$-invariant  $\mathrm{XXX}$ Heisenberg spin chain in the non-compact $s=-1/2$ representation. We examine the modified symmetry structure: the twist breaks the full $\mathfrak{sl}(2,\mathbb{R})$ algebra, leaving only a non-Cartan generator, which we use to study the spectral problem (in contrast to \cite{Borsato:2025smn}).  Notably, the boundary-twist is also built from this residual generator. Its eigenbasis departs from the conventional Fock-space picture of magnon-like excitations.  Importantly, in this new eigenbasis,  integrability techniques such as the Algebraic Bethe Ansatz (ABA) appear to be  not applicable in the standard way, because there is no suitable pseudovacuum state with non-zero charge under the residual symmetry to initiate it. However, we find that working particularly in the twisted-boundary formulation does enable an initial and non-trivial spectral analysis. 

In the continuum or thermodynamic limit (where the spin chain length $J\rightarrow \infty$), we derive the Landau-Lifshitz theory via the  coherent state representation  and show that the ground state energy shift matches the classical string result. This not only confirms a non-trivial deformation but also provides  a first string-spin agreement in this overlapping regime.  At finite length, we  study  the spectral problem at $J=2$, by reducing the twisted transfer matrix eigenvalue equation to a singular ordinary differential equation (ODE) in the symmetry eigenbasis. Using Frobenius-type methods, we compute the eigenvalue perturbatively in the deformation parameter from  recurrence relations. Importantly, this ODE analysis  obscures the higher-spin excitations by regularity conditions. Only the ground state explicitly appears and is   deformed, and we compute its energy using Baxter's $TQ$-equation. 
 
 We further initiate the exact finite-length spectral analysis by studying the asymptotic analytic structure of the $TQ$ and  $QQ$-system, providing additional support for the twisted-boundary formulation and setting the first steps to a full spectral curve description in this non-abelian, non-Cartan twisted context.

The paper is structured as follows. Section \ref{sec:HYB} reviews HYB deformations of $AdS_5\times S^5$,  introduces the minimal Jordanian background,  and describes its classical BMN-like solution and associated  Landau-Lifshitz limit. Section \ref{sec:chain} develops the Drinfel'd twisted framework,  including our gluing construction, and applies it to the Jordanian-twisted $\mathfrak{sl}(2,\mathbb{R})$ chain. 
In section \ref{sec:landau-lifshitz} we match the  energy of the spin chain to the classical string via coherent states in the thermodynamic Landau-Lifshitz limit, while section \ref{sec:small-chain} tackles the ground state spectrum of the length $J=2$ spin chain via Baxter’s equation. Section \ref{subsec:twisted_BCs_Baxter_eq} studies $Q$-function asymptotics at finite length and the structure of the $QQ$-system. We conclude in section \ref{sec:concl} with a discussion  and future directions. Appendix \ref{app::classical_bcs} discusses the classical limit of the twisted transfer matrix, and appendix \ref{app:dif_eq} describes the perturbative  method to solve singular ODEs used in the $J=2$ analysis.

\section{Jordanian  string sigma-models}\label{sec:HYB}
Here we review the essential elements of HYB deformations, 
and set the stage for a comparison with the proposed  spin chain formulation for its simplest relevant Jordanian subclass.

\subsection{Review of the action and twisted symmetries}\label{s:HYB-intgb}
The HYB deformations form a large family of integrable deformations of symmetric and semi-symmetric space sigma models. Their worldsheet action takes the form \cite{Klimcik:2008eq,Delduc:2013fga,Delduc:2013qra}
\[ \label{eq:S-WS-HYB}
S_{\tts{WS}} = -\frac{\sqrt{\lambda}}{4\pi} \int \mathrm{d} \tau\mathrm{d} \sigma \left(\frac{\sqrt{\vphantom{h}{|h|}} h^{\alpha\beta}-\epsilon^{\alpha\beta}}{2} \right) \STr \left(  J_\alpha \hat{d}_- (1-\eta R_g \hat{d}_-)^{-1} J_\beta \right) , 
\]
where $\tfrac{\sqrt{\lambda}}{4\pi}$ is the string tension,  $h^{\alpha\beta}$ is the worldsheet metric,  $\epsilon^{\alpha\beta}$ is the antisymmetric tensor with conventions $\epsilon^{\tau\sigma} = - 1$, and $\STr$ denotes the supertrace on a $\mathbb{Z}_4$-graded Lie superalgebra $\mathfrak{g} = \mathrm{Lie}(G)$, which we take in this paper to be $\mathfrak{g}=\mathfrak{psu}(2,2|4)$. We introducted the Maurer-Cartan form $J= g^{-1}dg$, with $g(\tau, \sigma)\in G$ the worldsheet embedding, and the operators $\hat{d}_{\pm} = \mp \frac{1}{2}P^{(1)}+P^{(2)} \pm\frac{1}{2}P^{(3)}$ that act with the projections $P^{(i)}$, $i=0,1,2,3$,  onto the corresponding   $\mathbb{Z}_4$-graded eigenspaces. Last but not least, the key object inducing the deformation is $R_g := \Ad_g^{-1} R \Ad_g$, scaled  by a real deformation parameter $\eta\in\mathbb{R}$, where $\Ad_g X := g X g^{-1}$ for $X\in\mathfrak{g}$, and $R:\mathfrak{g} \rightarrow \mathfrak{g}$ is a linear operator satisfying two properties:   $R$ is antisymmetric with respect to $\STr$ and it solves the classical Yang-Baxter equation (CYBE), i.e.~respectively
\[ \label{eq:R-AS-CYBE}
\STr (RX~Y) = - \STr(X~RY) , \qquad [RX,RY]-R([RX,Y]+[X,RY]) = 0  ,
\]
for all $X,Y \in \mathfrak{g}$. Such objects are in one-to-one correspondence (up to equivalence relations) with Drinfel'd twists that are continuously connected to the identity \cite{drinfeld1983constant,Giaquinto:1994jx}. In this context, they explicitly ensure that the sigma-model remains  classically Lax integrable \cite{Klimcik:2008eq,Delduc:2013fga,Delduc:2013qra,Delduc:2014kha}. 
We further assume that the undeformed model $(\eta=0)$ gives rise to a solution of the type II supergravity equations, as is the case for the $AdS_5\times S^5$ superstring that is realised from $\mathfrak{g}=\mathfrak{psu}(2,2|4)$. A further condition on $R$ is then needed for the deformed model to remain a type II supergravity solution, i.e.~$R$ must solve the linear ``unimodularity'' constraint \cite{Borsato:2016ose}
\[
[T_A , RT^A] = 0 ,
\]
where $\{T_A\}$ and $\{T^A\}$ with $A=1, \cdots ,\mathrm{dim}(G)$ are dual bases with respect to the supertrace.\footnote{Given a basis of generators $T_A$ for $\mathfrak{g}$,  with structure constants defined by $[T_A, T_B] = F_{AB}{}^C T_C$, let the invariant metric be $K_{AB} = \STr (T_A T_B)$, and its inverse $K^{AB}$ used to define the dual basis $T^A = K^{AB} T_B$.  Writing $R T^A = R_B{}^A T^B$, the unimodularity conditions become $R^{AB}F_{AB}{}^C =0$, with  $R^{AB} = R^A{}_{C} K^{CB}$.
 }

We can equivalently represent  $R$  via the two-fold graded wedge product $r = - \frac{1}{2} R^{AB} T_A \wedge T_B$. Our focus in this paper is on rank-2 unimodular Jordanian deformations,\footnote{The rank denotes the number of bosonic generators in the construction of the $r$ wedge product.} which are of the form  \cite{2004Tolstoy,vanTongeren:2019dlq,Borsato:2022ubq}
\[ \label{eq:def-r-jor}
r_{\tts{jor.}} = \mathsf{h} \wedge \mathsf{e} -  \frac{i}{2} (\mathsf{Q}_1 \wedge \mathsf{Q}_1 + \mathsf{Q}_2 \wedge \mathsf{Q}_2) ,
\]
where $\h, \e$ are bosonic and $\Q_1$, $\Q_2$ fermionic generators in $\mathfrak{g} = \mathfrak{psu}(2,2|4)$ that are such that $[\h,\e]=\e $, $[\Q_{\sfi},\e]=0$, $[\h,\Q_{\sfi}]=\frac{1}{2} (\Q_\sfi - \epsilon_{\sfi \sfj} a \Q_\sfj )$, with $a\in \mathbb{C}$ a free parameter, and $\{ \Q_\sfi , \Q_\sfj \} = - i \delta_{\sfi \sfj} \e$, for $\sfi, \sfj=1,2$. Note that $\e$ acts as a raising operator with respect to $\h$.\footnote{In order to preserve reality conditions, this requires that $\mathfrak{g}$ needs to have a non-compact $\mathfrak{sl}(2,\mathbb{R})$ subalgebra.} 
Importantly, for Jordanian models, the fermionic tail built from $\Q_{\sfi}$ is necessary to ensure unimodularity.  For $\mathfrak{g} = \mathfrak{psu}(2,2|4)$,   the inequivalent Jordanian $R$-operators (including higher-rank ones) admitting unimodular extensions  have been classified in \cite{Borsato:2022ubq}. More details can be found therein. 

Another important subclass of HYB models is based on the abelian $r$-matrices
\[\label{eq:def-r-ab}
r_{\tts{ab.}} = \mathsf{t}_1 \wedge \mathsf{t}_2 ,
\]
where $\mathsf{t}_1 , \mathsf{t}_2$ are commuting generators of $\mathfrak{psu}(2,2|4)$. These deformations are equivalent to TsT transformations \cite{Lunin:2005jy} over the isometries generated by $\mathsf{t}_1$ and $\mathsf{t}_2$ \cite{Osten:2016dvf} and they are trivially unimodular. 
Their integrability in the context of $AdS_5/CFT_4$ have been discussed  for various examples.  The simplest case is the Lunin-Maldacena background \cite{Lunin:2005jy} obtained from an $r$ wedge product over two Cartan charges of $SO(6)$, and thus deforming the $S^5$, dual to the $\beta$-deformation of $N=4$ SYM \cite{Leigh:1995ep}. Other interesting examples are the Hashimoto-Itzhaki-Maldacena-Russo background \cite{Hashimoto:1999ut,Maldacena:1999mh}, from two commuting generators in $SO(2,4)$, and the Schr\"odinger background \cite{Alishahiha:2003ru,Maldacena:2008wh,Herzog:2008wg,Adams:2008wt}, from one (null) generator in $SO(2,4)$ and one Cartan in $SO(6)$,   both deforming $AdS_5$ and thus introducing non-commutativity in the dual field theory \cite{Seiberg:1999vs} (for the latter in terms of a dipole deformation, see \cite{Bergman:2000cw,Alishahiha:2003ru}). Their integrability studies are based on abelian Drinfel'd-Reshetikhin (DR) twists \cite{Reshetikhin:1990ep}, see \cite{Beisert:2005if} and \cite{Guica:2017mtd} for the $\beta$ and dipole deformation respectively. 

\noindent {\bf Manifest symmetries ---} The residual (manifest) symmetries of the deformed sigma-model \eqref{eq:S-WS-HYB} are generated by the subalgebra $\mathfrak{k}=\mathrm{span}(T_{\bar{A}} ) \subset \mathfrak{psu}(2,2|4)$ defined by
\[
\ad_{T_{\bar{A}}} R (X) = R \ad_{T_{\bar{A}}} (X) , \qquad \forall X \in \mathfrak{g} ,
\]
with $\ad_X Y := [X,Y]$. This guarantees invariance of the action \eqref{eq:S-WS-HYB} under left multiplication
\[
g\rightarrow \exp (\epsilon T_{\bar{A}}) g ,
\]
with $\epsilon$ a global symmetry parameter. 
For all Jordanian $R$-operators  \eqref{eq:def-r-jor},  it then follows that $\e$ always generates a global symmetry.\footnote{It is straightforward to show this using $R(X) =  \STr_2 ( r(1\otimes X))$ for $X\in\mathfrak{g}$ as well as the above commutation relations. 
} 
This fact will be important in several parts  of our analysis. 

\noindent {\bf Classical integrability and twisted symmetries ---} 
Provided the conditions \eqref{eq:R-AS-CYBE} are met, HYB deformations preserve classical worldsheet integrability. This means that their equations of motion can be represented by a one-parameter family of flat  Lax connection one-forms ${\cal L}(z)$, which ensures the construction of an infinite family of conserved charges, and that these charges Poisson commute \cite{Klimcik:2008eq,Delduc:2013fga,Delduc:2013qra,Delduc:2014kha}. 
In conformal gauge, the Lax can be written as\footnote{In the rest of this section, we adapt conformal gauge to keep notation light. One may, however, translate the formulas back to prior gauge-fixing by replacing one-form components $V_\pm$ with $V_+ \rightarrow P^{\alpha\beta}_{(+)} V_\beta$ and $V_- \rightarrow P_{(-)\alpha}{}^\beta V_\beta$ with $ P^{\alpha\beta}_{(\pm)} = \tfrac{1}{2} (\sqrt{\vphantom{h}{|h|}} h^{\alpha\beta}\pm\epsilon^{\alpha\beta})$ while raising and lowering  worldsheet indices with $\sqrt{\vphantom{h}{|h|}} h^{\alpha\beta}$.}
\[
{\cal L}_\pm(z ) = A_\pm^{(0)} + z A_\pm^{(1)} +  z^{\pm 2} A_\pm^{(2)} + z^{-1} A_\pm^{(3)} , \qquad z\in \mathbb{C}, 
\]
where $A^{(i)}:= P^{(i)} A$ and
\[
A_\pm = \left( 1 \pm \eta R_g \hat{d}_- \right)^{-1} J_\pm . 
\]
An interesting property of \textit{Homogeneous} Yang-Baxter deformations is that, on-shell, they are classically equivalent to the undeformed sigma-model up to a twist in the boundary conditions 
\cite{Frolov:2005dj,Matsumoto:2015jja,Vicedo:2015pna,vanTongeren:2018wek,Borsato:2021fuy,Borsato:2022drc} (see also \cite{Kawaguchi:2013lba} for early considerations in the Jordanian context). This is due to the fact that the deformation can be understood as a non-local canonical transformation on the worldsheet \cite{Vicedo:2015pna}, which is precisely realised as the map
$J_\pm \rightarrow A_\pm$. 
Formally, up to gauge transformations generated by $\mathfrak{g}^{(0)}$, one may therefore identify
\[ \label{eq:Jt-to-A}
 \tilde{J}_\pm := \tilde{g}^{-1} \partial_\pm \tilde {g} \equiv h A_\pm h^{-1} - \partial_\pm h h^{-1} ,
\]
where $\tilde{g} \in G$ represents the collection of fields of the undeformed sigma-model with twisted-boundary conditions and $h\in G^{(0)} = SO(1,4) \times SO(5)$. 
Periodic-boundary solutions to the equations of motion of the fields $g$ of the deformed sigma-model \eqref{eq:S-WS-HYB}  are then mapped to twisted-boundary solutions of the  fields $\tilde{g}$ of the undeformed sigma-model, i.e.~\eqref{eq:S-WS-HYB} at $\eta=0$. 
We can translate between the two formulations with the ``twist field'' ${\cal F}(\tau, \sigma) \in F$ as 
\[ \label{eq:g-to-gt}
g={\cal F} \tilde{g} h ,
\]
with $ F \subset G$  the Lie group of $\mathfrak{f} =\mathrm{Im}(R)$. 
Explicitly, we can  write the twist of the boundary conditions of $\tilde{g}$  as \cite{Borsato:2022drc}
\[ \label{eq:G-tbc}
\tilde{G} (\tau, \sigma = 2\pi) = W \tilde{G} (\tau, \sigma =  0) W^T ,
\]
where $\tilde{G} = \tilde{g} K \tilde{g}^T$ is the gauge-invariant collection of fields, with $K$ the $G^{(0)} = SO(1,4) \times SO(5)$ invariant, and $W = {\cal F}^{-1}(\tau,2\pi) {\cal F}(\tau,0) \in F$ the object that encodes the twisting.

For Jordanian models with $R$-matrix \eqref{eq:def-r-jor}, an explicit local expression for $W$ is given by \cite{Borsato:2021fuy,Borsato:2022drc}
\[ \label{eq:W-F-jor}
W_{\tts{jor.}} = \exp \left(\mathbf{Q} (\h - \mathbf{q} \e ) \right) , \qquad {\cal F}_{\tts{jor.}} = \exp \left(  \tithtsce Y_{\h} \e  \right) \exp\left( \log \left(1- \tithtsce Y_{\e} \right) \h \right) ,
\]
with 
\[\label{eq:HYB_twist_charges}
\mathbf{Q} := \log \left( \frac{1-  \tithtsce  Y_{\e} ( \tau, 0)}{1-  \tithtsce  Y_{\e} ( \tau, 2\pi)} \right) , \qquad \mathbf{q} := \frac{Y_{\h} (\tau, 2\pi) - Y_{\h} (\tau, 0)}{Y_{\e} (\tau, 2\pi) - Y_{\e} (\tau, 0)} ,
\]
and $Y_{\e} = \STr (  \e Y) $, $Y_{\h} = \STr ( \h Y)$ projections of  the Lie-algebra value field\footnote{Here we rescale $Y(\tau,\sigma)$ with $\thtsc$ compared to \cite{Borsato:2021fuy,Borsato:2022drc} to match with conventional expressions for conserved Noether charges (as in e.g.~\eqref{eq:ab-noether}).} 
\[
Y (\tau, \sigma ) := \thtsc P^T\left(\int^\sigma_0 \mathrm{d}\sigma' \hat{d}_{\tilde{g}^{-1}} (\partial_\tau \tilde{g} \tilde{g}^{-1})(\tau , \sigma') + \int^\tau_0 \mathrm{d}\tau' \hat{d}_{\tilde{g}^{-1}} (\partial_\sigma \tilde{g} \tilde{g}^{-1})(\tau' , 0) \right) + Y(0,0) ,
\]
with $\hat{d}_{\tilde{g}^{-1}} := \mathrm{Ad}_{\tilde{g}} \hat{d}_- \Ad_{\tilde{g}}^{-1}$ and $P^T$ a projector on $\mathfrak{f}^\star$, the dual of $\mathfrak{f}=\mathrm{Im}(R)$. 
One may include additional contributions from fermionic degrees of freedom associated to the supercharges $\Q_{\mathsf{i}}$, but it was shown in \cite{Borsato:2022drc} that such twists are equivalent, after field redefinitions, to \eqref{eq:W-F-jor} for classical solutions (i.e.~ones with trivial fermionic dynamics). Furthermore, in on-shell sectors where  $\mathbf{Q}\neq 0$ (which will be the case for the solution that we will study in section \ref{s:string-sol}), an additional field redefinition can actually relate $W_{\tts{jor.}}$ to $W'_{\tts{jor.}}=\exp (\mathbf{Q} \h )$. Interestingly, we find especially the latter expression at large $\lambda$ to be consistent with the spin chain formulation (cf.~sec.~\ref{s:jor-chain}, \ref{subsec:twisted_BCs_Baxter_eq} and app.~\ref{app::classical_bcs}).

For abelian models with $R$-matrix \eqref{eq:def-r-ab}, the local expression for $W$ is much simpler and given by \cite{Borsato:2021fuy,Frolov:2005dj,vanTongeren:2018wek}
\[
W_{\tts{ab.}} = \exp \left(-  \tithtsce \left( \mathbf{\cal Q}_{1}  \mathsf{t}_2 - \mathbf{\cal Q}_{2} \mathsf{t}_1 \right)\right) ,  \qquad {\cal F}_{\tts{ab.}} = \exp \left(  \tithtsce \left( Y_{1}  \mathsf{t}_2 - Y_{2} \mathsf{t}_1 \right)\right) ,
\]
where $Y_{i} = \STr (  \mathsf{t}_i Y) $ and $\mathbf{\cal Q}_{1,2}$ are the Noether charges corresponding to the $\mathsf{t}_{1,2}$ isometries  
\[ \label{eq:ab-noether}
\mathbf{\cal Q}_{i} = \STr ( \mathsf{t}_{i}  (Y (\tau, 2\pi) - Y (\tau, 0) ) =  \thtsc \int^{2\pi}_0 \mathrm{d}\sigma \STr ( \mathsf{t}_{i} \Ad_{\tilde{g}} \tilde{J}_\tau^{(2)}  ) .
\]

Importantly, both $W_{\tts{jor.}}$ and $W_{\tts{ab.}}$ are constant and thus encode conserved quantities (charges) of the respective models. The Jordanian $\mathbf{Q}, \mathbf{q}$ charges do not, however, have an immediate interpretation as Noether charges for a continuous symmetry, in contrast to the $\mathbf{\cal Q}_{1,2}$ appearing in the abelian twists. Furthermore note that generally  the $\mathfrak{psu}(2,2|4)$ Noether symmetries of the undeformed model are explicitly broken by the twist in the boundary conditions. The subalgebra of residual symmetries $\tilde{\mathfrak{t}}=\mathrm{span}(T_{\tilde{A}})$ of the twisted-boundary model, which is determined by $[T_{\tilde{A}}, \mathfrak{f}]=0$, is itself a subalgebra of the residual symmetries $\mathfrak{t}$ of the deformed model, and their local Noether charges coincide on the on-shell mapping  \cite{Borsato:2021fuy,Borsato:2022drc}.

Next to the subset of manifest residual Noether symmetries, HYB deformations have  an infinite set of (hidden) twisted classical Yangian symmetries \cite{vanTongeren:2015uha}. The monodromy matrix that generates the  conserved charges can indeed be written as\footnote{To show the second equality one can use the fact that on-shell $A_\pm \equiv h^{-1}\tilde{J}_\pm h + h^{-1}\partial_\pm h = {\cal G} J_\pm {\cal G}^{-1} - \partial_\pm {\cal G}{\cal G}^{-1}$ with ${\cal G}= h^{-1}\tilde{g}^{-1} g =  g^{-1}{\cal F}g $.}
\[ \nonumber
{\cal T}(z) &=  \overleftarrow{{\cal P} \exp} \left(- \int^{2\pi}_0  \mathrm{d}\sigma {\cal L}_\sigma^g (z)  \right) 
= {\cal F}(\tau, 2\pi) {\cal T}_{\eta=0} (z) {\cal F}^{-1} (\tau, 0 )  
\]
where ${\cal L}^g (z) = g {\cal L}(z) g^{-1} - dg g^{-1}$ and ${\cal T}_{\eta=0} (z)$
is the monodromy matrix of the undeformed sigma-model built 
from the (gauge-transformed) Lax connection  in terms of projections of the currents $J_\pm$. 
We can thus understand the original ($\eta=0$)  classical Yangian symmetry  to be twisted by ${\cal F}$. At large $\lambda$ we get
\begin{alignat}{2}
{\cal F}_{\tts{jor.}} &=  1 -  \ithtsce (Y_{\e} \h - Y_{\h}\e ) + {\cal O}(\lambda^{-1}) &&= 1 -  \ithtsce \iota_Y r_{\tts{jor.}}  + {\cal O}(\lambda^{-1}),\\
{\cal F}_{\tts{ab.}} &= 1-  \ithtsce (Y_{2} \mathsf{t}_1 - Y_1 \mathsf{t}_2 ) + {\cal O}(\lambda^{-1}) &&= 1 -  \ithtsce \iota_Y r_{\tts{ab.}}  + {\cal O}(\lambda^{-1}),
\end{alignat}
with the contraction $\iota_Y$ defined as $\iota_Y r := \STr_2 (r (1\otimes Y)) $.
This precisely represents a classical Drinfel'd twist to leading order \cite{drinfeld1983constant,vanTongeren:2015uha}  with $\iota_Y$  the action of the operators (charges) on the classical fields $\tilde{g}$.  We thus see  explicitly how the deformed sigma-model inherits a twisted Hopf-algebra structure at the classical level. In a holographic interpretation, the twisting may be interpreted as a $\tfrac{1}{\sqrt{\lambda}}$ correction. 

At various instances in this paper, we will absorb the 't Hooft coupling $\lambda$ in the (sigma-model) deformation parameter $\eta$ as
\[ \label{eq:xi-to-eta}
\xi := \ithtsce .
\]
In particular, $\xi\in \mathbb{R}$ will then denote the deformation parameter in the Drinfel'd twist which will be deforming the relevant spin chain model, as we will discuss in section \ref{sec:chain}.

\subsection{Singling out the ``minimal'' Jordanian model} \label{s:min-jor}
In this section, we briefly analyse the dilaton $\Phi$ of all rank-2 unimodular Jordanian deformations of $\mathfrak{g}=\mathfrak{psu}(2,2|4)$. We focus on identifying those backgrounds for which $\Phi$ is a constant across target space, motivated from  the simple fact that this implies a constant string coupling $g_s $, such that the classical supergravity approximation from the worldsheet is valid.

For HYB models, the dilaton $\Phi$ is given by \cite{Borsato:2016ose}
\[ \label{eq:dil-HYB}
e^{-2 \Phi} = e^{-2 \Phi_0}~\mathrm{sdet} \left( 1+ \eta R_g \hat{d}_+ \right) .
\]
where $\Phi_0$ is the constant dilaton of the undeformed $AdS_5\times S^5$ background.
We will  compute $\Phi$ for all  rank-2 bosonic Jordanian models of the $\mathfrak{so}(2,4)$ subalgebra that admit unimodular extensions in $\mathfrak{psu}(2,2|4)$, as  listed in \cite{Borsato:2022ubq} (cf.~table 2 or 6 therein for notation and definitions).
For this purpose, we will take the following (bosonic) coset representatives: For $\bar{R}_{1}, \bar{R}_{1'},\bar{R}_{2}, \bar{R}_{2'},\bar{R}_{3}, \bar{R}_{4},\bar{R}_{5}, \bar{R}_{5'}, $ we use
\[ \label{eq:g-par-schr}
g = g_{\mathfrak{a}}  \cdot g_{\mathfrak{s}}, \qquad g_{\mathfrak{a}} = e^{T H_T + V H_V + \Theta H_\Theta} e^{P p_1} e^{\log(Z) D} ,
\]
where $g_{\mathfrak{s}}\in SO(6)$ parametrises the round $S^5$ sphere, 
and
\[
H_T = \frac{1}{2} (p_0 - k_0 - p_3 - k_3) , \qquad H_V = \e= p_0 + p_3  , \qquad H_{\Theta} = J_{12} ,
\]
are commuting generators of $\mathfrak{psu}(2,2|4)$, with $D$, $J_{ij}$, $p_{i}$, and $k_{i}$, $i=0,1,2,3$,  the standard dilatation, rotation, translation, and special conformal generators of the conformal $\mathfrak{so}(2,4)$ subalgebra.  For $\bar{R}_{6}, \bar{R}_{6'}$, we instead take  
\[
 g = g'_{\mathfrak{a}}  \cdot g_{\mathfrak{s}}, \qquad g'_{\mathfrak{a}} = e^{V H'_V + \Theta H_\Theta} e^{P_1 p_1 + P_2 p_3} e^{\log(Z) D} ,  
 \]
 with $H'_V = \e= p_0$.
Both $g_{\mathfrak{a}}$ and $g'_{\mathfrak{a}}$ 
are faithful representatives
of $SO(2,4)/SO(1,4)$, with local coordinates $X^m = (T, V, \Theta, P, Z)$ and 
$X^m = (V, \Theta, P_1, P_2, Z)$, $m=1,\ldots, 5$, respectively. 
They are adapted in such a way that shifting $V$ is a  manifest isometry ($\e$) of the corresponding backgrounds. 
We then obtain
\begin{alignat}{2}
\bar{R}_{1}, \bar{R}_{1'}~&: \qquad &&e^{-2(\Phi-\Phi_0)} = 1-\frac{\eta^2(1 + 2b)^2   \sin^2 2T}{Z^4} ,  \\
\bar{R}_{2}, \bar{R}_{2'}~&: \qquad &&e^{-2(\Phi-\Phi_0)} = 1- \frac{16 \eta^2  \cos^4 T }{Z^4} , \\
 \bar{R}_3, \bar{R}_4~&: \qquad &&e^{-2(\Phi-\Phi_0)} = 1-\frac{ \eta^2  \sin^2 2T}{Z^4} , \\
 \bar{R}_5, \bar{R}_{5'}~&: \qquad &&e^{-2(\Phi-\Phi_0)} = 1- \frac{16 \eta^2 b^2  }{Z^4} ,\\
 \bar{R}_6~&: \qquad &&e^{-2(\Phi-\Phi_0)} = \left(1- \frac{4 \eta^2 ((1+a^2)P_1^2 + P_2^2 + Z^2)}{Z^4} \right) \mathcal{D} , \\
  \bar{R}_{6'}~&: \qquad &&e^{-2(\Phi-\Phi_0)} = \left(1- \frac{4 \eta^2 (P_1^2 + P_2^2 + Z^2)}{Z^4} \right) \mathcal{D} ,
\end{alignat}
where $a, b$ are free, real parameters, and ${\cal D}$ is a complicated function depending on the sphere parametrisation, but which, importantly is independent of the coordinates $P_1,P_2$ as well as the free parameter $a$ entering the definition of the Jordanian $\bar{R}_6$-matrix.  

We thus conclude that only $\bar{R}_{1}, \bar{R}_{1'}$ on $b=-\frac{1}{2}$ and $\bar{R}_{5}, \bar{R}_{5'}$ on $b=0$ yield constant dilatons. All of these share the same bosonic $\h$ and $\e$ generator, namely
$\h = \frac{D-J_{03}}{2} + a J_{12}$ and $\e = p_0 + p_3$,
with $a=0$  for $\bar{R}_{1}, \bar{R}_{5}$, and generic $a$ for $\bar{R}_{1'}, \bar{R}_{5'}$. 
Notably, turning on $a$ corresponds to an additional TsT deformation along the residual bosonic isometries $J_{12}$ and $\e$ of the $a=0$ deformed background, see e.g.~\cite{Driezen:2024mcn} for more comments. Furthermore, the $a=0$ case preserves 12 supercharges, the maximal number found in \cite{Borsato:2022ubq}, while turning on $a$ breaks all supersymmetry. Both  cases are precisely the ones for which the spectrum has been studied semi-classically using spectral curve methods in  \cite{Borsato:2022drc} and \cite{Driezen:2024mcn}, respectively.

We therefore define the ``minimal'' Jordanian model to be the unimodular rank-2 model with 
\[ \label{eq:def-jor-min}
\h = \frac{D-J_{03}}{2} , \qquad \e = p_0 + p_3 ,
\]
ensuring both a valid supergravity background with constant dilaton, as well as maximal preserved supersymmetry. For its unimodular extension, see \cite{Borsato:2022ubq}.  Since we will focus on a classical bosonic string solution in what follows, let us give here  the resulting deformed target space metric and $B$-field extracted from  \eqref{eq:S-WS-HYB} with the above choice, and in the coordinates \eqref{eq:g-par-schr}. One gets 
\[ \label{eq:jor-G-B}
ds^2 &= \frac{dZ^2 + dP^2 + P^2 d\Theta^2 - 2 dT dV}{Z^2} - \frac{(4 Z^4+\eta^2) (Z^2+P^2)}{4 Z^6}dT^2 + ds_{S^5}^2 , \\
B &= \eta \frac{P dP \wedge dT}{2 Z^4}  - \eta d\left(\frac{dT}{4 Z^2}\right) ,
\]
with $ds_{S^5}^2$ the metric of the round $S^5$. 
The full type IIB supergravity background is also supported by a non-trivial $F_3$ and $F_5$ flux;  see e.g.~eq.~7 of \cite{Driezen:2024mcn}, after sending $a\rightarrow 0$ and $\eta \rightarrow \eta/\sqrt{2}$, for explicit expressions in these coordinates and  \cite{Kawaguchi:2014fca,Matsumoto:2014ubv,vanTongeren:2019dlq} for  original constructions.
The spacetime was shown to be geodesically complete in \cite{Borsato:2022drc} and exhibits features of a Schr\"odinger $Sch_2$ geometry \cite{Blau:2009gd}. The residual $AdS$ Noether symmetries of the deformed model are
\[ \label{eq:j-AdS-isoms}
\mathfrak{t}_{\mathfrak{a}} = \mathrm{span} (D+J_{03}, k_0+k_3 , p_0 - p_3 , p_0 + p_3 , J_{12}) \cong \mathfrak{sl}(2,\mathbb{R}) \oplus \mathfrak{u}(1)^2 ,
\] 
where the $\mathfrak{sl}(2,\mathbb{R})$ subalgebra encodes the non-relativistic CFT characteristics of Schr\"odinger-invariant field theories and is therefore a convenient language to describe the residual symmetry structure \cite{Nishida:2007pj,Son:2008ye}. The $\mathfrak{u}(1)^2$ part are the two central elements $H_V=\e=p_0+p_3$ and $H_\Theta = J_{12}$, while $H_T = \tfrac{1}{2} (p_0 - k_0 - p_3 - k_3) \in \mathfrak{sl}(2,\mathbb{R})$ is the only time-like Cartan element up to inner automorphisms.\footnote{On the other hand, the residual ($AdS$) Noether symmetries of the twisted-boundary model $\tilde{\mathfrak{t}}_{\mathfrak{a}}$ share   the same $\mathfrak{sl}(2,\mathbb{R})$ algebra and central $J_{12}$ with $\mathfrak{t}_{\mathfrak{a}}$, but it does not contain the central $\e=p_0+p_3$. We can instead understand the latter (at least to first order in $\lambda^{-1}$) as being manifested in the twist $W_{\tts{jor.}}$ \eqref{eq:W-F-jor} instead. \label{f:twisted-noether}} In the deformed bulk spacetime, the isometric coordinate $T$ thus serves as  global time (with conjugate energy $E$ generated by $H_T$),   while $\Theta$ is angular (associated to $AdS$ spin) and  $V$ is a non-compact null direction with conjugate null momentum $M$ (also ``central mass'' in the Schr\"odinger language). Interestingly, in the non-relativistic CFT setup, $D+J_{03}$ defines  non-relativistic scaling dimensions $\Delta$. The generator $k_0+k_3$ acts as a lowering operator and is used to define non-relativistic primary operators, while the generator $p_0-p_3$  defines a  state-operator map, relating  eigenstates of the global time Hamiltonian $H_T$ (bulk strings) to  operators diagonalised by $D+J_{03}$ \cite{Nishida:2007pj}. 
An important takeaway for the holographic spectral problem is thus that one should expect that the relevant symmetry generator  to be diagonalised is not the original dilatation operator $D$, but rather the twisted dilatation $D+J_{03}$.\footnote{Nevertheless, as emphasized in \cite{Guica:2017mtd} and references therein, in the related dipole model a genuine non-relativistic field theory interpretation  arises only after dimensional reduction along  non-local directions of the deformed   four-dimensional QFT dual, and the same should be expected for the Jordanian case.  }

\noindent {\bf Comments on extensions ---} Although it is outside the aim of this paper, one may also 
consider rank-2 bosonic Jordanian  models of the full $\mathfrak{so}(2,4) \oplus \mathfrak{so}(6)$ subalgebra admitting unimodular extensions, as well as 
the higher-rank unimodular Jordanian deformations given in \cite{Borsato:2022ubq}.\footnote{For these cases,   one cannot simply restrict to an $\mathfrak{sl}(2,\mathbb{R})$ subsector and in general one would instead have to match with calculations of the full Drinfel'd twisted $\mathfrak{psu}(2,2|4)$ spin chain.}   However, as explained in \cite{Borsato:2022ubq}, the extensions to $\mathfrak{so}(6)$ can  also be interpreted as additional TsT deformations on top of the rank-2 bosonic $\mathfrak{so}(2,4) $ solutions.\footnote{A key example would be  the Schr\"odinger/dipole deformation considered in \cite{Guica:2017mtd} which is obtained by a TsT along a null generator $\e=p_0+p_3$ and a  Cartan generator in $SO(6)$ (the $R$ charge).} For the higher-rank cases, we simply find, by
restricting to the bosonic sector, that the only rank-4 ones which would give rise to a constant dilaton are $R_7$ on $b=-\frac{1}{2}$ and $a=0$ and $R_9$ on $d_+ = b_- =0$ and $d_- = \frac{b_+}{1+b_+^2}$ with $b_+ \in \mathbb{R}$ free.\footnote{Using an inner automorphism, $R_7$ on $b=-\frac{1}{2}$ and $a=0$ can be seen as a special case of $R_9$ on $d_+ = b_- =0$ and $d_- = \frac{b_+}{1+b_+^2}$, i.e.~for $b_+=0$.} The only rank-6 cases with constant dilaton would be $R_{13}$ on $b=-\frac{1}{2}$ but $a\in \mathbb{R}$ free and $R_{14}$ without any further constraints on the parameters other than those listed in eq.~(2.14) of \cite{Borsato:2022ubq}. Interestingly,  all cases are ultimately built on  $\h = \frac{D-J_{03}}{2} $ and $\e = p_0 + p_3$, which supports the universality of the minimal Jordanian $R$-operator defined in \eqref{eq:def-jor-min}. 

\subsection{Classical solution in $AdS_3\times S^1$ and $SL(2,\mathbb{R})$ Landau-Lifshitz limit} \label{s:string-sol} 
After singling out the minimal Jordanian model \eqref{eq:def-jor-min}, let us now specify a point-like string solution in the bosonic target space background \eqref{eq:jor-G-B}. The simplest case  is  already non-trivial in an $AdS_3\times S^1$ subspace of $AdS_5 \times S^5$  \cite{Borsato:2022drc} (see also \cite{Guica:2017mtd})
\[ \label{eq:BMN-like-sol-per}
T = - \kappa \tau , \qquad V = \frac{\eta^2 }{4} m \tau , \qquad Z = \sqrt{\frac{\kappa}{m} } , \qquad \psi = \omega \tau ,
\]
where $\psi$ is an angle parametrising a big circle in $S^5$ and $\kappa, m, \omega$ are constant parameters. The other coordinates are trivial: in particular, $X_2  := \frac{P \cos\Theta}{Z} =0 $ and  $ X_3 := \frac{P \sin\Theta}{Z} = 0$ in addition to the vanishing of the other two isometric  angles ($\psi_1, \psi_2$) of $S^5$. Hence this solution has no $AdS$ spin ($S=0$) in $\Theta$ and only one angular momentum $J=J_3$ along  $\psi$ (with angular frequency $\omega$), while the other momenta vanish ($J_1=J_2=0$). 
The parameters are related to the Cartan Noether charges as\footnote{In contrast to classical solutions in undeformed $AdS_5$, which are characterised by the energy $E$ and two spins $S_1$, $S_2$, here the solutions are characterised by energy $E$, momentum $M$ and one spin $S=S_2$, as adapted to the Cartan subalgebra of the residual isometries \eqref{eq:j-AdS-isoms}.}
\[
E &= \thtsc \int^{2\pi}_0 \mathrm{d}\sigma \STr (H_T \mathrm{Ad}_g A_\tau^{(2)} ) = \sqrt{\lambda} \kappa, \\
M &= \thtsc \int^{2\pi}_0 \mathrm{d}\sigma \STr (H_V \mathrm{Ad}_g A_\tau^{(2)} ) = \sqrt{\lambda} m, \\
S &= \thtsc \int^{2\pi}_0 \mathrm{d}\sigma \STr (H_\Theta \mathrm{Ad}_g A_\tau^{(2)} ) = 0,  \\
J &= \thtsc \int^{2\pi}_0 \mathrm{d}\sigma \STr (H_\psi \mathrm{Ad}_g A_\tau^{(2)} ) = \sqrt{\lambda} \omega, 
\]
with $H_\psi$ the generator of rotations along $\psi$. The Virasoro constraint then relates them as
\[ \label{eq:Del-BMN-like}
\Delta = \sqrt{J^2 + \frac{\eta^2 M^2}{4}} = \sqrt{J^2 + \lambda\frac{\xi^2 M^2}{16\pi^2}} ,
\]
where we implemented $\Delta \equiv E$ and  \eqref{eq:xi-to-eta}, since we have in mind the holographic interpretation discussed in the previous section, where $\Delta$ is the (twisted) conformal dimension of the operator that should be dual to the solution \eqref{eq:BMN-like-sol-per} with global energy $E$.
Following \cite{Guica:2017mtd} we call this solution ``BMN-like'', as it deforms the BMN vacuum (BPS operator) $\Delta = J$. One may interpret it also as a higher-order  correction in $\lambda$. 

On-shell, the periodic BMN-like solution \eqref{eq:BMN-like-sol-per} is equivalent to a solution of the undeformed model with twisted-boundary conditions, as reviewed in section \ref{s:HYB-intgb}, propagating in the background \eqref{eq:jor-G-B} at $\eta=0$.  For the Jordanian deformation \eqref{eq:def-jor-min}, the mapping is non-trivial only in the $AdS$ sector. Following the transformations \eqref{eq:Jt-to-A} and \eqref{eq:g-to-gt} in the parametrisation \eqref{eq:g-par-schr} (with fields $X^m$ for $g_{\mathfrak{a}}$ and $\tilde{X}^m$ for $\tilde{g}_{\mathfrak{a}}$) one then finds \cite{Borsato:2022drc}
\[\label{eq:BMN-like-sol-tw}
\tilde{T} = -\kappa \tau , \qquad \tilde{V} = 0 , \qquad \tilde{Z} = \sqrt{\frac{\kappa}{m}} \exp \left(  - \frac{\eta}{2} m \sigma \right)  , \qquad \psi = \omega \tau ,
\]
with all other fields still vanishing. This solution satisfies the twisted-boundary conditions \eqref{eq:G-tbc}  with the twist charges $\mathbf{Q}$ and $\mathbf{q}$ defining $W_{\tts{jor.}}$ taking the values
\[
\mathbf{Q} = -2\pi \eta m = -  \xi M , \qquad \mathbf{q}=0 .\label{eq:sigma_model_charge}
\]
As in the deformed, periodic case, this solution is non-trivial only in an $AdS_3\times S^1$ subspace. In this formulation the Cartan Noether charges are $\tilde E=E=\sqrt{\lambda}\kappa$ and $\tilde{S} = S = 0$, and the charge $M$ appears in $\mathbf{Q}$ (cf.~also footnote \ref{f:twisted-noether}). The angular momenta are still $J_1=J_2=0$ and $J=J_3=\sqrt{\lambda} \omega$ and the Virasoro constraint remains the same. 


\noindent {\bf Landau-Lifshitz map ---} We now want to map the twisted-boundary model in the $S=J_1=J_2=0$  sector  to the $SL(2,\mathbb{R})$ Landau-Lifshitz (LL) Hamiltonian of \cite{Stefanski:2004cw} and evaluate it on the twisted solution \eqref{eq:BMN-like-sol-tw}. 
For this purpose, we found it useful to relate the global Jordanian-adapted coordinates $(T,V,\Theta , P, Z)$ for the $AdS_5$ metric in \eqref{eq:jor-G-B} at $\eta=0$ (also called Schr\"odinger coordinates) to the global cylindrical coordinates of $AdS_5$. This is most conveniently done through their relation to the  embedding coordinates in $\mathbb{R}^{2,4}$ 
\[
-X_0^2 +X_1^2+X_2^2 + X_3^2+X_4^2-X_5^2 = -1 , 
\]
 which was worked out in appendix A of \cite{Borsato:2022drc}, see eq.~(A.2) therein. The standard cylindrical parametrisation is defined as
 \[
 X_0 + i X_5 = \cosh\rho e^{i t} , \qquad X_1 + i X_4 = \sinh\rho \cos\theta e^{i \phi_1} , \qquad X_2 + i X_3 =  \sinh\rho \sin\theta e^{i \phi_1} .
 \]
The $SL(2,\mathbb{R})$ (or $S=J_1=J_2=0$) sector that we are interested in is  obtained by considering the consistent truncation $\theta=0$. Indeed, then $X_2=X_3=0$, as is the case for the BMN-like solution \eqref{eq:BMN-like-sol-per} or \eqref{eq:BMN-like-sol-tw}.  The $SL(2,\mathbb{R})$ Landau-Lifshitz variables, which define the two-dimensional elliptic hyperboloid $\eta^{ij} \ell_i \ell_j=-1$ with $\eta^{ij} = \mathrm{diag} (-1,+1,+1)$, $i,j=0,1,2$,  are then
\[
\ell_0 &=  \cosh 2\rho = X_0^2 + X_1^2 + X_4^2+X_5^2,\\
\ell_1 &= \sinh 2\rho \cos 2 \varphi = 2 X_0 X_1 + 2 X_4 X_5 , \label{eq:LL_strings_param}\\
\ell_2 &= \sinh 2\rho \sin 2\varphi = 2 X_1 X_5 - 2 X_0 X_4,
\]
where $\varphi:=\tfrac{1}{2} (t-\phi_1)$. 
In terms of the Jordanian-adapted variables (on the truncation $P=0$), they read
\[
\ell_0 &= \frac{3(Z^4+4 V^2+2)+(Z^4-4 V^2-2)\cos 2 T - 4 V Z^2 \sin 2 T}{8 Z^2} , \\
\ell_1 &= \frac{  (Z^4-4 V^2-2)\cos T \sin T+2 V Z^2 \cos 2 T}{\sqrt{2}Z^2}, \\ 
\ell_2 &= -\frac{(Z^4+4 V^2+2) + 3(Z^4 -4 V^2 -2)\cos 2 T - 12 V Z^2 \sin 2T}{8 Z^2} .
\]

The $SL(2,\mathbb{R})$ LL theory is  obtained by  eliminating the  coordinate $\psi$ associated to the (angular) momentum $J$ from the (undeformed) $AdS_3\times S^1$ sigma-model by worldsheet gauge-fixing, and taking the large charge $J \rightarrow \infty$ limit while keeping the effective coupling $\tilde{\lambda} = \tfrac{\lambda}{J^2}$  fixed.
In terms of the $\ell_i$ variables, the $SL(2,\mathbb{R})$ LL Hamiltonian  is  \cite{Stefanski:2004cw,Bellucci:2004qr} (see also the review \cite{Tseytlin:2004xa})
\[ \label{eq:LL-string-ham}
H_{\tts{S-LL}} =  \frac{\tilde{\lambda} J}{8} \int^{2\pi}_0 \frac{\mathrm{d}\sigma}{2\pi}  \eta^{ij} \partial_\sigma \ell_i \partial_\sigma \ell_j  ,
\] 
and on the twisted solution \eqref{eq:BMN-like-sol-tw} it evaluates to
\[\label{eq:LL_strings_final_result}
H_{\tts{S-LL}}  = \frac{\eta^2 M^2}{8J} = \frac{\lambda}{32 \pi^2} \frac{\xi^2 M^2}{J} .
\]
This expression matches with the shift $\Delta^{(1)}$ of the dimension $\Delta = J + \Delta^{(1)}$ of this solution \eqref{eq:Del-BMN-like} when expanded to first order in $1/J$ while keeping $\eta$ and $M$ fixed.

\section{Drinfel'd twisted spin chains}\label{sec:chain}

\subsection{Hopf algebra elements}
We start this section by recalling key  elements of a (quasi-triangular) Hopf algebra as a quantum group which are relevant for our construction. For more details we refer to  \cite{Chari:1994pz}.  In the standard case, the Hopf algebra ${\cal A} = U (\mathfrak{g} )$ is the universal enveloping algebra associated with a Lie algebra  $\mathfrak{g}$. The Hopf  structure extends the Lie algebra symmetry  to composite (multiparticle) systems via the coproduct $\Delta : {\cal A} \rightarrow {\cal A} \otimes {\cal A}$, which in the canonical case acts  on elements $X\in \mathfrak{g}$ as
\[
\Delta (X) = X \otimes 1 + 1 \otimes X, \qquad \Delta (1) = 1 \otimes 1 , 
\]
and is comultiplicative
\[
\Delta(XY) = \Delta(X)\Delta(Y), \qquad \text{for  } X, Y \in \mathcal{A}.
\]
Iterating the coproduct $J-1$ times extends the action of symmetries to the tensor product  $\mathcal{A}^{\otimes J}$
\[ \label{eq:can-cop}
\Delta^{J-1} ( X) = \sum_{j=1}^J X_j , \qquad X_j = 1^{\otimes j-1} \otimes X \otimes 1^{\otimes J-j} , \qquad X\in \mathfrak{g} ,
\]
which lends itself naturally to systems such as $J$-sited spin chains. 
In addition, the Hopf algebra is equipped with a co-unit $\epsilon : {\cal A} \rightarrow \mathbb{C}$   and an antipode map $s : {\cal A} \rightarrow {\cal A}$ acting as
\[
\epsilon(1) = 1 , \qquad \epsilon (X) = 0 , \qquad \epsilon (XY) = \epsilon(X)\epsilon(Y),  \qquad s(X) = -X .
\]
One also defines  the opposite coproduct $\Delta^{\tts{op.}} := P \circ \Delta$, where $P(X \otimes Y) = Y \otimes X$ is a permutation of tensor spaces.

In the  quasi-triangular case, the Hopf algebra is further equipped with a  universal $\R$-matrix, $\R : {\cal A} \otimes {\cal A} \rightarrow  {\cal A} \otimes {\cal A}$, which is an invertible element satisfying the   quasi-cocommutativity condition
\[
\Delta^{\tts{op.}} (X) = \R \Delta(X) \R^{-1} ,
\]
as well as quasi-triangularity relations with the standard leg-ordering convention
\[ \label{eq:quasi-triang}
(\Delta \otimes 1) (\R)  = \R_{13} \R_{23} , \qquad (1\otimes \Delta) (\R) = \R_{13} \R_{12}  .
\]
These  imply the famous Yang-Baxter equation (YBE)
\[ \label{eq:YBE}
{\cal R}_{ij} {\cal R}_{ik} {\cal R}_{jk} = {\cal R}_{jk}{\cal R}_{ik}{\cal R}_{ij} ,
\]
in which the subscripts $i,j,k$ label the factors in the  $J$-fold tensor product space ${\cal A}^{\otimes J}$ on which $\R$ acts.
Upon choosing a representation in which each copy of  ${\cal A}$  acts on a vector space $V$, the universal $\R$-matrix becomes an  operator $R$  on $V \otimes V$, typically depending on a free spectral parameter $u\in\mathbb{C}$. This yields the familiar $R_{ij}(u)$ and Lax operators of integrable physical models, which play a central role in solving such systems. 

\noindent {\bf Drinfel'd twists ---} Starting from a solution $\R$ to the YBE, one may apply certain transformations to generate new solutions. A key solution-generating technique is applying a Drinfel'd twist\footnote{Based on the conjecture of \cite{vanTongeren:2015uha}, we purposefully use the same notation ${\cal F}$ as for the twist in section \ref{s:HYB-intgb}, although here it will act on Hilbert space states rather than classical fields.} ${\cal F}$, which is an invertible element of $\mathrm{End}({\cal A}\otimes {\cal A})$ that is normalised $(\epsilon \otimes 1) {\cal F} = (1\otimes \epsilon) {\cal F}=  1$ and satisfies the cocycle condition  \cite{drinfeld1983constant}
\[ \label{eq:cocyle-cond}
({\cal F} \otimes 1) (\Delta \otimes 1) ({\cal F}) = (1 \otimes {\cal F}) (1\otimes {\Delta}) ({\cal F}) .
\]
The twisted quasi-triangular Hopf algebra   is then equipped with a twisted coproduct 
\[
\Delta_{\cal F} (X) = \F \Delta(X) \F^{-1} ,
\]
and twisted universal $\R$-matrix, which also satisfies  the YBE \eqref{eq:YBE},
\[
\R^{\cal F}= {\cal F}^{\tts{op.}} \R \F^{-1}  .
\]
Here we introduced the notation ${X}^{\tts{op.}}:= P X P$ for any endomorphism $X :  {\cal A}\otimes {\cal A} \rightarrow {\cal A}\otimes {\cal A}$. 
It will be convenient to further introduce
\begin{equation}\label{eq:MS-definitions}
\begin{alignedat}{2}
&X_{1 \cdots j-1, j}  := (\Delta^{j-2} \otimes 1) (X),  \qquad  &&X_{1,2\cdots j} := (1\otimes \Delta^{j-2})(X) ,  \\
&{X}^{\tts{op.}}_{1 \cdots j-1, j} := (\Delta^{j-2} \otimes 1)({X}^{\tts{op.}}) =: X_{j, 1 \cdots j-1} , \qquad &&{X}^{\tts{op.}}_{1,2\cdots j} :=(1\otimes \Delta^{j-2})(X^{\tts{op.}}) =: X_{2\cdots j ,1}
\end{alignedat}
\end{equation}
for any $X\in \mathrm{End} ( {\cal A}\otimes {\cal A})$. 
Iterating $J-1$ times   $\Delta_{\F}$   then gives the expression  \cite{Maillet:1996yy}
\[ \label{eq:tw-CP-iterated}
\Delta_{\F}^{J-1} (X) = \Omega_{1 \cdots J} \Delta^{J-1} (X) \Omega_{1 \cdots J}^{-1}  ,
\]
with the  global intertwiner $\Omega_{1 \cdots J} \in \mathrm{End}({\cal A}^{\otimes J})$ defined as
\[ \label{eq:gl-int-1}
\Omega_{1 \cdots J} := \F_{12} \F_{12,3} \cdots \F_{12 \cdots J-1, J}    .   
\] 
The cocycle condition further implies that we can alternatively write   \cite{Maillet:1996yy}
\[ \label{eq:gl-int-2}
\Omega_{1 \cdots J} = \F_{(J-1)J} \F_{(J-2), (J-1)J} \cdots \F_{1,23\cdots J}  .
\]
Useful identities for later finally are
\[ \label{eq:intertwiner-strips}
\Omega_{1 \cdots J} &= \Omega_{1 \cdots J-1} \ \F_{12 \cdots J-1, J} \\ &=\Omega_{2 \cdots J} \ \F_{1,23 \cdots J} \ ,
\]
which ``strip off'' intertwiners of certain lengths.

Having in mind deformations, we will consider twists continuously connected to the identity
\[
{\cal F} = 1\otimes 1 + \xi {\cal F}^{(1)} + {\cal O}(\xi^2) , 
\]
with $\xi\in\mathbb{R}$. The co-cycle condition \eqref{eq:cocyle-cond} implies at  ${\cal O}(\xi)$ that $\F^{(1)}$ must satisfy a comptability condition with $\Delta$. At ${\cal O}(\xi^2)$, its antisymmetric part $r_{12} = \tfrac{1}{2} ({\cal F}_{12}^{(1)} - {\cal F}_{21}^{(1)})$ satisfies the CYBE $[r_{12}, r_{13}]+[r_{12}, r_{23}]+[r_{13}, r_{23}]=0$ \cite{Giaquinto:1994jx}. This matches the defining properties   \eqref{eq:R-AS-CYBE} of integrable HYB deformations in tensor product form. 
As in section \ref{s:HYB-intgb}, we will consider two types of Drinfel'd twists. The abelian, or DR twists \cite{Reshetikhin:1990ep}, are built from abelian $r$-matrices \eqref{eq:def-r-ab}
\[ \label{eq:F-ab}
{\cal F}_{\tts{ab.}} = \exp\left ( \xi \ \mathsf{t}_1 \wedge \mathsf{t}_2 \right),  
\]
with $\mathsf{t}_{1}$, $\mathsf{t}_2$ commuting generators of $\mathfrak{g}$. These are arguably well-understood, including in the context of AdS/CFT, particularly when $\mathsf{t}_{1,2}$ are compact Cartan generators, see e.g.~\cite{Beisert:2005if,McLoughlin:2006cg,deLeeuw:2012hp,Kazakov:2018ugh,Guica:2017mtd,Meier:2023kzt,Meier:2023lku,Meier:2025tjq,Beisert:2024wqq}.
The non-abelian Jordanian twists can instead be represented as \cite{Gerstenhaber:1992,Ogievetsky:1992ph,Kulish2009}
\[ \label{eq:F-jor}
{\cal F}_{\tts{jor.}} = \exp \left( \h \otimes \sigma   \right) , \qquad \sigma:= \log (1+\xi \e ) ,
\]
where $\h,\e \in \mathfrak{g}$ with $[\h,\e]=\e$, as before. Jordanian twists  come in various forms \cite{Tolstoy:2008zz,Meljanac:2016njp},
all of which are related by co-boundary twists, which provide different representations of the same (twisted) Hopf algebra. This includes  $r$-symmetric versions \cite{Giaquinto:1994jx} for which $\F_{21}^{(1)}=-\F_{12}^{(1)}$ (as opposed to the above non-$r$-symmetric form). Also further extended Jordanian Drinfel'd twists exist \cite{Kulish:1998be,2004Tolstoy}. We will however not consider or need any of these different forms or generalisations here.

Our main focus in this paper  will  be the understanding of the spectrum of the simplest Jordanian-twisted spin chains that should correspond to the string state sector laid out in section \ref{s:string-sol}, i.e.~the $\mathfrak{sl}(2,\mathbb{R})$-invariant $\mathrm{XXX}_{-1/2}$ spin chain. Before doing that, we will give two essential aspects for this study: General expressions for the deformed monodromy and transfer matrix in section \ref{subsec:transfer_matrix_after_general_twists}, as well as an explicit and general map of a Drinfel'd twisted system to an undeformed system with twisted-boundary conditions for the one-site algebra operators in section \ref{s:SC-TBC}.

\subsection{Transfer matrix after general twists}\label{subsec:transfer_matrix_after_general_twists}
A central object underlying many of the  techniques used to solve integrable systems such as spin chains is the transfer matrix.  It is constructed by introducing an  auxiliary space, labelled ``$\mathsf{a}$'', in a representation $V_{\mathsf{a}}$ of $\mathfrak{g}$, and taking the iterated canonical coproduct of $R_{\mathsf{a} j} (u)$, where $j$ labels a physical site in a representation $V_j$. Specifically, in the ordering
\[ \label{eq:V-a-J-ordering}
V_{\mathsf{a}} \otimes V_1 \otimes \cdots \otimes V_j \otimes \cdots \otimes V_{J} \ ,
\]
one can define the {monodromy} matrix as
\[ \label{eq:monodromy-undef}
T_{\mathsf{a}} (u) := (1_{\mathsf{a}} \otimes  \Delta^{J-1}) (R(u)) 
= R_{\mathsf{a} J} (u) \cdots R_{\mathsf{a} 2} (u) R_{\mathsf{a} 1} (u) ,
\]
where \eqref{eq:quasi-triang} was used. The transfer matrix is then obtained by tracing over the auxiliary space
\[
\hat{\tau} (u) := \Tr_{\mathsf{a}} T_{\mathsf{a}} (u) ,
\]
and (by virtue of the RTT relations) it encodes a one-parameter family of commuting operators, typically including the momentum (shift) operator and Hamiltonian of closed  chains, at least when $\mathsf{a}$ is taken to be in the same representation as the physical sites. 

Let us now give a useful expression of the monodromy and transfer matrix after applying a Drinfel'd twist.
The deformed monodromy matrix is then constructed by iteratively applying $\Delta_{\F}$ to $R^{\F}$   
\[
T_{\mathsf{a}}^{\F} (u) := (1_{\mathsf{a}} \otimes  \Delta_{\cal F}^{J-1}) (R^{\cal F}(u))    = R^{\cal F}_{\mathsf{a} J} (u) \cdots R^{\cal F}_{\mathsf{a} 2} (u) R^{\cal F}_{\mathsf{a} 1} (u) .
\]
Using \eqref{eq:tw-CP-iterated}, we may rewrite this as a similarity transformation in the physical space by the global intertwiner
\[
T_{\mathsf{a}}^{\F} (u) = \Omega_{1 \cdots J} \ T_{\mathsf{a}}^{\star} (u)  \ \Omega_{1 \cdots J}^{-1} ,
\]
where $T_{\mathsf{a}}^{\star} (u) :=  (1_{\mathsf{a}} \otimes  \Delta^{J-1})(R^{\cal F}(u)) $. We will call $T_{\mathsf{a}}^{\star} (u) $ the monodromy matrix in the \textit{twisted-boundary} picture (for more justification of this name, see the next section) or just ``twisted monodromy''. Using the definition of $R^{\F}$, the compatibility of the coproduct, as well as the notation of \eqref{eq:MS-definitions}, one can simply write\footnote{Note that one should pay attention to the fact that we are working over the tensor-product space \eqref{eq:V-a-J-ordering} here.}
\[ \label{eq:twisted-monodromy}
T_{\mathsf{a}}^{\star} (u)  
&=  (1_{\mathsf{a}} \otimes  \Delta^{J-1})(\F^{\tts{op.}})  (1_{\mathsf{a}} \otimes  \Delta^{J-1})(R)  (1_{\mathsf{a}} \otimes  \Delta^{J-1}) (\F^{-1}) \\
&= \F_{ 1\cdots J, \mathsf{a} } \ T_{\mathsf{a}} (u) \ \F^{-1}_{\mathsf{a},1\cdots J} , 
\]
so that ultimately for any Drinfel'd twist we have
\[ \label{eq:deformed-monodromy}
T_{\mathsf{a}}^{\F} (u) = \Omega_{1 \cdots J}\ \F_{ 1\cdots J, \mathsf{a} } \ T_{\mathsf{a}} (u) \ \F^{-1}_{\mathsf{a},1\cdots J}  \ \Omega_{1 \cdots J}^{-1} ,
\]
without additional assumptions. We are not aware of a derivation as presented in this section, although all the ingredients are present in \cite{Maillet:1996yy}. The formula \eqref{eq:deformed-monodromy} is equivalent to eq.~(3.30) of \cite{Guica:2017mtd} when specified to the case of abelian DR twists (see also \cite{Ahn:2010ws,deLeeuw:2012hp} for alternative derivations) and eq.~(3.33) of \cite{Borsato:2025smn} for the case of Jordanian twists of the type \eqref{eq:F-jor}.  
In section \ref{s:jor-chain} we will however show that our \eqref{eq:deformed-monodromy} takes a particularly nice form in the case of Jordanian twists.

\subsection{Mapping to twisted-boundary conditions of one-site operators} \label{s:SC-TBC}
Closed Drinfel'd twisted   spin chains   are understood to have two equivalent formulations: \textit{(i)} a model of spins with \textit{deformed} interactions and \textit{periodic}-boundary conditions along a one-dimensional  chain, or \textit{(ii)} a model of spins with \textit{undeformed} interactions but with \textit{twisted}-boundary conditions. Indeed, consider e.g.~models with  nearest-neighbour interactions, such as the Heisenberg $\mathrm{XXX}$ spin chain, for which the Hamiltonian density $h \in \mathrm{End}( {\cal A}\otimes {\cal A})$ is a two-sited operator and with
\[
R_{12}(u=u_0) = P_{12} , \qquad h_{12} := P_{12} \left( \frac{\mathrm{d}}{\mathrm{d}u} R_{12}(u) \right)_{u=u_0} ,
\]
for some $u_0\in\mathbb{C}$. The deformed interactions are  then governed by
\[ \label{eq:Ham-def}
H^{\F} := \sum_{j=1}^J h^{\F}_{j,j+1} , \qquad h^{\F}_{j,j+1} := \F_{j,j+1} h_{j,j+1} \F_{j,j+1}^{-1} ,
\]
with periodic-boundary conditions $h^{\F}_{J,J+1} \equiv h^{\F}_{J,1}$. Under a similarity transformation $\Omega$ of the Hilbert space of states, this is equivalent to an undeformed interacting Hamiltonian 
\[ \label{eq:Ham-twist}
 H^\star := \Omega^{-1} H^{\F} \Omega =  \sum_{j=1}^{J-1} h_{j,j+1} + S^{-1} h_{J,1} S ,
\]
with twisted-boundary conditions $h_{J,J+1} = S^{-1} h_{J,1} S$. In the case of abelian and Jordanian twists of the type \eqref{eq:F-ab} and \eqref{eq:F-jor}, one can show that $S = \F_{J1}^{-1}\Omega$ and $\Omega = \Omega_{1\cdots J}$, the global intertwiner, see e.g.~\cite{Guica:2017mtd,Borsato:2025smn}.

In this section, we will instead propose a general method of deriving the twisted-boundary conditions for one-site operators on ${\cal A}=U(\mathfrak{g})$.\footnote{See  \cite{Bazhanov:2010ts} for obtaining twisted-boundary conditions of one-site operators in the case of constant twists using  the transformation of the deformed to the undeformed Hamiltonian. }
On the one hand, we have the deformed picture, where the action of one-site operators  $X\in {\cal A}$ is
\[
\Delta_{\F}^{J-1} (X) = \sum_{j=1}^J \Omega_{1 \cdots J} X_j \Omega_{1 \cdots J}^{-1} , 
\] 
with periodic-boundary $X_{J+i} \equiv X_i$. 
On the other hand, we will have the twisted-boundary picture where 
\[
\Delta^{J-1} (X) = \sum_{j=1}^J X_j  ,
\]
but with the price to pay that $X_{J+i} \not\equiv X_i$.
One may travel between the two formulations using the non-local (all-site) similarity transform $X_j \rightarrow X^\star_j \equiv \Omega^{-1}_{1 \cdots J} X_j \Omega_{1 \cdots J}$. 

The question that we want to answer now is: How do the periodic-boundary conditions map to an explicit formula for the twisted-boundary conditions?

In order to systematically implement the boundary twist induced by the Drinfel’d twist, we begin by extending the spin chain through the introduction of two auxiliary spaces: one before and one after the physical chain, labelled $\mathsf{a}$ and $J+\mathsf{a}$, respectively. That is, we consider the extended open chain 
\[ \label{eq:V-a-J+a-ordering}
V_{\mathsf{a}} \otimes V_1 \otimes \cdots \otimes V_j \otimes \cdots \otimes V_{J} \otimes V_{J+\mathsf{a}} \ .
\]
The purpose of this extension is to realise the eventual twisted periodicity as an operator insertion \textit{localised} at the boundary site $J+\sfa$.\footnote{We use the some notation $\sfa$ as in the previous section  to make a simple comparison with the twisted transfer matrix later on.}

\begin{figure}[h]
\centering
\includegraphics[scale=0.8]{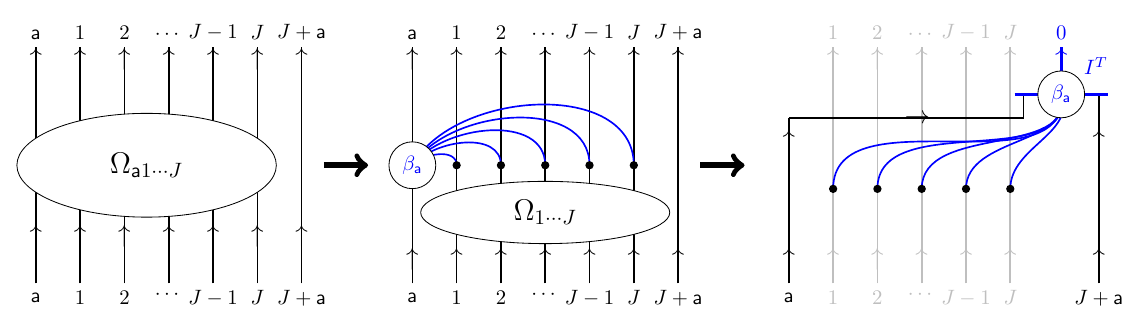}
\caption{Illustration of the gluing procedure leading to twisted-boundary conditions.}
\label{fig:glue}
\end{figure}

Before explicitly performing the procedure, let us first sketch the idea of each step, which we also illustrate in figure \ref{fig:glue}. \textit{(i)}
We will  deform the chain by applying the Drinfel’d twist on the chain from $\sfa$ to $J$.\footnote{Due to the (in general) non-abelian nature of Drinfel'd twists, the ordering in which we deform the chain
(from $\sfa$ to $J$, or from $1$ to $J+\sfa$) 
and map the effect to the boundary (at $J+\sfa$, or at $\sfa$) is important. We will comment more on this point later for the explicit example of Jordanian twists.} 
Since the twist $\F$ is a two-site operator, this will ensure  (at the end of the day) that  every physical site contributes to the total boundary deformation. In particular, the resulting boundary twist will be composed from multiple $\F$-factors of which one leg is consistently anchored at the auxiliary space $\sfa$ and the other leg connects to each physical site $j$, such that the entire physical chain contributes to its overall expression.
\textit{(ii)} To isolate the physical effect of the boundary twist, we then apply a similarity transformation that removes the ``internal'' twist factors---those connecting pairs of physical sites---and retains only the twist components linking $\sfa$ to the physical chain. This will result in a configuration where the twist is effectively localised at $J+\sfa$.
\textit{(iii)} Finally, to recover the canonical operator action up to a boundary twist, we glue together the  auxiliary spaces $\sfa$ and $J+\sfa$ via a suitable projection $I^T$. The resulting space, denoted ``0'', 
represents the defect site that 
carries the entire effect of the twisted periodicity. 
See  \eqref{eq:jor-expl-X-TBCs} and \eqref{eq:GLMZ-BCs} for explicit forms of the resulting boundary operator.

Let us now explicitly apply the procedure as described above. We consider twist the chain and consider  the deformed  action
\[
B^{\cal F}(X) := ( \Delta_{\F}^J \otimes 1_{J+\mathsf{a}} ) (  X \otimes 1_{J+\mathsf{a}} + 1 \otimes X_{J+\mathsf{a}}  ) .
\]
Using the definitions \eqref{eq:can-cop}, \eqref{eq:tw-CP-iterated} and the identities \eqref{eq:intertwiner-strips}, one can write
\[
B^{\cal F}(X) &= \Omega_{\mathsf{a} \cdots J} \left( X_{\mathsf{a}} + \sum_{j=1}^J X_j  \right) \Omega_{\mathsf{a} \cdots J}^{-1} + X_{J+\mathsf{a}}  \\
&= \Omega_{1 \cdots J} \beta_{\mathsf{a}} \left( X_{\mathsf{a}} +  \Delta^{J-1}(X)  \right) \beta_{\mathsf{a}}^{-1} \Omega_{1 \cdots J}^{-1} + X_{J+\mathsf{a}} ,
\]
where we lightened notation by introducing $\beta_{\mathsf{a}}:=\F_{\mathsf{a}, 1 \cdots J}$. 
Now,  we perform a similarity transformation by $\Omega_{1\cdots J} \beta_{J+\sfa}$ as $B^{\cal F}(X) \rightarrow B^\star (X) = (\Omega_{1\cdots J} \beta_{J+\sfa})^{-1} B^{\cal F}(X) (\Omega_{1\cdots J} \beta_{J+\sfa})$ so that
\[ \label{eq:b-star-similarity}
B^\star (X) =  \beta_{J+\sfa}^{-1} \beta_{\sfa} \left( X_{\sfa} + \Delta^{J-1}(X) \right) \beta_{\sfa}^{-1} \beta_{J+\sfa} + \beta_{J+\sfa}^{-1}  X_{J+\sfa} \beta_{J+\sfa} ,
\]
where $\beta_{J+\mathsf{a}}:= \F_{J+\mathsf{a}, 1 \cdots J}$. This has removed the internal physical twists present in $\Omega_{1\cdots J}$. 

We will now consider the gluing of the auxiliary spaces $V_{\sfa}$ and $V_{J+\sfa}$  and projection onto the single defect space $V_{0}$. For this purpose, let us introduce two maps\footnote{In an orthonormal basis  $\ket{e_a}$ of $V$, we can for instance define
\[
I := \sum_a \ket{e_a}_{\sfa} \otimes \ket{e_a}_{J+\sfa} \bra{e_a}_{0} , \qquad I^T :=  \sum_a \ket{e_a}_{0} \bra{e_a}_{\sfa} \otimes \bra{e_a}_{J+\sfa} .
\]
In principle, one may also introduce coefficients $\lambda_a$ in the definition of  $I$ and $I^T$ to allow for a weighted identification.
By construction we have $I^T \left(\ket{e_a}_{\sfa} \otimes \ket{e_b}_{J+\sfa}\right) = \delta_{ab} \ket{e_a} $, implementing the formal gluing identification by projecting onto the (diagonal) subspace where the two auxiliary sites agree. The opposite $I$ embeds a defect space vector $\psi_0 = \sum_a \langle e_a|\psi\rangle\ket{e_a}_0$ as a diagonal tensor, $I (\psi_0) = \sum_a \langle e_a|\psi\rangle \ket{e_a}_{\sfa}\otimes \ket{e_a}_{J+\sfa}$. We can interpret the latter as an entangled state between $V_{\sfa}$ and $V_{J+\sfa}$.}
\[
I : V_{0} \rightarrow V_{\sfa} \otimes V_{J+\sfa} , \qquad I^T :  V_{\sfa} \otimes V_{J+\sfa} \rightarrow V_{0} .
\]
We then assume the following gluing condition
\[ \label{eq:gluing-assumption}
I^T \beta_{\mathsf{a}} = I^T \beta_{J+\mathsf{a}} , \qquad \beta_{\mathsf{a}}  I = \beta_{J+\mathsf{a}}  I . 
\]
Concretely, this ensures that when we project  back to the single defect space $V_0$, the same overall twist $\beta_{\mathsf{a}}=\F_{\mathsf{a},1\cdots J}$ acts regardless of whether it was applied to the first or second auxiliary leg.
Performing the gluing on \eqref{eq:b-star-similarity} we then find the canonical coproduct  up to the defect term
\[
B(X) = I^T B^\star (X) I = I^T X_{\sfa} I + I^T \Delta^{J-1}(X) I + I^T   \beta_{J+\sfa}^{-1} X_{J+\sfa}  \beta_{J+\sfa} I .
\]
We thus have two copies of the operator algebra acting on the $0$-site: one un-twisted, appearing ``before''   site $1$, and one twisted, appearing ``after'' site $J$. This action can only be consistent when we identify the twisted-boundary condition
\[ \label{eq:X-TBCs-gen}
X_{J+ \sfa} \equiv  \beta_{\sfa} X_{\sfa} \beta^{-1}_{\sfa} .
\]
In other words, the operator on the defect site is the original  generator conjugated by the total  boundary-twist $\beta_{\mathsf{a}}=\F_{\mathsf{a},1\cdots J}$.
In the next section, we will give an explicit expression of \eqref{eq:X-TBCs-gen} for the  $\mathfrak{sl}(2,\mathbb{R})$ Jordanian twist \eqref{eq:F-jor},  and we will demonstrate that it passes several consistency checks. Nonetheless, at this stage, let us already mention two interesting observations:
Firstly, the total Jordanian boundary-twist  will be expressed in terms of a symmetry operator of the resulting model, and the boundary conditions will respect the $\mathfrak{sl}(2,\mathbb{R})$ algebra. {This would not have been the case if we (naively) pulled the twist from site   $1$ to $J+\sfa$ and measured at $\sfa$.\footnote{We would however find the same (consistent) formulas as given in \eqref{eq:X-TBCs-gen}  when pulling the twisting from  $J+\sfa$ to $1$ by means of the opposite coproduct and measuring at $\sfa$.}  We believe this difference is rooted in the underlying (leg-ordering) axioms of the quasi-triangular Hopf algebra and in the fact that the twist \eqref{eq:F-jor} is not $r$-symmetric (unlike abelian twists) such that its tensor legs are on unequal footing. 
It would be interesting to find a derivation that  makes this distinction more explicit.} 
Secondly, we note that the above formula \eqref{eq:X-TBCs-gen} is consistent with the form of the twisted monodromy matrix \eqref{eq:twisted-monodromy}, obtained from  the iterated coproduct of the twisted $R$-matrix (which encodes a permutation), after the similarity transformation. In both cases, the object that first  acts on the Hilbert space is $\beta^{-1}_{\sfa} = \F^{-1}_{\mathsf{a}, 1 \ldots J}$.

\subsection{Applied to the Jordanian-twisted $\mathfrak{sl}(2,\mathbb{R})$ spin chain} \label{s:jor-chain}
Given the BMN-like string solution of the Jordanian sigma-model presented in \ref{s:string-sol}, which is non-trivial only in an $AdS_3\times S^1$ subspace of $AdS_5\times S^5$, the relevant spin chain  we want to study and twist is the non-compact $\mathfrak{sl}(2,\mathbb{R})$-invariant $\mathrm{XXX}_{-1/2}$ spin chain \cite{Kulish:1981gi,Tarasov:1983cj,Beisert:2003jj,Beisert:2003yb} (see also \cite{Derkachov:2002tf,Kirch:2004mk}).
In this section we will clarify the residual symmetry of this model in both the deformed and twisted-boundary picture, and we will give explicit formulas for the twisted monodromy matrix and boundary conditions of the one-site algebra generators.

We  denote the generators of $\mathfrak{sl}(2,\mathbb{R})$ by $X=\{ \h, \e, \f\}$, with the commutation relations
\[ \label{eq:comm-rels-sl2}
[\h , \e] = \e , \qquad [\h,\f]=-\f , \qquad [\e,\f] = -2\h  .
\]
Here, $\h$ is the Cartan generator, while $\e$ and $\f$ are respectively raising and lowering operators with respect to the spectrum of $\h$. 
Unless  otherwise stated, from here on out we will take $\F \equiv \F_{\tts{jor.}}$ as given in \eqref{eq:F-jor}. 
The twisted coproducts of the $\mathfrak{sl}(2,\mathbb{R})$ generators can then be computed explicitly as
\begin{align}
\Delta_{\F} (\h ) &= \h \otimes \exp(-\sigma) + 1 \otimes \h ,\\
\label{eq:tw-cop-e}
\Delta_{\F} (\e) &= \e \otimes \exp (\sigma) + 1\otimes \e =\e \otimes 1 + 1\otimes \e +\xi \e \otimes \e,  \\
\Delta_{\F} (\f) &= \f \otimes \exp (-\sigma) + 1\otimes \f - 2\xi \h\otimes \h \exp(-\sigma) - \xi (\h-\h^2) \otimes (\exp (-\sigma)- \exp(-2\sigma) ) .  
\end{align}
From this, one finds on $U(\mathfrak{sl}(2))$ that
\begin{align}
\label{eq:primitive-sigma}
\Delta_{\F} (\sigma) &= \sigma\otimes 1 + 1 \otimes \sigma .
\end{align} 
meaning that $\sigma$ is primitive (or canonical) with respect to the twisted coproduct.
It furthermore implies  the following  factorisation identities for Jordanian twists of the type \eqref{eq:F-jor}
\begin{alignat}{2}
\label{eq:F-jor-fact-1}
(\Delta\otimes 1)(\F) &= \exp \left( \Delta (\h) \otimes \sigma \right) &&= \F_{13} \F_{23} , \\
\label{eq:F-jor-fact-2}
(1 \otimes \Delta_{\F})(\F) &= \exp \left( \h \otimes \Delta_{\F}(\sigma) \right) &&= \F_{12} \F_{13} ,
\end{alignat}
Eq.~\eqref{eq:F-jor-fact-1} allows us to simplify the global intertwiner \eqref{eq:gl-int-1} and write
\[
\Omega_{1 \cdots J} &= \F_{12} (\F_{13} \F_{23}) \cdots (\F_{1J} \F_{2J} \cdots \F_{(J-1)J} ) \\
&= (\F_{12} \F_{13} \cdots \F_{1J}) (\F_{23} \F_{24} \cdots \F_{2J}) \cdots  \F_{(J-1) J} .
\]
For abelian DR twists, where $[\F_{ij}, \F_{kl}]=0$ for all tensor sites, the above of course simplifies even further, as the ordering does not matter.

\noindent {\bf Residual symmetries ---} As mentioned earlier, the Hamiltonian of the Jordanian-twisted spin chain admits two equivalent formulations. In the deformed picture, one works with  $H^{\F}$ given in \eqref{eq:Ham-def}, while in the twisted-boundary picture one works with $H^{\star} = \Omega^{-1}H^{\F}\Omega$ given in \eqref{eq:Ham-twist}, with $S= \F_{J1} \Omega$ and $\Omega = \Omega_{1\cdots J}$, see \cite{Borsato:2025smn}.\footnote{The two formulations are thus related by the same non-local (all-site) similarity transform of the Hilbert space that relates also the deformed $T^{\F}_{\sfa}(u)$ and twisted-boundary $T^{\star}_{\sfa}(u)$ monodromy, as presented in sec.~\ref{subsec:transfer_matrix_after_general_twists}, as well as the periodic and twisted-boundary conditions of one-site operators (and states), as presented in sec.~\ref{s:SC-TBC}. } 
Importantly, the  universal enveloping Lie algebra $U(\mathfrak{sl}(2,\mathbb{R}))$  is represented differently in these cases: In the deformed picture,  the action of  $U(\mathfrak{sl}(2,\mathbb{R}))$  is twisted via the twisted coproduct $\Delta_{\F}$, while in the twisted-boundary picture  one uses  the canonical coproduct $\Delta$ to define the  algebra action. 

The residual $\mathfrak{sl}(2,\mathbb{R})$ symmetries of the  models may be derived following sec.~3.3 of \cite{Borsato:2025smn} and which we recap here. The undeformed periodic $\mathrm{XXX}_{-1/2}$ spin chain Hamiltonian is \cite{Kulish:1981gi,Tarasov:1983cj} (see also \cite{Derkachov:2002tf})  
\[
H=\sum_{j=1}^Jh_{j,j+1},\qquad h_{j,j+1}=\psi(\mathbb{J}_{j,j+1}+1)-\psi(0), \label{eq:sl(2)_hamiltonian}
\] 
with $\psi(x)$ the digamma function and $\mathbb{J}$ a two-site operator defined implicitly through $\mathbb{J}(\mathbb{J}+1)=\Delta(X) \cdot \Delta(X)=\mathbb{C}_{2}$, with $\mathbb{C}_{2}$ the two-site quadratic Casimir, and the dot product defined through the Cartan-Killing form.
This Hamiltonian is $\mathfrak{sl}(2,\mathbb{R})$-invariant 
\[
[\Delta^{J-1} (X) , h_{j,j+1}] = 0 , \qquad \forall X\in\mathfrak{sl}(2,\mathbb{R}) , ~~~j=1,\cdots J .
\]
For the deformed Hamiltonian ${H}_\F$ one then finds 
\[
[\Delta_{\F}^{J-1} (X) , H^{\F}] &=  \sum_{j=1}^{J-1}  \Omega[\Delta^{J-1}(X), h_{j,j+1}] \Omega^{-1} + [\Delta_{\F}^{J-1}(X), h^{\F}_{J,J+1}] \\
&=  [\Delta_{\F}^{J-1}(X), h^{\F}_{J,J+1}] ,
\]
since $h^{\F}_{j,j+1} = \Omega h_{j,j+1} \Omega^{-1}$ for $j=1, \cdots, J-1$  \cite{Borsato:2025smn}. Thus, the only potential  violation of $\mathfrak{sl}(2,\mathbb{R})$ symmetry arises at the end of the chain.
A generator $X \in \mathfrak{sl}(2,\mathbb{R})$ survives as an ordinary symmetry if there exists some $X' \in \mathfrak{sl}(2,\mathbb{R})$ such that, under the shift operator $U:= P_{12} \cdots P_{J-1,J}$, one has 
\[ \label{eq:res-symm}
U^{-1} \Delta_{\F}^{J-1} (X) U = \Delta_{\F}^{J-1} (X') .
\]
Indeed, in that case $ [\Delta_{\F}^{J-1}(X), h^{\F}_{J,J+1}] = U  [\Delta_{\F}^{J-1}(X'), h^{\F}_{J-1,J}] U^{-1} =0$ and so $\Delta_{\F}^{J-1}(X)$ commutes with the full Hamiltonian.  

Crucially,  the operator $\sigma = \log (1+\xi \e)\in U(\mathfrak{sl}(2,\mathbb{R}))$  is primitive with respect to $\Delta_\F$, and one can check immediately that it satisfies the shift-equivariance condition \eqref{eq:res-symm}. Although non-standard, it is   a residual symmetry  of both formulations
\[
\label{eq:charge_commutation}
[\Delta_{\F}^{J-1} (\sigma) , H^{\F}] =0 , \qquad \text{and} \qquad [\Delta^{J-1} (\sigma) , H^{\star}] =0  .
\]
Accordingly, in the deformed picture it would  be natural to diagonalise
\[
\widehat{Q} := \Delta_{\F}^{J-1} (\sigma) = \sum_{j=1}^J \sigma_j ,
\]
whereas in the twisted-boundary formulation, given the fact that   $\Delta^{J-1}(\sigma) = \log \left( 1+\xi \Delta^{J-1}(\e) \right)$,   it would be convenient to diagonalise the sum of raising operators
\[
\label{eq:M_charge}
\widehat{M} :=\Delta^{J-1}(\e)= \sum_{j=1}^J \e_j  .
\]  
However, non-trivial eigenstates of the $\widehat{Q}$-operator turn out to be involved to work with as they do not admit a regular expansion in $\xi$, while $\widehat{M}$-eigenstates have simple expressions.
One may also verify  explicitly that  $\e$  is a residual symmetry generator. In fact,  eq.~\eqref{eq:tw-cop-e}  implies
\[ \label{eq:e-symm-shift}
U \Delta_{\F}^{J-1} (\e) U^{-1} = U\left( \sum_{j=1}^J \e_j + \xi \sum_{j\neq j'}^J \e_j \e_{j'} + \cdots \xi^{J-1} \e_1 \e_2 \cdots \e_J  \right) U^{-1} = \Delta_{\F}^{J-1} (\e) ,
\]
such that $\e$ satisfies \eqref{eq:res-symm}. While $\Delta_{\F} (\e)$ is again non-standard, 
the spectrum of $\widehat{M}=\Delta^{J-1}(\e)$  defines an ordinary conserved charge $M$ in the twisted-boundary formulation.  
Importantly, the  other $\mathfrak{sl}(2,\mathbb{R})$ generators, the Cartan $\h$ and the lowering $\f$, do no satisfy \eqref{eq:res-symm}. This is consistent with the Jordanian sigma-model with \eqref{eq:def-r-jor}, which always has $\e$ as a residual symmetry.

In the upcoming sections, we will therefore abandon diagonalising the Cartan generator $\h$. 
This will represent a rather radical departure from more familiar spin chains, where the energy spectrum can be organised by its eigenvalues,  i.e.~in a Fock space of magnon-like excitations.   Instead,  we will work in an eigenbasis of the raising operator $\e$. 
These $\e$-eigenstates can be interpreted as (non-compact) coherent states with respect to the usual Fock space at each site, 
see e.g.~\cite{Barut:1970qf,Perelomov1977} for such a construction. 
The eigenbasis of $\widehat{M}$ is thus a tensor product of such states across the chain. While it aligns naturally with the residual symmetry structure, this basis is over-complete and continuous, rather than orthonormal and discrete. It will thus be  a significant change in how we label and organise our spin chain states.\footnote{Similar considerations were also needed in the dipole deformation of \cite{Guica:2017mtd}, realised via an abelian DR twist based on a root generator $\e\in\mathfrak{sl}(2,\mathbb{R})$, as well as a Cartan of the $SO(6)$  R-symmetry.}

\noindent {\bf Twisted transfer matrix and twisted-boundary conditions ---} We can now evaluate the general expressions for the twisted-monodromy matrix \eqref{eq:twisted-monodromy} and twisted-boundary condition  \eqref{eq:X-TBCs-gen}  explicitly for the Jordanian twist \eqref{eq:F-jor}.
In particular, we have
\begin{align}
\F_{1 \cdots J , \sfa} &= (1_{\sfa}\otimes \Delta^{J-1}) (\F^{\tts{op.}}) = \exp \left( \sigma_{\sfa} \otimes \Delta^{J-1} (\h) \right) , \\
\beta_{\sfa}= \F_{\sfa , 1 \cdots J}  &= (1_{\sfa}\otimes \Delta^{J-1} )(\F) =  \exp \left( \h_{\sfa} \otimes \Delta^{J-1}  ({\sigma}) \right)  .
\end{align}
From the above  discussion, it is then obvious that $\beta_{\sfa}$ is expressible in terms of the symmetry operator $\widehat{M}$, and thus we can write\footnote{Note that we  may also express the overall twist as $\F_{\sfa , 1 \cdots J} = \Ad^{-1}_{\Omega_{1\cdots J}}  \exp \left( \h_{\sfa} \otimes \Delta^{J-1}_{\F} ({\sigma}) \right) = \Omega^{-1}_{1\cdots J} \left(\prod_{j=1}^J \F_{\sfa j}\right)\Omega_{1\cdots J}$ such that  the expression \eqref{eq:jor-twisted-monodromy} clearly matches with eq.~(3.33) of \cite{Borsato:2025smn}. }
\[ \label{eq:jor-twisted-monodromy}
T^{\star}_\mathsf{a} (u) = \exp \left( \sigma_{\sfa} \otimes \Delta^{J-1} (\h) \right) T_{\sfa} (u)   \exp \left( -\h_{\sfa} \otimes  \log (1+\xi \widehat{M}) \right) ,
\]
and 
\[ \label{eq:jor-X-TBCs}
X_{J+\sfa} \equiv \exp \left( \h_{\sfa} \otimes  \log (1+\xi \widehat{M}) \right) X_{\sfa} \exp \left( -\h_{\sfa} \otimes \log (1+\xi \widehat{M})\right) ,
\]
for $X\in \mathfrak{sl}(2,\mathbb{R})$.  Note that here we see explicitly that the inclusion of the ``auxiliary'' defect site in the derivation of the boundary conditions is necessary such that the overall twist can be expressed in terms of the total symmetry operator $\widehat{M}$ on the full physical chain from site $1$ to  $J$.
Explicitly,  the twisted-boundary conditions  read
\begin{equation} \label{eq:jor-expl-X-TBCs}
    \e_{J+\sfa}\equiv \e_{\sfa} (1+\xi \widehat{M}), \qquad \h_{J+\sfa}=\h_\sfa , \qquad \f_{J+\sfa}\equiv \f_{\sfa}(1+\xi  \widehat{M})^{-1} ,
\end{equation}
which can be shown using the identity $\Ad_{e^X} = e^{\ad_X}$ for $X \in {\cal A}^{\otimes J}$.
An important  check of  \eqref{eq:jor-expl-X-TBCs} is that the $X_{J+\sfa}$ generators should satisfy the $\mathfrak{sl}(2,\mathbb{R})$ algebra \eqref{eq:comm-rels-sl2}. This is indeed the case here.
Note that when the physical spin chain is in an eigenstate of the operator $\widehat{M}$ we can replace in \eqref{eq:jor-expl-X-TBCs} its action with the corresponding eigenvalue $M$.

As per illustration, let us also compare to the DR twist considered in \cite{Guica:2017mtd}. In that case, 
one would find $\F^{\tts{ab.}}_{\tts{a}, 1 \cdots J} = \exp (\xi \e_{\sfa} \widehat{R}-\xi R_{\sfa} \widehat{M})$ with $\widehat{R}:=\Delta^{J-1} R_j$ the total $R$-charge. In the $\mathfrak{sl}(2,\mathbb{R})$ sector, this implies 
$X_{J+\sfa} = \exp (\xi \widehat{R} \e_\mathsf{a}) X_{\sfa} \exp (-\xi \widehat{R} \e_\mathsf{a})$
because  $R$ acts as trivially (as the identity) on $\mathsf{a}$. The boundary conditions \eqref{eq:X-TBCs-gen} then become
\[ \label{eq:GLMZ-BCs}
\e_{J+\sfa} \equiv \e_{\sfa}, \qquad \h_{J+\sfa} \equiv \h_{\sfa} - \xi \widehat{R} \e_{\sfa} , \qquad \f_{J+\sfa} \equiv \f_{\sfa} - 2 \xi \widehat{R} \h_{\sfa} + \xi^2 \widehat{R}^2 \e_{\sfa} ,
\]
with $\widehat{R} = J$ when evaluated on the physical  spin chain, which matches with    \cite{Guica:2017mtd}.

Several consistency checks thus support our  method presented in section \ref{s:SC-TBC} to deriving twisted-boundary conditions for Drinfel'd twisted spin chains. In particular, next to consistency with the transfer matrix,  the boundary conditions \eqref{eq:jor-X-TBCs} are given in terms of the residual symmetry operator $\widehat{M}$ and respect the $\mathfrak{sl}(2,\mathbb{R})$ algebra.  In appendix \ref{app::classical_bcs}, we  also analyse the classical limit of the twisted transfer matrix of \eqref{eq:jor-twisted-monodromy}. Its overall twisting (denoted by ${\cal W}$) is
structurally analogous  to the twist appearing in the boundary conditions of  the one-site operators. 
In particular, we show that the eigenvalues of the classical ${\cal W}$ depend only on the $\widehat{M}$-charge, with no contribution from the $\Delta^{J-1}(\h)$ operator on the physical sites, in analogue to the form of the boundary conditions in \eqref{eq:jor-X-TBCs}.
We  also comment in appendix \ref{app::classical_bcs} on  connections of ${\cal W}$ and \eqref{eq:jor-expl-X-TBCs}  to the classical sigma-model picture, where similar simplifications of the twisted-boundary conditions arise.
Finally, in section \ref{subsec:twisted_BCs_Baxter_eq} we  find consistency with the large $u$ asymptotics of the $Q$-functions of the underlying integrable system, which generally are known to encode the boundary conditions. 

\section{Landau-Lifshitz model from the twisted spin chain}\label{sec:landau-lifshitz} 

In section \ref{s:string-sol}, we considered a classical  twisted-string solution of the Jordanian  sigma-model in the $S=J_1=J_2=0$ sector, and mapped it to  the semi-classical $SL(2,{R})$ LL Hamiltonian. 
Here, we will show how the same LL Hamiltonian arises directly from the spin chain description described in section~\ref{s:jor-chain}. We begin by reviewing how the untwisted  $SL(2,{R})$ LL model  is obtained from  the classical continuum limit of the $\alg{sl}(2,{R})$ spin chain, following \cite{Stefanski:2004cw,Bellucci:2004qr} (generalising the seminal work \cite{Kruczenski:2003gt}).
Then, we show how the Jordanian Drinfel'd twist with \eqref{eq:jor-expl-X-TBCs} leads to an LL model with corresponding twisted-boundary conditions. Finally, we demonstrate that the classical energy of the LL configuration from the twisted spin chain matches that of the twisted-string solution on the sigma-model side, thus providing a first direct link between the two descriptions.

\subsection{$\mathfrak{sl}(2,\mathbb{R})$ coherent states and  Landau-Lifshitz limit}
 
\noindent {\bf Review of the bulk model ---} The classical limit of the  $\alg{sl}(2,\mathbb{R})$ spin chain is typically obtained via the coherent state formalism, which captures semiclassical dynamics by associating to each site an $\alg{sl}(2,\mathbb{R})$  coherent state,  defined as eigenstates of the lowering operator $\f$
\begin{equation}
    \ket{\textbf{n}}=\sqrt{1-|\vartheta|^2}e^{\vartheta \e }\ket{0}, \qquad     \bra{\textbf{n}}=\bra{0}e^{\bar{\vartheta} \f }\sqrt{1-|\vartheta|^2},\label{eq:coherent_states}
\end{equation}
with $\ket{0}$ the lowest-weight state with $\h\ket{0} = -\tfrac{1}{2}\ket{0}$ and $\f\ket{0}=0$. The parameter $\vartheta$  labels a point on a two-dimensional elliptic hyperboloid.\footnote{In particular, in terms of  the parametrisation  \eqref{eq:LL_strings_param} it is $\vartheta=-\tanh(\rho)e^{2 i \varphi}$. } 
In fact, it is a key property of these states  that the expectation values of the $\mathfrak{sl}(2,\mathbb{R})$ generators
\begin{equation} \label{eq:CS_key_property}
    n_0:=-2\bra{\textbf{n}}\h\ket{\textbf{n}}, \qquad n_+:=-2\bra{\textbf{n}}\e\ket{\textbf{n}}, , \qquad n_-:=-2\bra{\textbf{n}}\f\ket{\textbf{n}} , 
\end{equation}
which evaluate to 
\begin{equation}\label{eq:CS_components}
    n_0=\frac{1+|\vartheta|^2}{1-|\vartheta|^2}, \qquad n_+= -\frac{2\vartheta}{1-|\vartheta|^2}, \qquad n_-=-\frac{2\Bar{\vartheta}}{1-|\vartheta|^2} ,
\end{equation}
satisfy the   constraint
\begin{equation}\label{eq:CS_casimir}
    \vec{\textbf{n}} \cdot \vec{\textbf{n}}= \eta^{ij}n_in_j=-n_0^2+n_+n_-=-1 .
\end{equation}
The vector $\vec{\textbf{n}}:=(n_0,n_1,n_2)$ with $n_\pm=n_1\pm i n_2$ is thus constrained to the two-dimensional elliptic hyperboloid. 
Importantly, these coherent states are normalised, $\braket{\mathbf{n}}{\mathbf{n}}=1$, but they are non-orthogonal and  over-complete, see \cite{Barut:1970qf,Perelomov1977,Bellucci:2004qr} for more details.

To construct the classical spin chain Hamiltonian, one evaluates the quantum Hamiltonian density on a product of neighbouring coherent states, yielding \cite{Stefanski:2004cw,Bellucci:2004qr}  
\begin{equation}\label{eq:classical_H}
h^{\tts{cl.}}_{j,j+1}:=\bra{\textbf{n}_j\textbf{n}_{j+1}}h_{j,j+1}\ket{\textbf{n}_j\textbf{n}_{j+1}}=\log\left(1+\frac{(\vec{\textbf{n}}_j-\vec{\textbf{n}}_{j+1}) \cdot (\vec{\textbf{n}}_j-\vec{\textbf{n}}_{j+1})}{4}\right) ,
\end{equation}
where $\vec{\textbf{n}}_{j}=\{n_0,n_1,n_2\}_j$   and the dot product is defined  as in \eqref{eq:CS_casimir}. In the following we will drop the vector notation. 
To obtain the full classical action, however, one must use the coherent state path integral with \eqref{eq:classical_H} as the time evolution operator \cite{Kruczenski:2004kw,Bellucci:2004qr}.
This involves discretising total time $T$ into $N$ intervals  $\delta t=T /N$ and inserting resolutions of the identity in terms of coherent states at each step. Importantly, the overlap between (the non-orthogonal) coherent states then introduces a non-trivial phase which will give rise to a Wess-Zumino like term. In the time-continuum (classical)  limit $\delta t \rightarrow 0$ one then arrives at
$Z=\int [\mathrm{d}\mathbf{n}]e^{i S_{\tts{cl.}}}$ with
\[
S_{\tts{cl.}} = \int \mathrm{d}t   \sum_{j=1}^J \left(i\braket{\Dot{\textbf{n}}_{j}}{\textbf{n}_{j}} -\frac{\lambda}{8\pi^2} h^{\tts{cl.}}_{j,j+1}\right) ,
\]
and $\frac{\lambda}{8\pi^2}$ the (effective) spin chain coupling (which originally arises at one-loop  from the planar dilatation operator in ${\cal N}=4$ SYM). At this point, the semiclassical coherent states $\textbf{n}_{j} (t)$ have become dynamical variables that remain constrained to the hyperboloid. 

Passing to the field theory (long chain) continuum limit is achieved by letting the number of sites $J\rightarrow \infty$, while interpreting the site index $j$ as the discretisation of a spatial coordinate $\sigma = \tfrac{2\pi j}{J} \in [0,2\pi]$. The spin vector then becomes a field: $\mathbf{n}_j (t) \rightarrow \mathbf{n}(t,\sigma)$ and $\mathbf{n}_{j+1}-\mathbf{n}_j \rightarrow \frac{2\pi}{J} \partial_\sigma \mathbf{n}$. This gives 
\begin{equation} \label{eq:CS_continuous_limit}
    h^{\tts{cl.}}_{j,j+1}=\log\left(1+\frac{(\textbf{n}_j-\textbf{n}_{j+1})\cdot (\textbf{n}_j-\textbf{n}_{j+1})}{4}\right)= \frac{\pi^2}{J^2} \partial_{\sigma} \textbf{n} \cdot \partial_{\sigma} \textbf{n} +O( J^{-4}),
\end{equation}
so that taking $J\rightarrow \infty$ while keeping $\tilde{\lambda}=\lambda/J^2$  fixed yieds the LL sigma-model (for non-compact ferromagnets) \cite{Stefanski:2004cw,Kruczenski:2004cn,Kruczenski:2004kw}
\begin{equation}\label{eq:CS_Landau_Lifshitz_action}
   S_\tts{{C-LL}}=\frac{J}{2\pi} \int \mathrm{d}t  \int_0^{2\pi} \mathrm{d}\sigma \left( \mathcal{L}_{\tts{WZ}}(\textbf{n})-\frac{\tilde{\lambda}}{8  } \partial_{\sigma} \textbf{n} \cdot \partial_{\sigma} \textbf{n} +{\cal O}(J^{-1})\right) ,
\end{equation}
with the Wess-Zumino term
\begin{equation}\label{eq:CS_WZ_term}
   \mathcal{L}_{\tts{WZ}}(\textbf{n})= i \braket{\partial_t \textbf{n}}{\textbf{n}}=-\frac{1}{2}\int_0^1 \mathrm{d}z \ \varepsilon^{ijk} n_i \partial_z n_j \partial_t n_k , 
\end{equation}
and total Hamiltonian 
\begin{equation} \label{eq:CS_Hamiltonian}
    H_\tts{{C-LL}}=\frac{J \tilde{\lambda}}{16\pi} \int_{0}^{2\pi}  \mathrm{d}\sigma \left[ - (\partial_{\sigma} n_0)^2 + \partial_{\sigma} n_+ \partial_{\sigma} n_-  \right] ,
\end{equation}
which is equivalent to  the effective Hamiltonian \eqref{eq:LL-string-ham} of the string sigma-model in the $SL(2,\mathbb{R})$ sector  under the identification  $\ell_i = n_i$. 
The fields $\mathbf{n}(t,\sigma)$ satisfy the Poisson brackets
\[
    \{n_0(\sigma),n_\pm(\sigma')\}=\pm 2 i n_\pm(\sigma)\delta(\sigma-\sigma'),\qquad     \{n_+(\sigma),n_-(\sigma')\}=-4in_0(\sigma)\delta(\sigma-\sigma'), \label{eq:CS_Poisson_brackets}
\]
which are the  classical counterparts of the $\alg{sl}(2,\mathbb{R})$ commutation relations \eqref{eq:comm-rels-sl2} for the spin operators $\{\h,\e,\f\}$ under the relation \eqref{eq:CS_key_property} and $\{ \cdot , \cdot \} \rightarrow -i [\cdot, \cdot ]$.
The bulk equations of motion of the LL model \eqref{eq:CS_Landau_Lifshitz_action} finally read 
\[
\label{eq:LL_EOM_1}
   \partial_t n_0 &= \frac{i\tilde{\lambda}}{4}\left(  n_+ \partial^2_\sigma n_-  - n_-\partial^2_\sigma n_+\right),\qquad
   \partial_t n_\pm = \pm \frac{i\tilde{\lambda}}{2} \left( n_\pm \partial^2_\sigma n_0 - n_0\partial^2_\sigma n_\pm\right),
\]
which must be solved under the constraint \eqref{eq:CS_casimir} and, in the untwisted case,  periodic boundary conditions, as directly inherited from the periodic spin chain.

\noindent {\bf Jordanian twisted-boundary LL model ---}  
We will now take the LL limit of the  non-compact Jordanian $\mathfrak{sl}(2,\mathbb{R})$ spin chain. One could in principle consider doing this both in the deformed or the twisted-boundary picture. However, in the infinite-dimensional $s=-1/2$ representation of  $\alg{sl}(2,\mathbb{R})$, the Jordanian Drinfel'd twist \eqref{eq:F-jor} is an infinite series of differential operators and correspondingly it will be highly cumbersome to work with the over-complete and non-orthonormal coherent states.\footnote{This is   in contrast to the two-dimensional compact representation, for which the Jordanian twist truncates at first order in $\xi$, since then $\e^2=0$. A deformed LL model obtained from a Jordanian-type Drinfel'd twist in the compact representation was  studied in \cite{deLeeuw:2025sfs}.} We therefore prefer to work with the twisted-boundary picture, which in addition must also straightforwardly connect to the twisted sigma-model of section \ref{s:string-sol}. The LL action from the bulk spin chain then remains undeformed, but we should instead take the classical continuous limit of the twisted-boundary conditions \eqref{eq:jor-expl-X-TBCs}.
A similar route was followed for the non-diagonal DR twist considered in \cite{Guica:2017mtd}.

To evaluate the (quantum) twisted-boundary conditions in the coherent state representation, we also assign a state $\ket{\mathbf{n}_{\sfa}}$ to the defect space. We can then immediately evaluate   \eqref{eq:jor-expl-X-TBCs} in the product state $\ket{\mathbf{N}} := \ket{\mathbf{n}_{\sfa}} \otimes \bigotimes_{j=1}^J \ket{\mathbf{n}_{j}}$.
 Defining
\[
(n_0)_{J+\mathsf{a}} := -2 \bra{\mathbf{N}} \h_{J+\sfa} \ket{\mathbf{N}} , \quad (n_+)_{J+\mathsf{a}} := -2 \bra{\mathbf{N}} \e_{J+\sfa} \ket{\mathbf{N}} , \quad (n_-)_{J+\mathsf{a}} := -2 \bra{\mathbf{N}} \f_{J+\sfa} \ket{\mathbf{N}} ,
\]
then gives
\[
(n_0)_{J+\mathsf{a}} = (n_0)_{\mathsf{a}} , \qquad (n_\pm)_{J+\mathsf{a}} =(n_\pm)_{\mathsf{a}} \left( 1+ \xi M_{\tts{cl.}} \right)^{\pm 1} ,
\]
with the classical value $M_{\tts{cl.}}$ of the symmetry operator $\widehat{M}$ defined as
\[
M_{\tts{cl.}} := \sum_{j=1}^J \bra{\mathbf{n}_j} \e_j \ket{\mathbf{n}_j}  = -\frac{1}{2}\sum_{i=1}^J (n_{+})_j ,
\]
where we used that, by construction, the coherent states satisfy $\bra{\mathbf{n}_j} \e_j^s = \left(-\tfrac{n_+}{2}\right)^s \bra{\mathbf{n}_j} $ for any power $s$. Recall furthermore that $\braket{\mathbf{n}}{\mathbf{n}}=1$ at each site.
In the classical and (two-dimensional) field theory continuum limit this then finally becomes
\[ \label{eq:CS_momentum_constraint}
M_{\tts{C-LL}}:= -\frac{J}{4\pi}\int^{2\pi}_0 \mathrm{d}\sigma' n_+(t, \sigma') ,
\]
so that the LL twisted-boundary conditions read
\[ \label{eq:BC_ns}
    n_0(t,2\pi)&=n_0(t,0), \qquad 
    n_\pm(t,2\pi)=n_\pm(t,0)\left(1+\xi M_{\tts{C-LL}} \right)^{\pm 1} .
\]
Since $\widehat{M}$ was a residual symmetry of the   spin chain (recall eq.~\eqref{eq:charge_commutation} or \eqref{eq:e-symm-shift}) and governs the twisted-boundary conditions, it makes sense to also demand that  solutions to the LL equations of motion   must be such that $M_{\tts{C-LL}}$ remains conserved. This will in fact require to study only a subsector of solutions:
demanding $\partial_t M_{\tts{C-LL}} = 0 $ for $\xi M_{\tts{C-LL}}\neq 0$  introduces  the following additional constraint 
\[ \label{eq:LL-M-constraint}
n_0 (t,0) \partial_\sigma n_+ (t,0 )- n_+(0) \partial_\sigma n_0 (0) = 0 .
\]
It seems that the continuum limit thus spoiled the manifest $M$-conservation of this charge. We believe it would be interesting to understand how to restore it  in the LL theory, and expect this to be related to appropriate global treatment of the hyperboloid fields and WZ term. This may also provide an understanding for the apparent loss of $\e$-symmetry of the Jordanian string sigma-model in the twisted picture, cf.~footnote \ref{f:twisted-noether}. We will however find that for the simplest possible ground state deformations, \eqref{eq:LL-M-constraint} is trivially satisfied.

\subsection{Solutions and energy shift}\label{subsec:LL_spectrum}

We are now interested in  twisted  solutions of the equations of motions \eqref{eq:LL_EOM_1}, subject to the hyperbolic constraint  \eqref{eq:CS_casimir}, with fixed (minimal) energy and momentum $M_{\tts{C-LL}}$. 

In the undeformed, periodic theory, the simplest ground state  is the constant configuration $n_0=1$ and $n_\pm=0$,
which trivially has zero energy $H_{\tts{C-LL}}=0$ and momentum $M_{\tts{C-LL}}=0$. However, 
we can also consider ground state configurations carrying non-zero momentum $M_{\tts{C-LL}}=M$. 
To maintain vanishing energy while allowing non-zero $M$, we take a constant configuration with $n_+ = -\tfrac{2M}{J}$ as required by \eqref{eq:CS_momentum_constraint}. Imposing the hyperbolic constraint \eqref{eq:CS_casimir}  then leads to two distinct possibilities (up to overall factors). Firstly, we can have
\begin{equation}\label{eq:CS_GS_M_1}
n_0=1, \qquad  n_+= -\frac{2M}{J}, \qquad  n_-=0 , 
\end{equation}
which has a smooth  $M\rightarrow0$ limit to the trivial vacuum. This is in contrast to a second  solution with momentum $M$,  given by
\begin{equation} \label{eq:CS_GS_M_2}
n_0=0, \qquad  n_+= -\frac{2M}{J}, \qquad  n_-=\frac{J}{2M}.
\end{equation}
As we will see, the solution \eqref{eq:CS_GS_M_2} is the relevant undeformed solution in the case of the Jordanian deformation, while \eqref{eq:CS_GS_M_1} was the appropriate undeformed solution in the dipole-deformed case \cite{Guica:2017mtd}. 
In both cases, the fields $n_\pm$ are generically complex-valued for real $M$, necessitating an analytic continuation of the LL model in which $n_+$ and $n_-$ are treated as independent complex variables constrained to lie on the complexified hyperboloid. 
Despite the presence of complex fields, the energy and consequently   anomalous scaling dimensions remain real (zero in this case).

Now we turn to the Jordanian twisted LL model and look for deformations of the ground state  solutions with non-zero  momentum $M$, which are subjected to the twisted-boundary conditions \eqref{eq:BC_ns}.
For this purpose, we consider the ansatz $n_0 (t, \sigma ) = c \in \mathbb{R}$ constant,  which is justified by the solutions \eqref{eq:CS_GS_M_1} and \eqref{eq:CS_GS_M_2} and the fact that $n_0$ remains periodic in the twisted model. 
Demanding non-zero momentum, and thus non-zero $n_+$, we can furthermore use the hyperboloid constraint \eqref{eq:CS_casimir} to remove $n_-$ from the system of equations as $n_-= \frac{c^2-1}{n_+}$. 
The equation of motion for  $n_0$  \eqref{eq:LL_EOM_1} then becomes
\begin{equation}
  (c^2-1) \left( (\partial_\sigma n_+)^2-n_+ \partial_\sigma^2n_+ \right)= 0 ,\label{eq:spec_sol_2}
\end{equation}
which is solved by $c = 1$ or
\begin{equation}
    n_+(\sigma,t)=e^{\sigma \alpha_1(t)}\alpha_2(t),\label{eq:spec_sol_3}
\end{equation}
where $\alpha_{1,2}(t)$ are two a-priori arbitrary functions of $t$. In the case $c = 1$ there is no consistent twisted solution of the full system of equations of motion as well as \eqref{eq:CS_momentum_constraint} and the constraint \eqref{eq:LL-M-constraint}. We thus continue with \eqref{eq:spec_sol_3}. Imposing the twisted-boundary conditions for $n_+$ \eqref{eq:BC_ns} then fixes
\begin{equation}\label{eq:spec_sol_4}
    \alpha_1(t)=\frac{\log(1+\xi M_{\tts{C-LL}})}{2\pi} + i m, \qquad m \in \mathbb{Z} ,
\end{equation}
while requiring fixed $M_{\tts{C-LL}}=M$ for the momentum from \eqref{eq:CS_momentum_constraint} imposes
\begin{equation}
    \alpha_2(t)=-\frac{2 \log(1+\xi M) + 4\pi i m}{\xi J},\label{eq:spec_sol_5}
\end{equation} 
such that the solution for $n_+$  has no $t$-dependence.
To have a well-defined $\xi\rightarrow0$ limit in which we recover $n_+=-\tfrac{2M}{J}$  as above we furthermore need to demand $m=0$. 
Inserting the resulting solution for $n_+$ in the remaining  equations of motions \eqref{eq:LL_EOM_1} for non-zero $\xi$ and $M$  then demands $c=0$. Ultimately we thus have 
\begin{equation} \label{eq:jordanian_ground_state}
  n_0(t,\sigma) = 0 , \qquad n_+(t, \sigma)=-\frac{2 \log(1+\xi M)}{\xi J}(1+\xi M)^{\sfrac{\sigma}{2\pi}} , \qquad n_-(t,\sigma) =  -\frac{1}{n_+(t,\sigma)} ,
\end{equation}
coinciding with \eqref{eq:CS_GS_M_2} in the undeformed limit. 
The constraint \eqref{eq:LL-M-constraint} and remaining twisted-boundary conditions for $n_0$ and $n_-$ are furthermore trivially satisfied. 
Evaluating \eqref{eq:CS_Hamiltonian} on this solution then gives the following non-trivial energy shift of the ground state 
\begin{equation}\label{eq:Jordanian_ground_state_energy}
    H_\tts{{C-LL}}=\frac{J \tilde{\lambda}}{32\pi^2}\log\left(1+\xi M\right)^2=\frac{J \tilde{\lambda}}{32\pi^2}\log\left(1+ \frac{2 \pi \eta}{\tilde{\lambda}^{\sfrac{1}{2}} J} M\right)^2= \frac{\eta^2 M^2}{8J} + {\cal O}(J^{-2}) .
\end{equation}
where we employed the identification \eqref{eq:xi-to-eta} and expanded the result at large $J$ but fixed $\eta$, $M$ and $\tilde{\lambda}$. This matches precisely the classical string theory result given in  \eqref{eq:LL_strings_final_result} corresponding to the energy shift of the twisted scaling dimension \eqref{eq:Del-BMN-like} expanded at large $J$ but fixed $\eta,M$. Here we thus have a non-trivial overlap between the energy of the sigma-model BMN-like solution and  the twisted-boundary spin chain description in the coherent state and supergravity approximation.

It is interesting to note that the undeformed ground state configuration does not have a well-defined $M\rightarrow 0$ limit. Related observations have been found at the Jordanian sigma-model side, where no  solutions were found with vanishing $M$. 
Furthermore, the corresponding spectral curve of the BMN-like solution has two branch cuts whose  size is proportional to $\xi M$ \cite{Borsato:2022drc},  and attempts to find no-cut but twisted solutions did not materialise.

\section{Spectrum of the $J=2$ twisted spin chain}\label{sec:small-chain} 

In this section, we study the spectral problem of the Jordanian-twisted $\mathfrak{sl}(2,\mathbb{R})$ spin chain  at length $J=2$ by  directly diagonalising the transfer matrix and using Baxter's $TQ$-equation. 

The traditional method to determine the spectrum is the Algebraic Bethe Ansatz (ABA) \cite{Faddeev:1996iy}, which involves constructing  the monodromy matrix using a two-dimensional ($s=+1/2$) auxiliary space representation  of $\mathfrak{sl}(2,\mathbb{R})$. From its expression,  one  derives the Bethe equations that encode the spectrum of the corresponding transfer matrix. 
While the Hamiltonian of the non-compact $\mathfrak{sl}(2,\mathbb{R})$ spin chain will not be included in its spectral expansion, the eigenvectors of the transfer matrix constructed from the ABA can still be shown to diagonalise the Hamiltonian \cite{Faddeev:1994zg,Derkachov:2002tf}.
In addition, the transfer matrix eigenvalues can  be used as input to Baxter’s equation, from which  physical quantities such as the energy  can be  extracted, see e.g.~\cite{Faddeev:1994zg}.

In our case, however, we cannot directly apply the ABA. 
A key aspect  of this work is to take the  non-compact spin chain  in an eigenstate of the residual $\widehat{M}$ symmetry operator. In this basis, a suitable reference (or ``pseudovacuum'') state required for the ABA is not   available. Without  the ABA, we will instead try to directly solve the eigenvalue problem of the transfer matrix, aiming to uncover at least some of the ``integrable data''. For this purpose, we restrict to the small $J=2$ chain: Since the physical sites are in 
in a non-compact representation, the $\alg{sl}(2,\mathbb{R})$ generators act as differential operators, and  
the eigenvalue problem becomes a PDE with as many independent variables as the number of sites of the spin chain. Although this rapidly becomes intractable for longer chains,\footnote{This difficulty was   one of the  motivations behind the construction of the quantum Separation of Variables method for  non-compact $\alg{sl}(2,\mathbb{R})$ spin chains \cite{Derkachov:2002tf}.} it remains manageable for $J=2$. 

Working in an   $\widehat{M}$ eigenbasis is in  contrast  to  standard approaches,  where one works in an eigenstate of the Cartan of $\mathfrak{sl}(2,\mathbb{R})$. 
In this section, we will therefore first show how to obtain the transfer matrix eigenvalue of the undeformed ($J=2$) chain in the $\widehat{M}$ basis instead of the $\Delta(\h)$ basis. 
This will provide useful structure and benchmark for the Jordanian-twisted case, to which we then apply the identical procedure.  Subsequently, we will substitute its deformed transfer-eigenvalue into Baxter’s expression for the ground state energy.

Finally, note  that one might also consider a more hands-on approach and directly diagonalise the deformed or twisted-boundary Hamiltonian of resp.~\eqref{eq:Ham-def} and \eqref{eq:Ham-twist}. However, this is not immediately straightforward in the $\widehat{M}$ eigenbasis, since the $\Delta^{J-1}(\e)$ operator does not explicitly appear in either Hamiltonian. We will report more on this aspect elsewhere \cite{Driezen:2025izd}.

\subsection{Eigenvalue problem of the undeformed chain}\label{subsec:small-chain-undef} 

As usual, to construct the monodromy and transfer matrix, we take  the finite-dimensional $s=+1/2$ representation of $\mathfrak{sl}(2,\mathbb{R})$ for the auxiliary space $\sfa$. In this case, the generators satisfying the commutation relations \eqref{eq:comm-rels-sl2}, can be represented as in \eqref{eq:rep-2D-sl2R}.
Consider then the $J=2$ monodromy matrix built from the R-matrix (or Lax matrix) defined as
\begin{equation} \label{eq:Lax_matrix} 
    R_{\mathsf{a}j}(u):=u 1_{\mathsf{a}j}+ 2 i X_{\mathsf{a}}\cdot \textbf{S}_{j}
    =\begin{pmatrix}
      u+i \h_{{j}} & - i\f_{{j}}\\
      i \e_{{j}} & u-i \h_{{j}}
    \end{pmatrix},   
\end{equation}
where  the dot product  is defined through  the Cartan-Killing form, the operators $X_{\sfa}=\{\h_{\sfa},\e_{\sfa},\f_{\sfa}\}$ are as in \eqref{eq:rep-2D-sl2R}, while $\textbf{S}_j=\{\h_j,\e_j,\f_j\}$ are the physical spin operators at site $j$ in the  $s=-1/2$ representation,  which we represent as\footnote{This representation  has a quadratic Casimir \eqref{eq:one-site-Casimir} equal to $-\frac{1}{4}$.}
\[
\e_j=-\partial_j,\qquad \h_j=-z_j\partial_j-\frac{1}{2}, \qquad \f_j=-z_j^2\partial_j -z_j, \label{eq:sl_2_explicit_rep}
\]
with $z_{j}$ the representation space parameters at each site $j=1,2$.
In general, one then writes
\begin{equation}
    T_\mathsf{a}(u)=R_{\mathsf{a}2}(u)R_{\mathsf{a}1}(u)=\begin{pmatrix}
                A(u) & B(u)\\
                C(u) & D(u)
            \end{pmatrix},
    \qquad \hat{\tau}(u)=\Tr_\mathsf{a}(T_\mathsf{a}(u))=A(u)+D(u), \label{eq:ABA_undef}
\end{equation}
where $A,B,C,D$ are built out of the differential spin $\textbf{S}$ operators.  In the undeformed case, we have
\[ \label{eq:undef-transfer}
\hat{\tau}(u) = 2u^2 + \e_2 \f_1 + \f_2 \e_1 - 2 \h_2 \h_1 .
\]

The differential operator structure of the transfer matrix implies that we should work within a functional representation of $\mathfrak{sl}(2,R)$, where states are realised as functions of the continuous site parameters $z_1$ and $z_2$. As per illustration,  the undeformed case with $s=-1/2$, where one can use an eigenbasis of the total Cartan operator $\Delta(\h)=\h_1+\h_2$, there is a highest-weight module with basis states
\begin{equation}\label{eq:usual_states}
    \ket{k_1,k_2}:=z_1^{k_1} z_2^{k_2}, \qquad k_1,k_2 \in \mathbb{N} ,
\end{equation}
with eigenvalue $-1-k_1-k_2$ and with $\ket{0,0}$ annihilated by $\Delta(\e)$. 
These are thus regular polynomials in $z_j$  and that have a discrete spectrum under $\Delta(\h)$. Note that they are also easily generalisable to longer chains. 
One can then straightforwardly apply the ABA on the above basis. In fact for $J=2$, the state $\ket{0,0}$ is  a suitable  pseudo-vacuum, satisfying $C(u)\ket{0,0}=0$.\footnote{Under automorphisms of the algebra, this  is of course equivalent to a $B(u)$ annihilation.} Since  $\hat{\tau}(u)$ commutes with $\Delta(\h)$,  one can diagonalise it sector by sector in $\Delta(\h)$ eigenvalues \cite{Beisert:2003jj}. 

Under the Jordanian twist, however, this eigensystem becomes non-diagonalisable \cite{Borsato:2025smn}, 
as $\Delta(\h)$ is not a symmetry, and the weight structure is obscured in terms of generalised eigenvectors. 
The above  states \eqref{eq:usual_states} are obviously not  eigenstates of $\widehat{M} = \e_1 + \e_2$, which is a raising operator with respect to $\Delta(\h)$. In certain deformations (e.g.~the abelian dipole \cite{Guica:2017mtd} or Jordanian twists), $\widehat{M}$ is the only conserved quantity in the $\mathfrak{sl}(2,\mathbb{R})$ sector, making it natural and advantageous  to diagonalise the transfer matrix in an $\widehat{M}$ eigenbasis from the start. 

Hence, we instead consider eigenfunctions of the (non-compact) differential operator $\widehat{M}=\e_1+\e_2=-\partial_1-\partial_2$, which take the form
\begin{equation}
    \Psi_M(z_1, z_2) := e^{-M x} g(z), \quad \text{with}\quad x:=\frac{z_1+z_2}{2},\quad \text{and} \quad z:=z_2-z_1,\label{eq:J=2_M_eigenstates}
\end{equation}
and have eigenvalue $M$. As we will see, these functional states will generally not be polynomial. Eigenstates of a root generator such as here naturally arise in the discrete series of $\mathfrak{sl}(2,R)$, see e.g.~\cite{Barut:1970qf,Perelomov1977}, but they can exist also in the principal series.\footnote{For the eigenvalue problem, it will be enough to simply consider \eqref{eq:J=2_M_eigenstates}, but it would be interesting to have a more systematic identification of the underlying module structure to clarify the spectral behaviour of the twisted model. }

 One can check that there are no consistent and non-trivial solutions of $C(u)\Psi_M = 0$ (or $B(u)\Psi_M=0$) in the Hilbert space of eigenfunctions, so that the ABA is not straightforwardly applicable.  
Instead we aim to directly solve
\begin{equation}
    \hat{\tau}(u)\Psi_M=\tau(u)\Psi_M,\label{eq:J=2_EV_eq_undef_full}
\end{equation}
with  $\tau(u)$ the eigenvalue of the transfer matrix.  
Since $\hat{\tau}(u)$ is a  polynomial in the spectral parameter $u$, we can rewrite this as an eigenvalue equation for each term in the $u$ expansion (i.e.~3 equations at $J=2$). Writing $\tau(u)=\alpha u^2 + \beta u + \chi$ we then immediately see from \eqref{eq:undef-transfer} that $\alpha=2$ and $\beta=0$, while the constant $u^0$ term yields the following ODE for $g(z)$ 
\[ \label{eq:J=2_EV_eq_undef_chi}
4 z^2 g''(z) + 8 z g'(z) + (4 \chi + 2-M^2 z^2) g(z) = 0 .
\]
Its general solution is given by
\begin{equation}
    g(z)=c_1 \mathrm{SJ}\left(-\frac{1}{2}\left(1+ i\sqrt{1+4 \chi}\right);-\frac{i}{2} M z\right) +c_2 \mathrm{SY}\left(-\frac{1}{2}\left(1+ i\sqrt{1+4 \chi} \right);-\frac{i}{2} M z\right),\label{eq:spherical_bessel}
\end{equation}
where $\mathrm{SJ}(n;y)$ and $\mathrm{SY}(n;y)$ are the spherical Bessel functions of the first and second kind and $c_{1,2}$ are integration constants. 
To ensure the well-definedness of the eigenfunction $\Psi_M$ and the action of spin operators, we further impose regularity and single-valuedness of $\Psi_M$ across the representation space. The solution involving $\mathrm{SY}$  diverges at $z=0$ and we thus excluded it, setting $c_2=0$ (and effectively one can then also set $c_1=1$). The remaining function $\mathrm{SJ}(n;y)$ is regular at $y=0$, but it is single-valued only when $n \in \mathbb{N}$. Imposing this condition then leads to the quantisation
\begin{equation}
    \chi=-\frac{1}{2}-n(n+1), \qquad n\in\mathbb{N}  .  \label{eq:regularity_undef} 
\end{equation}

Since the transfer matrix includes conserved charges in its spectral expansion, the eigenvalue $\chi$ is a conserved charge associated to the $u^0$ operator in $\hat{\tau}(u)$. Importantly, this operator can be written as
\begin{equation}
    \hat{q}_2:=\e_2 \f_1 + \f_2 \e_1 - 2 \h_2 \h_1 = - \frac{1}{2}-\mathbb{C}_2, \label{eq:J=2_two_site_casimir_undef}
\end{equation}
where $\mathbb{C}_2:=\Delta(\textbf{S}^2)=\Delta(\textbf{S})\cdot \Delta(\textbf{S})$ is the two-site Casimir and we used $s=-1/2$. The integer $n$ appearing in the solution \eqref{eq:regularity_undef} is thus simply the quantisation of  total spin,  $\mathbb{C}_2=n(n+1)$. 
Recalling the expressions of the Hamiltonian of the undeformed chain, given in
\eqref{eq:sl(2)_hamiltonian}, and using the fact that on each module $\mathbb{C}_2=\mathbb{J}(\mathbb{J}+1)=n(n+1)$ one has $\mathbb{J}=n$, 
the ground state  of the spin chain clearly corresponds to $n=0$, for which $\chi =-\tfrac{1}{2}$.
The corresponding eigenfunction with non-vanishing $M$  is 
\begin{equation}
    \Psi_{\tts{GS},M}(z)= e^{-M x} \mathrm{SJ}\left(0;-\frac{i}{2}M z\right)=\frac{2}{M z} e^{-M x} \sinh\left(\frac{M z}{2}\right) . \label{eq:J=2_GS_undef}
\end{equation}
We can
expand it around $z=0$ to find
\[ \label{eq:Psi-undef-exp}
\Psi_{\tts{GS},M}(z)= e^{-M x} \left( 1+ \frac{M^2z^2}{24}+\frac{M^4z^4}{1920}+\frac{M^6z^6}{322560} + {\cal O}(z^7) \right) , 
\]
which will provide an important consistency check for the Jordanian case that we  discuss now.

\subsection{Eigenvalue problem of  Jordanian-twisted spin chain}\label{subsec:small-chain-Jordanian}  
We now proceed analogously for the Jordanian twisted spin chain.
It will be convenient to do so in the twisted-boundary picture, where the non-local similarity transformation $\Omega_{1 \cdots J}$ has been absorbed in a rotation of the Hilbert space. 
The corresponding twisted monodromy is given in \eqref{eq:jor-twisted-monodromy}, with importantly  $\widehat{M}$ acting initially on the states.   
We write  
\begin{equation}
    T^\star_\mathsf{a}(u)=\begin{pmatrix}
                1& \xi\Delta(\h)\\
                0&1
              \end{pmatrix} 
              \begin{pmatrix}
                A(u) & B(u)\\
                C(u) & D(u)
              \end{pmatrix}
              \begin{pmatrix}
                (1+\xi \widehat{M})^{-1/2}&0\\
                0&(1+\xi \widehat{M})^{1/2}
              \end{pmatrix},\label{eq:twisted_monodromy_J=2}
\end{equation}
and
\begin{equation}
    \hat{\tau}_\xi(u):= \Tr T^\star_\mathsf{a}(u) = \left(A(u)+\xi \Delta(\h) C(u)\right)(1+\xi \widehat{M})^{-1/2} + D(u) (1+\xi \widehat{M})^{1/2}\label{eq:twisted_transfer_matrix_J=2}, 
\end{equation}
with $A,B,C,D$ as obtained from \eqref{eq:Lax_matrix} for two sites, as before.    
We then again consider the corresponding eigenvalue problem on the $\widehat{M}$ eigenstates $\Psi_M$ defined in \eqref{eq:J=2_M_eigenstates}. The initial action of $(1+\xi \widehat{M})^{\pm 1/2}$ can  be immediately replaced by the constants $(1+\xi M)^{\pm 1/2}$. As before, we then expand both the twisted transfer matrix and its eigenvalue in the spectral parameter and solve the eigenvalue equation order by order in $u$. Writing $\tau_\xi (u)=\alpha_\xi u^2+\beta_\xi u +\chi_\xi$,  one then finds at ${\cal O}(u^2)$ an algebraic equation that immediately sets
\begin{equation}
     \alpha_\xi =\frac{2+ \xi M}{\sqrt{1+ \xi M}} . \label{eq:J=2_deformed_EV_eq_u^2}
\end{equation}
The ${\cal O}(u^2)$ term of the transfer matrix eigenvalue thus clearly gets a $\xi M$ correction. On the other hand, there  is no ${\cal O}(u^1)$ term, and thus $\beta_\xi =0$ as in the undeformed case. Finally, the ${\cal O}(u^0)$ term yields a differential equation for $g(z)$ of the form
\begin{equation} 
    p_1(z) g(z) + p_2(z) g'(z) + p_3(z) g''(z) - 8 \xi z^2 g'''(z)=0, \label{eq:J=2_deformed_EV_eq_u^0}
\end{equation}
where 
\[ \label{eq:p-s}
    p_1(z)&:=4 + M (2 \xi - M z (-2 \xi + z (2 + \xi M))) + 
 8 \sqrt{1 + \xi M} \chi_\xi,\\
    p_2(z)&:=2 (8 z -4 \xi + \xi M z (4 + M z) ),\\
    p_3(z)&:=4 z (2 z - 6 \xi + M z \xi ).
\]
For non-zero $\xi$, this differential equation is  significantly more complicated than in the undeformed case: it is now a 3rd-order ODE, rather than 2nd-order, due to the additional action of $\Delta (h)$ induced by the Jordanian twist. 
The ODE has a singularity of the second kind at $z=0$ (that is, a $z^2$ coefficient in front of the $g'''(z)$ term), and exhibits a turning point behaviour around $\xi = 0$, where the order of the ODE changes. 
In appendix \ref{app:dif_eq}, we give a general explanation of how to  ``perturbatively'' construct  solutions to such singular ODEs as formal series around $z=0$, using our case as an illustrative example. 
As it turns out, the form of the solution can be written as
\begin{equation}
g(z)=g_1(z)+\log(z)g_2(z)+\log^2(z)g_3(z),\label{eq:J=2_dif_eq_}
\end{equation}
with $g_{1,2,3}(z)$ regular expansions around $z=0$, whose coefficients are determined through recurrence relations found from solving eq.~\eqref{eq:app_frob_recurrence} with \eqref{eq:app_leading_non_resonant}, and with the values $g_{1,2,3}(0)$ playing the role of integration constants. We have computed the expansion coefficients explicitly until ${\cal O}(z^8)$ in \textsc{Mathematica}.  As in the undeformed case, we again need to impose regularity and single-valuedness at $z=0$, such that we have to set $g_2(0)=g_3(0)=0$. Recursively,  we can then immediately conclude that this imposes $g_2(z)=g_3(z)=0$. We are thus left with a single regular expansion around $z=0$, given by $g_1(z)$, to which we can apply the usual Frobenius methods. We will set the last integration constant to $g_1(0)=1$. 

At this stage, we already observe   an interesting  turning point behaviour in the differential equation: In the undeformed case, the indicial equation  which determines the leading power $p$ of the solution to  \eqref{eq:J=2_EV_eq_undef_chi} in the context of Frobenius' method reads 
\[
4 + 8 p (1 + p) + 8 \chi =0 . \label{eq:indicial_undef}
\]
Requiring regularity  corresponds to demanding $p\in \mathbb{N}$, which precisely  coincides with the quantisation condition for $\chi$ given in \eqref{eq:regularity_undef}. In other words, the power $p$ coincides with the energy excitation number $n$ above the ground state.  However, in the deformed case, the indicial equation for \eqref{eq:J=2_deformed_EV_eq_u^0} reads
\[
-8\xi p^3=0, \label{eq:indicial_def}
\]
where  $p$ will be the leading exponent of $g_1(z)$ in \eqref{eq:J=2_dif_eq_}. For generic and fixed $\xi$, the only admissable Frobenius index is  $p=0$, meaning that the deformed eigenfunction $g_1(z)$ must behave as a constant around $z=0$, and analytic continuation to excited states with $p>0$ is obstructed in this analysis, even  as $\xi \rightarrow 0$. 
In other words, within the Frobenius treatment of the eigenvalue equation at fixed $\xi$,   we find that the Jordanian deformation connects smoothly only to the undeformed \textit{ground state} solution, and not to the undeformed excited states of higher total spin $n=p>0$. 
This is an immediate consequence of the discrete order change of the ODE induced by the additional action of the twist. 
While our analysis thus suggests a lifting of the excited stated after the twist,
this should more properly be interpreted as a limitation of the perturbative Frobenius method for turning point problems,  and the lack of a non-perturbative closed form analysis of the ODE.  We will give a few further comments on these points in a discussion at the end of this section.

Let us  now solve along the lines of Frobenius, considering the ansatz 
\[
g_1(z)=\sum_{n=0}^\infty g_nz^n,
\]
and substitute it into  the ODE \eqref{eq:J=2_deformed_EV_eq_u^0}. After some straightforward algebra, we then obtain the recurrence relation 
\[ \label{eq:frobenius_recurrence}
g_n={}& \frac{1}{8 \xi n^3 }\Bigg(   \left(2\left(2 n^2-2 n+1\right) (2+\xi M )+8 \sqrt{1+\xi M } \chi _{\xi }\right)g_{n-1}  \\
& +2 \xi (n-1) M^2  g_{n-2} -\left(M^2  (2+\xi M )\right) g_{n-3}  \Bigg), 
\]
which is induced by  $g_0= 1$ and $g_{-1}=g_{-2}=0$.
The first three terms explicitly read
\[
    g_1(z)={}&1 + \frac{z}{4\xi} \Bigg(2 +\xi M + 4 \sqrt{1+\xi M} \chi_\xi\Bigg)  \\
                 {}+& \frac{z^2}{128\xi^2} \Bigg(9 \xi ^2 M^2+24 \chi_\xi \sqrt{1+\xi  M} (2+\xi  M)+(1+ \xi  M)(20+16\chi_\xi^2)\Bigg)  +{\cal O}(z^3) ,\label{eq:J=2_g1}
\]
and one can in principle obtain arbitrarily many expansion coefficients recursively from \eqref{eq:frobenius_recurrence}.

We have thus obtained a perturbative solution to the differential equation as a regular expansion around $z=0$. However, in contrast to the undeformed case, we did not yet obtain  a constraint on the eigenvalue $\chi_\xi$. 
Such a constraint arises by requiring the solution to have a smooth limit as
$\xi\rightarrow0$: Since $\xi$ plays the role of a deformation parameter, we must also demand that $g_1(z)$ remains  regular in this limit.\footnote{We have verified that swapping the order of demanding regularity in $z$ and $\xi$ did not influence our conclusions.} However, looking at the expansion \eqref{eq:J=2_g1}, we observe that it is effectively an  expansion in   ${z}/{\xi}$,
with coefficients depending polynomially on $\chi_\xi$. As a result,  regularity at $\xi=0$ 
requires that $\chi_\xi$ itself admits a regular expansion in $\xi$, of the form
\begin{equation}
    \chi_\xi=\sum_{n=0}^\infty c_n \xi^n ,\label{eq:J=2_chi_expansion}
\end{equation}
with the coefficients $c_n$ fixed order by order in an expansion of small $\xi$. 

Consider therefore the ${\cal O}(z)$ term in \eqref{eq:J=2_g1}, whose expansion around $\xi=0$ reads
\begin{equation}\label{eq:J=2_c0}
    \frac{1}{4\xi}\Bigg(2 +\xi M + 4 \sqrt{1+\xi M} \chi_\xi\Bigg)=\frac{2+4 c_0}{4\xi}+\frac{M(1+2c_0)+4 c_1}{4}+{\cal O}(\xi) . 
\end{equation}
In order for this expression to be regular at $\xi=0$, the pole must vanish, which imposes $c_0=-\tfrac{1}{2}$.
This  corresponds precisely to the undeformed ground state solution, i.e.~\eqref{eq:regularity_undef} at $n=0$,  as expected from the analysis of the indicial equation, eq.~\eqref{eq:indicial_undef}.   

Similarly, expanding the  $O(z^2)$ term of \eqref{eq:J=2_g1} gives  
\begin{align}
    \frac{c_1}{4\xi}+\frac{3 M^2 + 16 M c_1 + 8 c_1^2+16 c_2}{64}+{\cal O}(\xi),\label{eq:J=2_c1}
\end{align}
which sets $c_1=0$. 
One can proceed analogously, order by order in $z$, to fix (in principle) arbitrary coefficients of $\chi_\xi$. Explicitly, using the solution of $g_1(z)$ to  ${\cal O}(z^8)$, we can fix $\chi_\xi$ to   ${\cal O}(\xi^7)$, giving
\begin{equation}
    \chi_\xi=-\frac{1}{2}-\frac{\xi ^2 M^2}{48}+\frac{\xi ^3 M^3}{48}-\frac{667 \xi ^4 M^4}{34560}+\frac{307 \xi ^5 M^5}{17280}-\frac{286211 \xi ^6 M^6}{17418240}+{\cal O}(\xi
   ^7).\label{eq:J=2_chi_def}
\end{equation}
Notice in particular that this is an expansion in $\xi M$. This can in fact be expected from the string relation \eqref{eq:Del-BMN-like}. In particular, for eigenstates with vanishing momenta $M$, the spectrum will thus not change. The same behaviour is observed in \cite{Guica:2017mtd}.
Finally, inserting the expansion for $\chi_\xi$  into the  regular solution \eqref{eq:J=2_g1} yields 
\[
    g_1(z)={}& 1+ \frac{M^2z^2}{24}+\frac{M^4z^4}{1920}+\frac{M^6z^6}{322560} +{\cal O}(\xi^0, z^7)\\
           &+ \xi M\left(\frac{Mz}{24}+\frac{M^3 z^3}{960}+\frac{M^5z^5}{107520}+{\cal O}(\xi^0, z^6)\right) \\
           &+\xi^2 M^2\left(- \frac{M z}{48}  + \frac{M^2 z^2}{1080} - \frac{M^3 z^3}{1920} + \frac{M^4 z^4}{60480} + {\cal O}(\xi^0, z^5)\right)+{\cal O}(\xi^3).\label{eq:J=2_deformed_g}
\] 
This expansion is organised as a double expansion in $\xi$ and $z$, 
where the highest power of $z$ at each order in $\xi$ is constrained by the structure of the original $g_1(z)$ expansion in $z/\xi$ and the imposed regularity at $\xi=0$. 
For example, the ${\cal O}(\xi^0, z^7)$ term depends on the as-yet undetermined coefficient $c_7$  of \eqref{eq:J=2_chi_expansion}, and similarly for higher order terms.

Crucially, the ${\cal O}(\xi^0)$ part of 
\eqref{eq:J=2_deformed_g} exactly matches the expansion of the undeformed ground state eigenfunction \eqref{eq:Psi-undef-exp}. This  provides a non-trivial consistency check because the ${\cal O}(\xi^0)$ contributions in \eqref{eq:J=2_deformed_g} arise cumulatively from all orders in the expansion of 
 $\chi_\xi$, such as the constant terms in \eqref{eq:J=2_c0} and \eqref{eq:J=2_c1}. In particular, the coefficients $c_n$ in \eqref{eq:J=2_chi_expansion} were not chosen to reproduce the undeformed solution---rather, they are uniquely fixed by the requirement of regularity at $\xi=0$.

In summary, we have solved  the eigenvalue problem of the $J=2$  Jordanian spin chain perturbatively in the twisted-boundary picture, with the ${\cal O}(u^0)$ term of the transfer matrix eigenvalue as an expansion in $\xi M$
\begin{equation}
   \tau_\xi(u)= u^2\frac{2+ \xi M}{\sqrt{1+ \xi M}}    -\frac{1}{2}-\frac{\xi ^2 M^2}{48}+\frac{\xi ^3 M^3}{48}-\frac{667 \xi ^4 M^4}{34560}+\frac{307 \xi ^5 M^5}{17280}+O(\xi^6) 
   ,\label{eq:J=2_deformed_eigenvalue}
\end{equation}
as well as the eigenfunction
\begin{equation}
    \Psi_M(z_1,z_2)=e^{-M\frac{z_1+z_2}{2}}g_1(z_2-z_1),\label{eq:J=2_deformed_eignefunction}
\end{equation}
with $g_1(z)$ given as an expansion in $\xi M$ and $z$ in \eqref{eq:J=2_deformed_g}. Both expansions can be computed in principle to arbitrary order, as they all result from the recurrence relations  \eqref{eq:frobenius_recurrence},  but the complete solution should be understood as  a formal series  (an infinite-degree polynomial). We however do not exclude the existence of a closed form solution.

As in the undeformed case, the $u^0$ coefficient of the transfer matrix defines an operator that commutes with the Hamiltonian. In the twisted setting, we denote this operator as $\hat{q}_{2,\xi}$,  with eigenvalue $\chi_\xi$ \eqref{eq:J=2_chi_def} on the $\widehat{M}$ eigenstates. Its explicit form is
\begin{equation}
\hat{q}_{2,\xi}=\Big(\e_1\f_2-\h_1\h_2+\xi\Delta(\h)(\e_1\h_2+\h_1\e_2)\Big)(1+\xi \Delta(\e))^{-1/2}+\Big(\e_2\f_1 -\h_1\h_2\Big)(1+\xi \Delta(\e))^{1/2}, \label{eq:J=2_twisted_casimir?}
\end{equation}
which reduces to the undeformed operator $\hat{q}_2 = -\tfrac{1}{2} - \mathbb{C}_2$ \eqref{eq:J=2_two_site_casimir_undef} in the  limit $\xi \rightarrow 0$, with $\mathbb{C}_2$ the two-site Casimir. 
 One may thus think of it as  containing a ``twisted two-site Casimir'' that knows about the boundary conditions, although its  precise physical or algebraic interpretation is not immediately obvious to us. 

\noindent {\bf Discussion --- }We conclude with a discussion of our results in this section and the previous one. First  let us relate to the recent work \cite{Borsato:2025smn}, which studied the the spectral problem of the same model considered here. To match our setup with theirs, one must perform the algebra autormorphism $\h\rightarrow - J^3$, $\e\rightarrow  J^{-}$, $\f\rightarrow  J^{+}$ and also $ \xi \rightarrow 2\xi$. 
Under this transformation, the lowering operator $\Delta^{J-1}(J^-)$ becomes a residual symmetry. One can then consider its  eigenstates in the $J=2$ sector, in which case \eqref{eq:J=2_M_eigenstates} is modified  to 
\[ \label{eq:Jmin-eigenstates}
\Psi_M=\frac{1}{z_1z_2}\exp^{\frac{M}{2}\left(\frac{1}{z_1}+\frac{1}{z_2}\right)}g\left(\frac{1}{z_2}-\frac{1}{z_1}\right) ,
\]
where  the function $g(z)$ with $z=\frac{1}{z_2}-\frac{1}{z_1}$ satisfies the same differential equation \eqref{eq:J=2_dif_eq_}, with the same deformed eigenvalue \eqref{eq:J=2_chi_def}. 

In \cite{Borsato:2025smn},  the ABA for the Jordanian-twisted $J=2$ spin chain was applied in the deformed picture by considering the usual $\alg{sl}(2,\mathbb{R})$ states as defined in \eqref{eq:usual_states} (but in the lowest-weight equivalent representation). Although $\Delta(\h)$ is not a residual symmetry of the chain, it was observed that the maximal value of $\Delta(\h)$ within a subsector is preserved. This is due to the form of the  deformed transfer matrix, which is constructed using a twisted $R$-matrix, where the Jordanian twist acts with the lowering operator $J^-$ first  on the physical chain.
In any given subsector with a fixed maximal value of $\Delta(\h)$, 
it was  then found that the deformation  contributes only to the strictly upper-triangular part of the  transfer matrix, 
and that its spectrum  is not deformed. Similar arguments led to the same conclusion for the  Hamiltonian of the spin chain.

In contrast, our analysis finds that the eigenvalue of the transfer matrix does depend on $\xi$, showing that  the Jordanian twist  \textit{does} affect the spectrum.  Crucially, our computation differs from the one of \cite{Borsato:2025smn} by a key aspect mentioned throughout, i.e.~we work with $\widehat{M}=\Delta(\e)$ eigenstates of a residual symmetry. For $M\neq 0$, these states exist only in the non-compact representation of $\alg{sl}(2,\mathbb{R})$, and do not admit a finite (maximal) value of $\Delta(\h)$. As a result, we cannot restrict the eigenvalue problem to an algebraic problem with finite size matrices, because   non-trivial $\Delta^{J-1}(J^-)$ eigenstates \eqref{eq:Jmin-eigenstates} are  formal infinite series over the Fock basis states $\ket{k_1,k_2}$ as in \eqref{eq:usual_states} (but in the lowest-weight equivalent representation). The only  $\Delta^{J-1}(J^-)$ eigenstates that are finite linear combinations of $\ket{k_1,k_2}$ can only be those with an $M=0$ eigenvalue: Since $\Delta^{J-1}(J^-)$ is a lowering operator, one could consider
\[
\Psi = \sum_{k_1=0,k_2=0}^{N_{\tts{max.}}} c_{k_1,k_2} \ket{k_1,k_2} ,
\]
with some coefficients $c_{k_1,k_2}$ and a finite upper cut-off $N_{\tts{max}.}$, but acting with $J^-$ will of course never give the same states back; Hence,  demanding $\Delta^{J-1}(J^-)\Psi = M \Psi$ immediately forces $M=0$ or $\Psi=0$. A trivial example is  the lowest-weight $\ket{0,0}$, which trivially has $M=0$, as well as $\ket{0,1}-\ket{1,0}$, etc. 
As we found in this section, the eigenvalue of the transfer matrix \eqref{eq:J=2_deformed_eigenvalue} is crucially a function of the combination $\xi M$, and thus states in a sector of vanishing charge $M=0$ do not exhibit a deformed spectrum. 

Similar conclusions were reached in the case of the non-diagonal abelian twist (the dipole deformation) of \cite{Guica:2017mtd}. Although the Jordanian twist considered here is more intricate due to it non-abelian nature,  both deformations preserve the  same  residual symmetries in the  $\alg{sl}(2,\mathbb{R})$ subsector. 
While the explicit form of  the eigenstate and eigenvalue  thus differ,  the dependence on the combination $\xi M$ is common. As mentioned, this is consistent with the classical string relation \eqref{eq:Del-BMN-like}  (which is in fact the same in both models). Conversely, a truncated analysis at fixed maximal value of $\Delta(\h)$ would also yield an undeformed spectrum in the dipole case.

Nevertheless, it remains unclear whether the  Jordanian  twisted Hamiltonian is fully diagonalisable over the complete Hilbert space. 
In the undeformed case, the spectrum  is organised via the two-site Casimir and by working in either the $\widehat{M}$ or $\Delta(\h)$ eigenbases,  both of which eventually label the total spin modules. In the deformed case, however, only $\widehat{M}$ remains, while $\mathbb{C}_2$ and $\Delta(\h)$ are non-conserved.\footnote{It would however be interesting to explore whether a twisted version of the Casimir \eqref{eq:J=2_twisted_casimir?} could play a role in the deformed spectrum.}  Within ODE analysis the Jordanian deformation connects smoothly only to the undeformed ground state. No  Frobenius expansion led to excited (higher-spin) solutions.
We believe that obtaining  a closed form solution to the eigenvalue differential equation would help to understand whether we can recover  the full spectrum.\footnote{A closed form to the ODE in the dipole case was obtained in \cite{Guica:2017mtd}, which showed a smooth connection to the undeformed higher-spin states labelled by $S$ therein. }
While we do obtain a consistent subsector of  eigenfunctions with $\widehat{M}$-charge, it is in particular not clear whether these form a complete basis for the Hilbert space or if non-diagonalisable sectors (Jordan blocks) persist. 
It would thus also be valuable to investigate whether  an alternative basis or  algebraic perspective can reincorporate the ``missing'' excitations (like $\ket{0,1}+\ket{1,0}$, etc.) in the deformed setting. 

All in all, it is obvious that the undeformed limit is  a subtle point.\footnote{A possible reason for this may be the equivalence of Jordanian sigma-models with a non-abelian T-duality transformation over the   Borel subalgebra of $\mathfrak{sl}(2,R)$ after deformation-dependent field redefinitions,   and it would be interesting to understand that also in this context.} Interestingly,  a similar discontinuity was in fact observed at the sigma-model side, where the diagonalisation of the monodromy matrix of the Jordanian sigma-model does not connect smoothly to  the undeformed case due to a change of order in the asymptotic expansion of the quasimomenta \cite{Borsato:2021fuy,Borsato:2022drc}.   

To end, let us mention that it is not straightforward to extract the energy spectrum of the spin chain from the twisted Hamiltonian to arbitrarily high order in $\xi$---in fact, its action  on the $\widehat{M}$ eigenstates is complicated by the non-local intertwiner present in its boundary term. 
In the dipole deformation of \cite{Guica:2017mtd} a way out of this was to use the light-ray representation of \cite{Belitsky:2004sc} extracted from the corresponding non-commutative dipole deformation of $N=4$ SYM obtained via the Seiberg-Witten map. Unlike their case, such a formulation for the Jordanian twist is an a-priori open problem. 
Alternatively, however, one can study the  Baxter's equation representation of the spin chain.

\subsection{Energy from Baxter's equation}\label{subsec:J=2_qSoV}

In this section, we compute the energy of the deformed ground state of the Jordanian-twisted $J=2$  spin chain using  our result \eqref{eq:J=2_deformed_eigenvalue}. As we find ourselves in the situation where the traditional Bethe ansatz techniques cannot provide a full description of the model, we have  to resort to more advanced methods for this purpose. One such approach is the Quantum Separation of Variables (qSoV), pioneered by Sklyanin \cite{Sklyanin:1991ss},  which has been successfully applied to many cases where the ABA is not applicable \cite{Derkachov:2002tf,Kirch:2004mk,Gaudin:1992ci}. While a complete implementation of the qSoV formalism is beyond the scope of this work, it  nonetheless provides a practical way to extract the ground state energy,  and  may give indications to future applications of qSoV.

For $\alg{sl}(2,\mathbb{R})$ based systems, the central object is  the following Baxter TQ-relation
\[
  \Delta^{+}(u) Q(u-i)+\Delta^{-}(u) Q(u+i)=\tau(u) Q(u),
\]
where we used the notation $f^{[k]}(u):=f(u+i k s)$ and $f^\pm(u)=f^{[\pm 1]}(u)$. Here $\tau(u)$ is the transfer matrix eigenvalue, $\Delta(u)$  a model specific coefficient, and $Q(u)$ is Baxter's $Q$-function  \cite{Baxter:1971sapm,Baxter:1972aop},  defined up to  multiplication by $i$-periodic functions in $u$.  In the undeformed, $\alg{sl}(2,\mathbb{R})$ $\mathrm{XXX}_s$  spin chain with $s=-1/2$,\footnote{Note that compared to \cite{Derkachov:2002tf,Kirch:2004mk}, where the Casimir is $C=s(s-1)$, one should send $s\rightarrow -s$.} Baxter's equation reads \cite{Derkachov:2002tf,Kirch:2004mk} 
\[ \label{eq:TQ-XXX}
(u-i/2)^JQ(u-i) +(u+i/2)^J Q(u+i) =\tau(u) Q(u) .
\]
This form was originally obtained from reformulations of the Bethe Ansatz equations, but it can also be derived directly via representation theory of Yangian algebras, where it arises as consequences of fusion relations,
see \cite{Bazhanov:2010ts,Bazhanov:2010jq} and \cite{Frenkel:2013uda,Felder_2017,Zhang:2018sej}. 
Importantly, Drinfel'd twists preserve the level-0 algebra  and the regularity of the $R$-matrix.
It has also be  shown explicitly that the form \eqref{eq:TQ-XXX} remains valid for chains deformed by 
abelian twisted-boundary conditions \cite{Bazhanov:2010jq,Frassek:2011aa,Kazakov:2018ugh,Guica:2017mtd}, but with modified analytic structure for $\tau(u)$ and $Q(u)$.
All of these models can be formulated as closed undeformed spin chains which are Yangian-invariant in the bulk, while their boundary conditions break the Yangian symmetry.\footnote{This also includes the undeformed XXX-chain with {periodic} boundary conditions  \cite{Loebbert:2016cdm}.} It is therefore expected that Baxter's equation is formally of the same form in all models based on  rational $\alg{sl}(2,\mathbb{R})$  $R$-matrices, i.e.~with the same coefficients $\Delta(u)$, even when the boundary breaks the symmetry. While we are not aware of a general algebraic proof,   \cite{Frenkel:2013uda,Felder_2017,Zhang:2018sej} and references therein support this expectation.

In the case of the Jordanian-twisted chain, we are as of yet not in a position to derive  both $Q(u)$ and $\tau(u)$ from first principles (that is, in the absence of ABA and a known complete analytic structure in $u$). Instead, we reverse-engineer: As argued, we assume that  the $TQ$-equation \eqref{eq:TQ-XXX} retains its form in the twisted-boundary picture, i.e.~with the same coefficients $\Delta(u)$, but with modified analytic structure of  $\tau_\xi (u)$ and $Q_\xi(u)$
\[ \label{eq:TQ-XXX-def}
(u-i/2)^JQ_\xi(u-i) +(u+i/2)^J Q_\xi(u+i) =\tau_\xi(u) Q_\xi(u) .
\]
We then determine $Q_\xi(u)$ from the explicit result \eqref{eq:J=2_deformed_eigenvalue} of the transfer matrix $\tau_\xi (u)$ for the deformed ground state. 
The Baxter equation thus serves here merely as a reformulation of the result and may offer a starting point for generalisations. We will however use that the derived $Q_\xi(u)$-functions around $u=0$ typically  encode the energy of the $\mathrm{XXX}_{-1/2}$ Hamiltonian as (see e.g.~\cite{Faddeev:1994zg})
\begin{equation}
    E=\frac{i\lambda}{8\pi^2}\partial_u\left(\log\frac{Q_\xi(u+i/2)}{Q_\xi(u-i/2)}\right)_{u=0}\label{eq:q_function_energy} ,
\end{equation}
while the $u\rightarrow \infty$ asymptotic regime  encodes the boundary conditions (to which we will come back to  in the next section).

By demanding regularity at $u=0$ we  take a general polynomial ansatz for
the  $Q_\xi$-function, where the coefficients may depend on $\xi$. Following the structure of the eigenvalue \eqref{eq:J=2_deformed_eigenvalue}, which is a polynomial of degree two in $u$ with coefficients that are an infinite series in $\xi M$,  we can thus take 
\begin{equation}\label{eq:q_function_ansatz}
    Q_{\xi}(u) = \sum_{k=0}^{\infty} \sum_{n=0}^k c_{n,k} u^n (\xi M)^k .
\end{equation}
One can then simply solve Baxter's equation as an expansion in $\xi$ for the unknown coefficients $c_{n,k}$. A solution for the deformed ground state then reads\footnote{Here, for simplicity of the presentation, we have set all remaining free coefficients of the polynomial ansatz to $c_{n,k}=1$, since their explicit value did not affect the resulting energy $E_{\tts{GS}}$ given below. }
\[\label{eq:q_function_ground_state}
    Q_{\xi}(u)={}& 1 + \xi M  + \xi^2 M^2\left(1 -  \frac{u^2}{24}\right)  + 
  \xi^3 M^3 + \xi^4 M^4 \left(1 - \frac{329 u^2}{8640} + \frac{ u^4}{1920}\right) \\ &+\xi^5  M^5 \left(1- \frac{31 u^2}{8640} - \frac{u^4}{1920} \right) +{\cal O}(\xi^6 M^6),
\]
with energy 
\begin{equation}
    E_{\tts{GS}}=\frac{\lambda}{\pi^2}\left( \frac{\xi^2M^2}{96}-\frac{\xi^3M^3}{96}+\frac{\xi^4M^4 131}{13824} - \frac{\xi^5M^5 59}{6912 }+{\cal O}(\xi^6 M^6)\right)\label{eq:q_function_ground_state_energy}.
\end{equation}
We can in principle compute these to arbitrary order in $\xi$. 
As before, we clearly see a deformation that depends only on the combination $\xi M$. Although, with a lacking Seiberg-Witten map and gauge theory interpretation, we did not benchmark this result (as how it was done in \cite{Guica:2017mtd}), we hope it provides useful input for future studies. We will come back to this in a follow-up \cite{Driezen:2025izd}.

\section{Asymptotics and twists of the $Q$-functions at all length}\label{subsec:twisted_BCs_Baxter_eq} 
As we were able to diagonalise the transfer matrix only for the $J=2$ case, a complete construction of the $Q_\xi$-functions for arbitrary chain length $J$ is not straightforward at this point. 
In particular, we lack a direct derivation of the full spectrum or the corresponding analytic structure that would normally support Baxter's equation. 
Nevertheless, following the same logic as in the previous section, we can still extract meaningful information from 
the asymptotic $u\rightarrow\infty$ structure of the twisted transfer matrix \eqref{eq:jor-twisted-monodromy} at any $J$. See \eqref{eq:twisted_transfer_matrix_J=2}  for its formal expression in the $+1/2$ representation of $\sfa$, but generalised to arbitrary physical sites. In particular, we will focus here on the $u\rightarrow \infty$ regime of the $Q$-functions, which typically encodes  spectral charges and twisted-boundary conditions of the chain.

Let us thus consider  \eqref{eq:twisted_transfer_matrix_J=2} at any length $J\geq2$, acting on $\widehat{M}$ eigenstates. By induction, one can show that the asymptotics of its eigenvalue around $u\rightarrow\infty$ reads\footnote{One can prove this  starting from the inductive hypothesis that the elements $A(u),C(u),D(u)$ of  the undeformed monodromy matrix $T(u)=R_{\sfa J} \cdots R_{\sfa 1}$ at length $J$ read 
\[
    A(u)&=u^J+i u^{J-1} \Delta^{J-1}(\h)+{\cal O}(u^{J-2}), \\
    C(u)&=i u^{J-1} \widehat{M}+ {\cal O}(u^{J-2}), \\
    D(u)&=u^J -i u^{J-1}\Delta^{J-1}(\h)+\ {\cal O}(u^{J-2}),
\]
which is easily verified for $J=1,2$ using the definition of the $R$-matrix \eqref{eq:Lax_matrix}, and $J'=J+1$ when assuming the above holds for general $J$.  } 
\[
\hat{\tau}_\xi(u) =\frac{2+\xi \widehat{M}}{\sqrt{1+\xi \widehat{M}}} u^J  +{\cal O}(u^{J-2}), \qquad \tau_\xi(u) =\frac{2+\xi M}{\sqrt{1+\xi M}} u^J  +{\cal O}(u^{J-2}),\label{eq:all_length_transfer_matrix.} 
\]
where crucially, after a non-trivial cancellation, there is no remaining $\Delta^{J-1}(\h)$ nor ${\cal O}(u^{J-1})$  contribution. Asymptotically this implies constant twisted behaviour, as we will explain. This is in stark contrast to the dipole deformation \cite{Guica:2017mtd}, where the twist introduces a non-trivial contribution at ${\cal O}(u^{J-1})$, which ultimately  results  in involved non-analytic (in particular non-powerlike, square-root exponential) behaviour  of the $Q_\xi$-functions, cf.~eq.~(6.17) therein. 

The  leading and sub-leading terms in the asymptotic expansion of ${\tau}_\xi(u)$ 
are  sufficient to extract the (effective) leading $u\rightarrow\infty$ asymptotics of the $Q_\xi$-function. Noting that, up to an overall factor, $\tau_\xi(u)$ has the same asymptotic structure as its undeformed counterpart, it is natural to consider the following ansatz  (see also \cite{Levkovich-Maslyuk:2019awk,Bazhanov:2010ts}) around $u\rightarrow \infty$
\[
    Q_\xi(u)  =e^{u \phi} \left[ \text{power-series} \right]
    = e^{u \phi} \left(u^\alpha +\beta u^{\alpha-1}+{\cal O}\left(u^{\alpha-2}\right)\right),\label{eq:Q_function_ansatz_asymptotic}
\]
where $\phi$, $\alpha$ and $\beta$ are parameters to be fixed from Baxter's equation \eqref{eq:TQ-XXX-def}. 
After cancelling an overall $e^{u \phi}$ factor, the leading behaviour of \eqref{eq:TQ-XXX-def} at $u\rightarrow\infty$  reads
\begin{equation}\label{eq:Baxter_equation_asymptotic}
\left( e^{i\phi}  +e^{-i\phi} - \frac{2+\xi M}{\sqrt{1+\xi M}} \right) u^{J+\alpha}+{\cal O}\left(u^{J+\alpha-1}\right) = 0 , 
\end{equation}
which gives a simple equation for $\phi$, solved by
\begin{equation}\label{eq:asymptotic_coefficient}
    \phi=\pm \frac{i}{2} \ln{(1+\xi M)} + 2\pi n, \quad \forall n \in \mathbb{N}. 
\end{equation}
The sub-leading term then reads
\begin{equation}\label{eq:subleading_Baxter}
     \left(  e^{i\phi} \left( \frac{iJ}{2} +i\alpha+\beta \right)-e^{-i\phi}\left(\frac{iJ}{2} +i\alpha -\beta \right)- \frac{2+\xi M}{\sqrt{1+\xi M}} \beta \right)u^{J+\alpha -1} + {\cal O}\left(u^{J+\alpha -2}\right)= 0 . 
\end{equation}
Upon the result for $\phi$ in   \eqref{eq:asymptotic_coefficient} the   terms proportional to $\beta$ cancel, so that this equation is satisfied only for $\alpha=-\frac{J}{2}$.
In conclusion, depending on the  sign for $\phi$, we  have two possible choices for  the leading asymptotics of the $Q_\xi(u)$-function, that we denote by $\textbf{Q}_{1,2}$. Explicitly\footnote{Note that we do not reproduce the $Q$-asymptotics of the undeformed model, which depends on  the integer Cartan charge $S_1$ of $\mathfrak{sl}(2,\mathbb{R})$ that corresponds to an AdS spin in the $SL(2,\mathbb{R})$ sector of the string sigma-model. This  may be expected: First, the Baxter equation is a difference  equation which does not necessarily have smooth dependence on continuous parameters such as $\xi$ or $M$. This is what happens for abelian twists too, see e.g.~\cite{Guica:2017mtd,Kazakov:2018ugh}. Second, as explained in sections \ref{s:min-jor} (also \ref{s:string-sol}) and \ref{s:jor-chain}, the Jordanian deformation explicitly breaks the symmetry associated to $S_1$. The fact that such a charge does not appear in the $Q$-asymptotics also happens in
the classical spectral curve of the Jordanian string \cite{Borsato:2022drc}, where the asymptotics of the quasimomenta are determined only by the twist charge, energy $E$ and a spin $S_2$ (one outside of the $SL(2,\mathbb{R})$ sector studied here), rather than $E, S_1, S_2$.
}
\[
    \textbf{Q}_{1}(u) &= (1+\xi M)^{ \frac{i}{2}u} \left( u^{-\frac{J}{2}} + {\cal O}\left(u^{-\frac{J}{2}-1}\right)\right), \\
    \textbf{Q}_{2}(u) &= (1+\xi M)^{- \frac{i}{2}u} \left( u^{-\frac{J}{2}} + {\cal O}\left(u^{-\frac{J}{2}-1}\right)\right) .     \label{eq:asymptotic_q_function}
\]
Baxter's $Q$-function is then generically a linear combination of $\textbf{Q}_{1}$ and $\textbf{Q}_{2}$.

Baxter ${Q}$-functions are not isolated objects: they satisfy additional algebraic constraints known as the 
$\textbf{QQ}$-relations
which encode the same integrable (TQ) structure from a more algebraic point of view \cite{Levkovich-Maslyuk:2019awk}. 
In the case of $\mathfrak{sl}(2,\mathbb{R})$-symmetric models, the $\textbf{QQ}$-relations read
\[
\textbf{Q}_1^{+}\textbf{Q}_2^{-}-\textbf{Q}_1^{-}\textbf{Q}_2^{+}=\textbf{Q}_0\textbf{Q}_{12}, \label{eq:QQ_relations}
\]
where we recall, for $s=-1/2$, $f^{\pm}:=f(u \mp \frac{i}{2})$. 
The functions $\textbf{Q}_0$ and $\textbf{Q}_{12}$ are additional, auxiliary $Q$-functions encoding details of our model. One can however generically set $\textbf{Q}_{12}=1$ by a rescaling (also referred to as gauge transformation \cite{Levkovich-Maslyuk:2019awk,Kazakov:2015efa}).
Using the asymptotic behaviour of $\textbf{Q}_{1}$ and $\textbf{Q}_{2}$, we  can thus also determine $\textbf{Q}_0$ at large $u$
\[
\textbf{Q}_0\sim u^{-J} \frac{\xi M}{\sqrt{1+\xi M}}=  u^{-J}\left(\sqrt{1+\xi M}-\frac{1}{\sqrt{1+\xi M}}\right),\quad  u\rightarrow\infty.\label{eq:Q0_asymptotics}
\]

Interestingly, the asymptotics of $\textbf{Q}_0$ and the overall exponential behaviour of $\textbf{Q}_1$ and $\textbf{Q}_2$ match with the expressions obtained for spin chains with twisted-boundary conditions determined by a constant twist matrix $\beta_{\sfa}=e^{2i \phi \h_\sfa}$ \cite{Bazhanov:2010ts,Bazhanov:2010jq,Frassek:2011aa,Levkovich-Maslyuk:2019awk,Kazakov:2018ugh}, which are of the form 
\[
\e_{J+\sfa}=\e_\sfa (1+\xi M),\qquad \h_{J+\sfa}=\h_\sfa, \qquad \f_{J+\sfa}=\f_{\sfa}(1+\xi M)^{-1} , \label{eq:BCs_from_Q_functions}
\]
as were derived from the twisted $\mathrm{XXX}$-Hamiltonian in \cite{Bazhanov:2010ts}, cf.~(2.16) and (2.21) therein.  
While the subleading behaviour of our $Q$-functions should differ from the case of constant twists, it is nevertheless interesting to note that the above expression coincides with the Jordanian twisted-boundary conditions \eqref{eq:jor-expl-X-TBCs} once  $\widehat{M}$ is replaced by its eigenvalue. 
We consider this as  additional support of \eqref{eq:jor-expl-X-TBCs} and more generally \eqref{eq:X-TBCs-gen}. 
Furthermore, the twist charge $\phi $ has an expression that is very reminiscent of the sigma-model twist charge  \eqref{eq:HYB_twist_charges}  which  appears in the leading asymptotics of the classical spectral curve of the Jordanian string  \cite{Borsato:2022drc} (at least at large $\lambda$ and after field redefinitions, see the discussion in that section). At last, we note that the expression $\beta_{\sfa}=e^{2i \phi \h_\sfa}$ also coincides with the classical twist of the twisted monodromy  \eqref{eq:classical-limit-W-transfer}. 

The asymptotic resemblance to a chain with constant twist is interesting. 
However, even without explicitly analysing the subleading behaviour, which should  be twisted by genuine operators, we know that $Q$-functions derived for constant twists are not compatible with our story. In fact, spin chains with constant twists such as $\beta_{\sfa}=e^{2i \phi \h_\sfa}$
in \cite{Kazakov:2018ugh} would preserve only a $U(1)$ symmetry associated to the Cartan operator $\Delta^{J-1}(\h)$, while the $\widehat{M}=\Delta^{J-1}(\e)$ symmetry would be broken, yet exactly the opposite happens for the Jordanian chain.

\section{Conclusions}\label{sec:concl}

In this work, we studied the proposed spin chain description of  a Homogeneous Yang-Baxter deformation of the $AdS_5\times S^5$ superstring sigma-model of Jordanian type, focusing on its formulation via Drinfel'd twists of the associated $R$-matrix.
By analysing the dilaton profiles of all possible Jordanian deformations that yield type IIB supergravity backgrounds, we first identified a ``minimal'' Jordanian deformation\footnote{The only other possible Jordanian backgrounds with constant dilaton profiles are necessarily non-trivial in  subalgebras of $\mathfrak{psu}(2,2|4)$ that are of higher-rank than $\mathfrak{sl}(2,\mathbb{R})$. In general it would be interesting to extend our analysis to the full $\mathfrak{psu}(2,2|4)$, especially considering the unimodular fermionic extension of the minimdal model, and investigate  higher-rank cases as well. } with constant dilaton, a property that suggests a valid supergravity approximation, and that is non-trivial only in a  $SL(2, \mathbb{R})$ sector.  
To understand the string-spin correspondence in this setup, we studied a BMN-like string solution that is non-trivial only in an $AdS_3\times S^1$ subspace of $AdS_5\times S^5$. The relevant spin chain in this case is the Jordanian Drinfel'd twist of the $\mathfrak{sl}(2,\mathbb{R})$-invariant $\mathrm{XXX}_{-1/2}$ Heisenberg spin chain. 

Beyond this motivating example, we approached the topic of closed Drinfel'd twisted spin chains more generally, deriving universal formulas for their monodromy and transfer matrices. We further formalised the expectation that all closed Drinfel'd twisted models are equivalent to undeformed models with twisted-boundary conditions, by proposing a new method to derive the twist on one-site algebra generators when traversing the spin chain. 
Our framework reproduces   the abelian Drinfel'd-Reshetikhin case (see e.g.~\cite{Bazhanov:2010ts,Ahn:2010ws,deLeeuw:2012hp,Guica:2017mtd}) 
and passes several non-trivial  checks. 
Crucially, we extended this understanding to the non-abelian Jordanian case and identified its boundary-twist operator, which is consistent both with the classical string expression of \cite{Borsato:2021fuy,Borsato:2022drc} at large 't Hooft coupling  and the asymptotics of the  $Q$-system of the length $J$ spin chain.  To our knowledge, the Jordanian twisted-boundary conditions we obtain are the first for  Drinfel'd twists that are neither (successively) abelian nor $r$-symmetric. Such twists belong to an equivalence class connected via co-boundary Drinfel'd twists \cite{Kulish2009,Tolstoy:2008zz,Meljanac:2016njp}. While these transformations are expected to preserve physical observables, it would be interesting to understand how they affect the gluing  realisation of the spin chain and the sigma-model formulation.

In the rest of the paper we used the twisted-boundary formulation to initiate the spectral analysis of the 
Jordanian  $\mathfrak{sl}(2,\mathbb{R})$-$\mathrm{XXX}_{-1/2}$  spin chain. A central aspect is that the  twist breaks the full $\mathfrak{sl}(2,\mathbb{R})$ symmetry, and in particular its Cartan generator. 
Instead, we used the residual symmetry generated  by  $\widehat{M}=\Delta^{J-1}(\e)$, with $\e$ a root generator of $\mathfrak{sl}(2,\mathbb{R})$, to study the spectrum. 
Using the coherent state representation, 
we obtained a first string-spin agreement  by comparing the ground state energy shift in the twisted spin chain with the classical energy shift of the BMN-like string solution in  the continuum (large $J$) limit.
At short lengths ($J=2$),  we computed the spectrum of the twisted transfer matrix 
perturbatively in the representation space parameter and in the deformation parameter. 
Interestingly, within our   differential analysis  we found that  the Jordanian twist connects only the ground state of the undeformed chain, while the magnon excitations associated to higher total spin appear to be hidden in the deformed spectrum. 
It would be very interesting to find an explicit understanding on how the magnons are repackaged in the twisted picture. 
We eventually used the $J=2$ transfer matrix eigenvalue of the deformed ground state as an input to Baxter's equation to obtain its energy. 

In the zero-charge sector of $\widehat{M}$,  we observed that the transfer matrix spectrum remains undeformed. 
This appears consistent with the work \cite{Borsato:2025smn} on the same model, where a trivial spectrum was obtained when relying on a Fock space decomposition, which should effectively truncate part of the Hilbert space to $\widehat{M}$-charge $M=0$. This supports the interpretation of  Drinfel'd twists as  deformations that only affect sectors of the spectrum with non-zero charge under the  symmetry used to construct them. 
However, we are not yet in a position to propose a change of basis of the full Fock space to eigenfunctions of the Hamiltonian. In fact, without a smooth connection to the magnon states,  it is unclear that our deformed  $\widehat{M}$-eigenfunctions span the whole Hilbert space. It is thus possible that the Hamiltonian remains non-diagonalisable in other sectors of the Hilbert space, as supported by \cite{Borsato:2025smn}. 

Looking forward, 
the rather simple asymptotic behaviour of Baxter's $Q$-functions and  $\mathbf{QQ}$-system at general length $J$ 
suggests  a possible  treatment via the Quantum Spectral Curve and Separation of Variables program.  
A crucial next step is to determine the full analytic structure of the $Q$-functions and to obtain closed-form expressions in $\xi$, at all length $J$.
Understanding the fate of the magnon-states and their Bethe root structure would prove key for this. We leave this for  future work.
It would then be particularly interesting to  take the large $J$ limit of the spectrum and compare it with the classical string prediction and its semiclassical corrections, as computed in \cite{Borsato:2022drc}, to further support   the string-spin correspondence in the overlapping continuum/semi-classical regime and the use of Baxter's equation. 

Another intriguing open problem is the formulation  of an integrable (quantum) S-matrix   for the Jordanian string sigma-model. Recent work has observed particle production involving massive asymptotic states  \cite{Borsato:2024sru}, which appears to challenge the use of  integrable methods based on factorised S-matrices. 
This  analysis was performed
after adopting an ``alternative'' light-cone gauge-fixing based on  the residual symmetries \cite{Borsato:2023oru}.
However, our results, along with  those of \cite{Borsato:2025smn},  suggest that
the notion of a fixed number of asymptotic particles is not well-defined in the deformed setting. After adopting the Hilbert space according to residual symmetries, the 
discrete excited  states   are not straightforwardly realised in the deformed model. This  
supports the interpretation that Jordanian deformations fundamentally alter the structure of the asymptotic Hilbert space. Additionally, \cite{Borsato:2025smn} explicitly observed  that the original Fock states in the undeformed chain mix after the deformation.  
In the recent \cite{deLeeuw:2025sfs}   production processes involving massless and soft particles was also found for a  deformed spin chain  of Jordanian type in the continuum limit. There,  it was proposed that a consistent scattering framework may exist by absorbing soft particles into the asymptotic Hilbert space via Faddeev–Kulish coherent states \cite{Kulish:1970ut}.
It would be interesting to explore 
whether such a construction can accommodate the massive particle production observed in the Jordanian sigma-model using  coherent $\widehat{M}$-eigenstates as we employed here. 

More broadly, the approach developed in this work may  provide new insights into the construction of a dual deformed gauge theory,  in which the spin chain Hamiltonian should arise as the one-loop anomalous dimension of a twisted  scaling operator, and  the relevant  basis of gauge-invariant operators.  
This would not only enhance our understanding of non-AdS holography  and its possible non-relativistic field theory realisations,  but also provide valuable structural input for the $Q$-system, similar to what has been achieved in previous integrability constructions. It is still a widely open problem for Jordanian models due to technical complications from non-commutative calculus,  but  significant progress has been made in the case of (successive) abelian twists of the Poincar\'e subalgebra of $AdS$ in \cite{Meier:2023kzt,Meier:2023lku,Meier:2025tjq}.

\vspace{.6cm}
\noindent {\bf Acknowledgements ---} 
We thank Zoltan Bajnok,  Niklas Beisert, Riccardo Borsato, Simon Ekhammar, Giovanni Felder, Egor Im, Davide Lai, Marius de Leeuw, Fedor Levkovich-Maslyuk, and Ana L.~Retore for useful discussions. 
We  thank Niklas Beisert and Riccardo Borsato for comments on the draft.
We also gratefully acknowledge the participants and  organisers of  the workshops \textit{Higher-$d$ Integrability} (Favignana 2025) and \textit{Integrability in Low Supersymmetry Theories} (Trani 2024), 
where part of this work was discussed and performed.
The work of S.D.~is supported by the Swiss National
Science Foundation through the SPF fellowship TMPFP2$\_224600$ and both A.M.~and S.D.~acknowledge support by the NCCR SwissMAP.

\appendix

\section{Comments on the classical limit of twisted transfer matrix}\label{app::classical_bcs}

In this appendix, we discuss the classical limit of the twisted transfer matrix \eqref{eq:jor-twisted-monodromy}, and comment on its relation to the twisted-boundary conditions, as well as connections to the would-be sigma-model dual of the Jordanian-twisted spin chain considered in this paper. 

For this purpose, it  will be sufficient to take the auxiliary space $\sfa$ in the two-dimensional spin-$1/2$ representation of $\mathfrak{sl}(2,\mathbb{R})$, with\footnote{The $\mathfrak{sl}(2,\mathbb{R})$ Casimir 
\[ \label{eq:one-site-Casimir}
C= \h^2 - \tfrac{1}{2}(\e \f + \f \e) = s ( s+1)
\]
is equal to $\tfrac{3}{4}$ in this representation (i.e.~$s=\tfrac{1}{2}$).}
\[ \label{eq:rep-2D-sl2R}
\h_{\sfa} = \frac{1}{2} \begin{pmatrix}
1 & 0 \\ 0 & -1
\end{pmatrix} , \qquad \e_{\sfa} = \begin{pmatrix}
0 & 1 \\ 0 & 0 
\end{pmatrix} , \qquad \f_{\sfa} = \begin{pmatrix}
0 & 0 \\ -1 & 0
\end{pmatrix} .
\]
Consider then the twisted monodromy \eqref{eq:jor-twisted-monodromy}  
\begin{equation} \label{eq:app_twisted_monodromy}
T^{\star}_\mathsf{a}(u)=\begin{pmatrix}
            1&\xi \Delta^{J-1}(\h)\\
            0&1
          \end{pmatrix}
          \begin{pmatrix}
            A(u)&B(u)\\
            C(u)&D(u)
          \end{pmatrix}
          \begin{pmatrix}
            (1+\xi \widehat{M})^{-1/2}&0\\
            0&(1+\xi \widehat{M})^{1/2}
          \end{pmatrix},         
\end{equation}
and the twisted transfer matrix $\hat{\tau}_{\xi}(u):=\Tr(T^{\star}_\mathsf{a}(u))$. 

To start the spectral analysis (particularly in the context of the ABA), one typically aims to rewrite $\hat{\tau}_{\xi}(u)$ as  
\begin{equation} \label{eq:app_usual_twisted}
    \hat{\tau}_{\xi}(u)= \Tr_\mathsf{a}({\cal W}_{\sfa} T_\mathsf{a}(u)),
\end{equation}
where we call ${\cal W}_{\sfa}$ the twist matrix, which acts on the auxiliary space.
When ${\cal W}_{\sfa}$ is diagonalisable, the change in the (Bethe) spectrum of the twisted transfer matrix can then be directly read off in terms of the eigenvalues of ${\cal W}_{\sfa}$ (see e.g.~\cite{DEVEGA1990229,deLeeuw:2012hp,Belliard:2018pvg}). 

However, to relate the Jordanian twisted case \eqref{eq:app_twisted_monodromy} to \eqref{eq:app_usual_twisted}, observe that this is only possible when $\hat{\tau}_{\xi}(u)$ acts on (energy) eigenstates $\Psi_M$ of the $\widehat{M}=\Delta^{J-1}(\e)$ operator, as we also do throughout the main text, so that 
 $\widehat{M} \Psi_M = M \Psi_M$. On such states, we can write
\begin{equation} \label{eq:app_twisted_monodromy_eigenstates}
{\cal W}_{\sfa} = \begin{pmatrix}
            (1+\xi M)^{-1/2}&\xi (1+\xi M)^{-1/2}\Delta^{J-1}(\h)\\
            0&(1+\xi M)^{1/2}
          \end{pmatrix}   .    
\end{equation}
However, this object still contains the operator $\Delta^{J-1}(\h)$, so ${\cal W}_{\sfa}$ remains non-diagonalisable as an operator on the full spin chain.
In the classical limit, we may instead replace
$\Delta^{J-1}(\h)$  by its expectation value in the state $\Psi_M$, i.e.~$H:=\bra{\Psi_M}  \Delta^{J-1}(\h) \ket{\Psi_M}$. This turns the twist ${\cal W}_{\sfa}$ into an ordinary $2 \times 2$ matrix, which can now be trivially diagonalised by a change of basis in the auxiliary space. Its eigenvalues  are $(1+\xi M)^{\pm 1/2}$,  and the corresponding eigenvectors 
\[ \label{eq:app_twist_eigenvector}
\textbf{v}_\mathsf{a}^-=(1,0),\qquad \textbf{v}_{\sfa}^+=(M^{-1} H,1) ,
\]
resulting exactly in
\[ \label{eq:classical-limit-W-transfer}
{\cal W}_{\sfa} = e^{-\h_{\sfa} \log (1+\xi M)} .
\]
This is consistent with the twisted-boundary conditions \eqref{eq:jor-expl-X-TBCs} of the spin operators derived in section \ref{s:SC-TBC}, as well as the asymptotics of the $Q$-functions discussed in section \ref{subsec:twisted_BCs_Baxter_eq}, which have been used for the study of the semi-classical coherent states in section \ref{sec:landau-lifshitz}. In particular, these boundary conditions are governed by the symmetry operator $\widehat{M}$ but not by $\Delta^{J-1}(\h)$. 

Interestingly,   for $M=0$ eigenstates, the eigenvectors  \eqref{eq:app_twist_eigenvector} become ill-defined. In that case,  the  classical twist ${\cal W}_{\sfa}$ obtains a Jordan block structure and is not diagonalisable. 
This aligns with the observation that the spectrum is undeformed in such sectors, 
as seen from 
the $\xi M$ expansion in \eqref{eq:J=2_deformed_eigenvalue} (see also \cite{Borsato:2025smn}). 

Finally, it is interesting to note that a similar classical freedom to diagonalise ${\cal W}_{\sfa}$  appears in the Jordanian  sigma-model analysis. 
There, in the undeformed but twisted formulation, the sigma-model fields obey twisted-boundary conditions governed by $W$ as in \eqref{eq:W-F-jor}. This same object  appears in the conserved monodromy of the twisted model, whose diagonalised form has an asymptotic expansion in terms of local charges (the sigma-model equivalent of \eqref{eq:app_usual_twisted}, see \cite{Borsato:2021fuy,Borsato:2022drc,Driezen:2024mcn} for more details).
A sigma-model field redefinition in fact  allows one to remove the twist charge $\mathbf{q}$ in $W$ (which may be seen as the classical sigma-model analogue of $\Delta^{J-1}(\h)$). This results in a diagonal twist $W' = \exp ( \mathbf{Q} \h)$ which depends only on the conserved charge  $\mathbf{Q}$
(which may be seen as the sigma-model analogue of  $\Delta^{J-1}(\e)$ at large $\lambda$). 
See eq.~\eqref{eq:W-F-jor} and the surrounding discussions below, as well as section 4.3 and 7 of \cite{Borsato:2022drc} for further details. 
Importantly, this diagonalisation is only possible in sectors of classical sigma-model solutions where $\mathbf{Q}\neq 0$. 
When $\mathbf{Q}=0$ (which we may understand as $M=0$ eigenstates of the spin chain), the twist is not diagonalisable but instead becomes of Jordan form. Similarly, the twisted-boundary conditions \eqref{eq:jor-expl-X-TBCs} are consistent with $W' = \exp ( \mathbf{Q} \h)$ at large $\lambda$. 
 
It would be interesting to understand whether these transformations---field redefinitions in the sigma-model and basis changes of the auxiliary space in the spin chain---can be interpreted as Hopf algebra isomorphisms, possibly induced by co-boundary Drinfel'd twists \cite{Kulish2009,Tolstoy:2008zz,Meljanac:2016njp}.

\section{Singular ODEs and their formal resolution}\label{app:dif_eq} 
In this appendix, we present a method for obtaining the formal solution $g(z)$ to ODEs of the form \eqref{eq:J=2_deformed_EV_eq_u^0} which are singular at some point $z=\lambda$. A standard reference  for this purpose is the book \cite{wasow1965asymptotic}, but we will follow the algorithm and terminology of \cite{Barkatou01011999} here.

The first step is to recast \eqref{eq:J=2_deformed_EV_eq_u^0} as a first order ODE in vector form. We let $\textbf{v}(z):=(g(z),g'(z),g''(z))^T$ and rewrite \eqref{eq:J=2_deformed_EV_eq_u^0}  as the generic system 
\begin{equation}
    \frac{d\textbf{v}(z)}{dz}=\frac{A(z)}{(z-\lambda)^{1+p[A]}}\textbf{v}(z),\label{eq:app_vector_eq}
\end{equation}
that one also denotes formally as $[A]$. Here, $p[A]$ is an integer  called  the ``Poincaré'' rank of $[A]$, determined by requiring  that  the matrix $A(z)$ is analytic at $z=\lambda$, i.e.~it has a regular polynomial expansion, with a constant leading term.  When $p[A]<0$, the point $z=\lambda$ is a regular point. When  $p[A]=0$,  it is a singularity of the first kind, and for $p[A]\geq 1$,  it is a singularity of the second kind. 
In our case, we have $\lambda=0$ and
\begin{equation}
    A(z)=\sum_{n=0}^2 A_n z^n
    =\frac{1}{8\xi}\begin{pmatrix}
            0&1&0\\
            0&0&1\\
            p_1(z) &p_2(z) &p_3(z) 
         \end{pmatrix}
         ,\label{eq:app_matrix_A}
\end{equation}
with $p_i(z)$ as in eqs.~\eqref{eq:p-s}, and $p[A]=1$.
Explicitly, the coefficient matrices are
\begin{equation}
\begin{gathered}
A_0 = \begin{pmatrix} 0 & 0 & 0 \\ 0 & 0 & 0 \\ \frac{2+\xi M + 16 \sqrt{1+\xi M} \chi_\xi}{4\xi} & -1 & 0 \end{pmatrix}, \qquad
A_1 = \begin{pmatrix} 0 & 0 & 0 \\ 0 & 0 & 0 \\ \frac{M^2}{4} & \frac{2 + \xi M}{\xi} & -3 \end{pmatrix}, \\
A_2 = \begin{pmatrix} 0 & 1 & 0 \\ 0 & 0 & 1 \\ -\frac{M^2 (2+\xi M)}{8 \xi}  & \frac{M^2}{4} & \frac{2+\xi M}{2\xi} \end{pmatrix} .
\end{gathered}
\end{equation}

\noindent {\bf Regular and irregular singularities ---}
Having singularities of the  ``first'' and ``second'' kind classifies the form of the differential system, but it does not determine the actual analytical behaviour of its solution. There are strong theorems of asymptotic analysis \cite{wasow1965asymptotic} that guarantee that the generic differential system \eqref{eq:app_vector_eq} always admits a formal solution around such singularities, at least as an asymptotic trans-series. 
Depending on the nature of the asymptotic expansion of the formal solution near $z=\lambda$, one further distinguishes between regular and irregular singular points.

\textit{\textbf{Regular singularities:}} In this case, the formal solution is of the form
\begin{equation}
    \textbf{v}(z) =P(z-\lambda) (z-\lambda)^\Lambda \textbf{c},\label{eq:app_regular_singular}
\end{equation}
with  $P(z-\lambda)$  a regular asymptotic matrix expansion around $z=\lambda$, $\Lambda$ a constant matrix and $\textbf{c}$ a vector of integration constants. These solutions can be obtained by (matrix versions of) Frobenius' method.    

\textit{\textbf{Irregular singularities:}} In this case, the formal solution is of the form
\begin{equation}
    \textbf{v}(z)=P(t)t^\Lambda e^{q(1/t)}\textbf{c},\label{eq:app_irregular_singular}
\end{equation}
where $P$, $\Lambda$ and $\textbf{c}$ are as above,  $t=(z-\lambda)^{1/s}$, with $s\in\mathbb{N}_0$ called the branching exponent, and $q(1/t)$ is a polynomial in $1/t$, called the exponential part, capturing the instantonic behaviour.

Singularities of the first kind are always regular, while singularities of the second kind can be either regular or irregular singularities. Our case is precisely of the second kind, thereby making the analysis more subtle. 
However, as we will now explain, one can also play with ``gauge transformations'' of the system to try to map singularities of the second kind to singularities of the first kind. To simplify the upcoming formulas,  we will take $\lambda=0$ from now on, but having non-zero $\lambda$ should be easily reinstateable. 

\noindent {\bf Gauge transformations ---} In certain cases, it is possible to simplify the singular structure of the system $[A]$ by means of a so-called ``gauge transformation''. Given  \eqref{eq:app_vector_eq}, one can consider the transformation  $\textbf{v}(z)=T(z)\textbf{w}(z)$ with $T(z)$ an invertible matrix, and map it to  a new, equivalent system
\begin{equation} \label{eq:app_gauge_transformation}
     \frac{d\textbf{w}(z)}{dz}=\frac{B(z)}{z^{1+p[B]}}\textbf{w}(z),   \qquad B(z):= z^{1+p[B]} \left(\frac{T^{-1}(z)A(z)T(z)}{z^{1+p[A]}}  - T^{-1}(z) T'(z)\right)   .
\end{equation}
with $p[B]$ again such that $B(z) = \sum_{n=0}^N B_n z^n$ for some positive integer $N$. 
We will denote this system  as $[B]=T[A]$. Importantly, while both systems are equivalent in terms of their solutions, the Poincaré rank is generally not invariant under such transformations. Specifically,  a $T(z)$ with appropriately chosen behaviour around $z=0$ can raise or lower the rank by modifying the singular structure via its derivative   $T'(z)$. Constant $T(z)$ do, however, leave the  rank invariant. 

\noindent {\bf Moser reducibility ---} Given the generic system [A]  in \eqref{eq:app_vector_eq}, there  exists a criterion, due to Moser,  for the reducibility of the system  to one with lower Poincaré rank. In particular,  
\begin{equation}
    \exists ~T(z)\quad \text{such that} \quad p[T[A]]<p[A] \quad \Longleftrightarrow \quad z^{\mathrm{rank}(A_0)}\det(\alpha{1}-A_0 z^{-1}-A_1)\vert_{z=0}=0 , \label{eq:app_Moser_reducibility}
\end{equation}
for some generic $\alpha$. 
As it turns out, our system \eqref{eq:app_matrix_A} has ${\mathrm{rank}(A_0)}=1$ and thus  satisfies this criteria. Since  $p[A]=1$, we are  guaranteed that there exists a $T(z)$ such that $p[T[A]]=0$, giving us a singularity at $z=0$ of the first kind, and  a regular solution of the form \eqref{eq:app_regular_singular}. 
In practice, one can find such a $T(z)$   by first doing a constant gauge transformation to put $A_0$ in its Jordan form, and  then by performing a ``shearing transform'' $T(z)=\mathrm{diag}(z^{k_1},z^{k_2},z^{k_3})$ with $k_i$ integers chosen such that the rank reduces. After this, one usually does  an additional constant gauge transform to  put also the leading term $B_0$ of $[B]=T[A]$ in Jordan form.  The system can then be solved by a  Frobenius-like method. In our case, we find that  $k_1=0$ and $k_2=k_3=1$ works. 

\noindent {\bf Matrix Frobenius method (or Sauvage's method) ---}
Since we are in the regular singular case, we will use a Frobenius-like method, known as Sauvage's method, in the matrix case. Looking at the form of the solution \eqref{eq:app_regular_singular}, the idea is to build the coefficients of the expansion of $P(z)$ around $z=0$ using recurrence relations. However, before proceeding, we must distinguish two cases. Consider a system with a singularity of the first kind 
\begin{equation}
 \frac{d\textbf{w}(z)}{dz}=\frac{B(z)}{z}\textbf{w}(z),\label{eq:app_vector_eq_frob}
\end{equation}
and  assume without loss of generality that $B_0$ is in Jordan form.

\textit{\textbf{Resonant case:}} If some of the eigenvalues of $B_0$ differ with another by a positive integer, we are in the so-called resonant case. We then need to perform additional shearing transforms $T(z)=\mathrm{diag}(1,...,1,z,1...1)$ with $z$ in the entry corresponding to the eigenvalue that differs from the other by a positive integer. This will essentially reduce this eigenvalue by $1$, and we repeat this 
until we are out of the resonant case.
In our case, after reducing the system from the second to the first kind and transforming the leading $B_0$ matrix to Jordan form, we find  
\begin{equation}
    B_0=\begin{pmatrix}
        -2&1&0\\
        0&-2&0\\
        0&0&-1
        \end{pmatrix},\label{eq:app_leading_resonant}
\end{equation}
so that we are precisely in the resonant case. We transform to a non-resonant case by performing the shearing transform $T(z)=\mathrm{diag}(1,1,z)$.

\textit{\textbf{Non-resonant case:}} When no eigenvalue differs from another by a positive integer, we can finally solve the system. Consider again the generic system \eqref{eq:app_vector_eq_frob} but with $B_0$ now such that one is in the non-resonant case. The solution is then given by
\begin{equation}\label{eq:app_frob_sol}
    \textbf{w}(z)=P(z)z^{B_0}\textbf{c}, \quad\text{with   } P(z)=\sum_{n=0}^\infty P_n z^n ,
\end{equation}
with, as usual, $\textbf{c}$ a vector of integration constants, and the coefficients of $P(z)$  given by the solutions of
\begin{equation}
    P_0={1},\qquad (B_0-n{1})P_n-P_nB_0=-\sum_{l=0}^{n-1}B_{n-l}P_l \label{eq:app_frob_recurrence}.
\end{equation}
The condition that no eigenvalue differs by positive integer is what guarantees that \eqref{eq:app_frob_recurrence} has a unique solution for all $n$. It is a Sylvester equation, which has well studied solutions, and one can solve it  through recurrence relations using, e.g., the \textsc{LyapunovSolve} method in \textsc{Mathematica}. 

After transforming the resonant case \eqref{eq:app_leading_resonant} with the shearing transform, we find the following  leading matrix in Jordan form
\begin{equation}
    B_0=\begin{pmatrix}
        -2&1&0\\
        0&-2&1\\
        0&0&-2
        \end{pmatrix},\label{eq:app_leading_non_resonant}
\end{equation}
which is a three-dimensional Jordan block. This is what causes both the $\log(z)$ and $\log^2(z)$ in \eqref{eq:J=2_dif_eq_} to appear. The $-2$ eigenvalues of $B_0$ further imply that \eqref{eq:app_frob_sol} leads with $z^{-2}$, which is eventually  shifted to $z^0$ by the successive   shearing and Jordan transforms that bring us back to the original system $[A]$, reproducing finally  \eqref{eq:J=2_dif_eq_}.

\addtocontents{toc}{\protect\setcounter{tocdepth}{-1}}
\begin{bibtex}[\jobname]

@book{Chari:1994pz,
    author = "Chari, V. and Pressley, A.",
    title = "{A Guide To Quantum Groups}",
    year = "1994",
    publisher      = "Cambridge University Press",
}

@article{Arutyunov:2009ga,
author = "Arutyunov, Gleb and Frolov, Sergey",
title = "{Foundations of the $AdS_5 \times S^5$ Superstring. Part I}",
eprint = "0901.4937",
archivePrefix = "arXiv",
primaryClass = "hep-th",
reportNumber = "ITP-UU-09-05, SPIN-09-05, TCD-MATH-09-06, HMI-09-03",
doi = "10.1088/1751-8113/42/25/254003",
journal = "J. Phys. A",
volume = "42",
pages = "254003",
year = "2009"
}

@book{wasow1965asymptotic,
  added-at = {2014-03-10T23:45:53.000+0100},
  author = {Wasow, Wolfgang},
  biburl = {https://www.bibsonomy.org/bibtex/20405958db68b29a0445e5f2e42573561/peter.ralph},
  interhash = {82ac86c40b1b85e07b8e102a2ed14982},
  intrahash = {0405958db68b29a0445e5f2e42573561},
  keywords = {ODE asymptotic_expansions reference},
  mrclass = {34.50},
  mrnumber = {0203188 (34 \#3041)},
  mrreviewer = {N. D. Kazarinoff},
  pages = {ix+362},
  publisher = {Interscience Publishers John Wiley \& Sons, Inc., New              York-London-Sydney},
  series = {Pure and Applied Mathematics, Vol. XIV},
  timestamp = {2014-03-10T23:45:53.000+0100},
  title = {Asymptotic expansions for ordinary differential equations},
  year = 1965
}

@article{Barkatou01011999,
author = {M.A. Barkatou and G. Chen},
title = {Computing the exponential part of a formal fundamental matrix solution of a linear difference system},
journal = {Journal of Difference Equations and Applications},
volume = {5},
number = {2},
pages = {117--142},
year = {1999},
publisher = {Taylor \& Francis},
doi = {10.1080/10236199908808176},
eprint = { https://doi.org/10.1080/10236199908808176}

}

@article{Delduc:2013qra,
    author = "Delduc, Francois and Magro, Marc and Vicedo, Benoit",
    title = "{An integrable deformation of the $AdS_5 \times S^5$ superstring action}",
    eprint = "1309.5850",
    archivePrefix = "arXiv",
    primaryClass = "hep-th",
    doi = "10.1103/PhysRevLett.112.051601",
    journal = "Phys. Rev. Lett.",
    volume = "112",
    number = "5",
    pages = "051601",
    year = "2014"
}

@article{Delduc:2014kha,
    author = "Delduc, Francois and Magro, Marc and Vicedo, Benoit",
    title = "{Derivation of the action and symmetries of the $q$-deformed $AdS_{5} \times S^{5}$ superstring}",
    eprint = "1406.6286",
    archivePrefix = "arXiv",
    primaryClass = "hep-th",
    doi = "10.1007/JHEP10(2014)132",
    journal = "JHEP",
    volume = "10",
    pages = "132",
    year = "2014"
}

@article{Klimcik:2008eq,
    author = "Klimcik, Ctirad",
    title = "{On integrability of the Yang-Baxter sigma-model}",
    eprint = "0802.3518",
    archivePrefix = "arXiv",
    primaryClass = "hep-th",
    doi = "10.1063/1.3116242",
    journal = "J. Math. Phys.",
    volume = "50",
    pages = "043508",
    year = "2009"
}

@article{Delduc:2013fga,
    author = "Delduc, Francois and Magro, Marc and Vicedo, Benoit",
    title = "{On classical $q$-deformations of integrable sigma-models}",
    eprint = "1308.3581",
    archivePrefix = "arXiv",
    primaryClass = "hep-th",
    doi = "10.1007/JHEP11(2013)192",
    journal = "JHEP",
    volume = "11",
    pages = "192",
    year = "2013"
}

@article{Borsato:2022drc,
    author = "Borsato, Riccardo and Driezen, Sibylle and Nieto Garc\'\i{}a, Juan Miguel and Wyss, Leander",
    title = "{Semiclassical spectrum of a Jordanian deformation of AdS5\texttimes{}S5}",
    eprint = "2207.14748",
    archivePrefix = "arXiv",
    primaryClass = "hep-th",
    reportNumber = "DMUS-MP-22/09",
    doi = "10.1103/PhysRevD.106.066015",
    journal = "Phys. Rev. D",
    volume = "106",
    number = "6",
    pages = "066015",
    year = "2022"
}

@article{Borsato:2021fuy,
    author = "Borsato, Riccardo and Driezen, Sibylle and Miramontes, J. Luis",
    title = "{Homogeneous Yang-Baxter deformations as undeformed yet twisted models}",
    eprint = "2112.12025",
    archivePrefix = "arXiv",
    primaryClass = "hep-th",
    doi = "10.1007/JHEP04(2022)053",
    journal = "JHEP",
    volume = "04",
    pages = "053",
    year = "2022"
}

@article{vanTongeren:2018wek,
    author = "Van Tongeren, Stijn J.",
    title = "{On Yang\textendash{}Baxter models, twist operators, and boundary conditions}",
    eprint = "1804.05680",
    archivePrefix = "arXiv",
    primaryClass = "hep-th",
    doi = "10.1088/1751-8121/aac8eb",
    journal = "J. Phys. A",
    volume = "51",
    pages = "305401",
    year = "2018"
}

@article{Belitsky:2004sc,
    author = "Belitsky, A. V. and Derkachov, Sergey E. and Korchemsky, G. P. and Manashov, A. N.",
    title = "{Dilatation operator in (super-)Yang-Mills theories on the light-cone}",
    eprint = "hep-th/0409120",
    archivePrefix = "arXiv",
    reportNumber = "LPT-ORSAY-04-60",
    doi = "10.1016/j.nuclphysb.2004.11.034",
    journal = "Nucl. Phys. B",
    volume = "708",
    pages = "115--193",
    year = "2005"
}

@article{Bellucci:2004qr,
    author = "Bellucci, S. and Casteill, P. -Y. and Morales, J. F. and Sochichiu, Corneliu",
    title = "{SL(2) spin chain and spinning strings on AdS(5) x S**5}",
    eprint = "hep-th/0409086",
    archivePrefix = "arXiv",
    doi = "10.1016/j.nuclphysb.2004.11.020",
    journal = "Nucl. Phys. B",
    volume = "707",
    pages = "303--320",
    year = "2005"
}

@article{Borsato:2025smn,
    author = "Borsato, Riccardo and Fern{\'a}ndez, Miguel Garc{\'\i}a",
    title = "{Jordanian deformation of the non-compact and $\mathfrak{sl}_2 $-invariant $XXX_{-1/2}$ spin-chain}",
    eprint = "2503.24223",
    archivePrefix = "arXiv",
    primaryClass = "hep-th",
    doi = "10.1007/JHEP08(2025)074",
    journal = "JHEP",
    volume = "08",
    pages = "074",
    year = "2025"
}

@article{Derkachov:2002tf,
    author = "Derkachov, Sergey E. and Korchemsky, G. P. and Manashov, A. N.",
    title = "{Separation of variables for the quantum SL(2,R) spin chain}",
    eprint = "hep-th/0210216",
    archivePrefix = "arXiv",
    reportNumber = "LPT-ORSAY-02-88, RUB-TP2-13-02",
    doi = "10.1088/1126-6708/2003/07/047",
    journal = "JHEP",
    volume = "07",
    pages = "047",
    year = "2003"
}
@article{Gromov:2016rrp,
    author = "Gromov, Nikolay and Levkovich-Maslyuk, Fedor",
    title = "{Quark-anti-quark potential in $ \mathcal{N} =$ 4 SYM}",
    eprint = "1601.05679",
    archivePrefix = "arXiv",
    primaryClass = "hep-th",
    reportNumber = "NORDITA-2016-134",
    doi = "10.1007/JHEP12(2016)122",
    journal = "JHEP",
    volume = "12",
    pages = "122",
    year = "2016"
}

@article{Perelomov1977,
author = "A.~M.~Perelomov",
title = "Generalized coherent states and some of their applications",
doi = "10.1070/PU1977v020n09ABEH005459",
journal = "Soviet Physics Uspekhi",
volume = "20",
number = "9",
pages = "703",
year = "1977"
}

@article{Kruczenski:2004cn,
      author         = "Kruczenski, Martin and Tseytlin, Arkady A.",
      title          = "{Semiclassical relativistic strings in S$^5$ and long
                        coherent operators in N=4 SYM theory}",
      journal        = "JHEP",
      volume         = "0409",
      pages          = "038",
      doi            = "10.1088/1126-6708/2004/09/038",
      year           = "2004",
      eprint         = "hep-th/0406189",
      archivePrefix  = "arXiv",
      primaryClass   = "hep-th",
      reportNumber   = "BRX-TH-543",
      SLACcitation   = "
}

@article{Kruczenski:2004kw,
      author         = "Kruczenski, M. and Ryzhov, A.V. and Tseytlin, Arkady A.",
      title          = "{Large spin limit of AdS$_5 \times$S$^5$ string theory and
                        low-energy expansion of ferromagnetic spin chains}",
      journal        = "Nucl.Phys.",
      volume         = "B692",
      pages          = "3-49",
      doi            = "10.1016/j.nuclphysb.2004.05.028",
      year           = "2004",
      eprint         = "hep-th/0403120",
      archivePrefix  = "arXiv",
      primaryClass   = "hep-th",
      reportNumber   = "BRX-TH-537",
      SLACcitation   = "
}
@article{Guica:2017mtd,
      author         = "Guica, Monica and Levkovich-Maslyuk, Fedor and Zarembo,
                        Konstantin",
      title          = "{Integrability in dipole-deformed
                        ${\mathcal{N}=4}$ super Yang–Mills}",
      journal        = "J. Phys.",
      volume         = "A50",
      year           = "2017",
      number         = "39",
      pages          = "394001",
      doi            = "10.1088/1751-8121/aa8491",
      eprint         = "1706.07957",
      archivePrefix  = "arXiv",
      primaryClass   = "hep-th",
      reportNumber   = "NORDITA-2017-061, UUITP-18-17",
      SLACcitation   = "
}

@article{Beisert:2003jj,
      author         = "Beisert, Niklas",
      title          = "{The complete one loop dilatation operator of N=4
                        superYang-Mills theory}",
      journal        = "Nucl.Phys.",
      volume         = "B676",
      pages          = "3-42",
      doi            = "10.1016/j.nuclphysb.2003.10.019",
      year           = "2004",
      eprint         = "hep-th/0307015",
      archivePrefix  = "arXiv",
      primaryClass   = "hep-th",
      reportNumber   = "AEI-2003-056",
      SLACcitation   = "
}
@article{Ouyang:2017yko,
    author = "Ouyang, Hao",
    title = "{Semiclassical spectrum for BMN string in $Sch_5\times S^5$}",
    eprint = "1709.06844",
    archivePrefix = "arXiv",
    primaryClass = "hep-th",
    doi = "10.1007/JHEP12(2017)126",
    journal = "JHEP",
    volume = "12",
    pages = "126",
    year = "2017"
}

@article{Maillet:1996yy,
    author = "Maillet, J. M. and Sanchez de Santos, J.",
    title = "{Drinfel'd twists and algebraic Bethe ansatz}",
    eprint = "q-alg/9612012",
    archivePrefix = "arXiv",
    reportNumber = "ENSLAPP-L-601-96, US-FT-47-96, ENSLAPP-601-96, ISBN:-978-0-8218-2133-6",
    month = "12",
    year = "1996"
}

@Inbook{Kulish2009,
author="Kulish, Petr",
title="Twist Deformations of Quantum Integrable Spin Chains",
bookTitle="Noncommutative Spacetimes: Symmetries in Noncommutative Geometry and Field Theory",
year="2009",
publisher="Springer Berlin Heidelberg",
pages="167--190",
abstract="Twist deformations of spacetime lead to deformed field theories with twisted symmetries. Twisted symmetries are quantum group symmetries. Most integrable spin systems have dynamical symmetries related to appropriate quantum groups. We discuss the changes of the properties of these systems under twist transformations of quantum groups. A main example is the isotropic Heisenberg spin chain and the jordanian twist of the universal enveloping algebra of{\$}{\$}sl(2){\$}{\$}. It is shown that the spectrum of the XXX label XXX model spin chain is preserved under the twist deformation while the structure of the eigenstates depends on the choice of boundary conditions. Another example is provided by abelian twists, these give physical deformations of closed spin chains corresponding to higher rank Lie algebras, e.g., {\$}{\$}gl(n){\$}{\$}. The energy spectrum of these integrable models is changed and correspondingly their eigenvectors.",
isbn="978-3-540-89793-4",
doi="10.1007/978-3-540-89793-4_9",
url="https://doi.org/10.1007/978-3-540-89793-4\_9"
}

@article{deLeeuw:2025sfs,
    author = "de Leeuw, Marius and Fontanella, Andrea and Nieto Garc\'\i{}a, Juan Miguel",
    title = "{An integrable deformed Landau-Lifshitz model with particle production?}",
    eprint = "2506.13598",
    archivePrefix = "arXiv",
    primaryClass = "hep-th",
    month = "6",
    year = "2025"
}

@inproceedings{Tseytlin:2004xa,
    author = "Tseytlin, Arkady A.",
    title = "{Semiclassical strings and AdS/CFT}",
    booktitle = "{NATO Advanced Study Institute and EC Summer School on String Theory: From Gauge Interactions to Cosmology}",
    eprint = "hep-th/0409296",
    archivePrefix = "arXiv",
    pages = "265--290",
    month = "9",
    year = "2004"
}

@article{Stefanski:2004cw,
    author = "Stefanski, Jr., B. and Tseytlin, Arkady A.",
    title = "{Large spin limits of AdS/CFT and generalized Landau-Lifshitz equations}",
    eprint = "hep-th/0404133",
    archivePrefix = "arXiv",
    reportNumber = "IMPERIAL-TP-3-04-10",
    doi = "10.1088/1126-6708/2004/05/042",
    journal = "JHEP",
    volume = "05",
    pages = "042",
    year = "2004"
}

@article{2004Tolstoy,
   author = {Tolstoy, V. N.},
    title = "{Chains of extended Jordanian twists for Lie superalgebras}",
  journal = {ArXiv Mathematics e-prints},
  volume= { },
  pages = { },
   eprint = {math/0402433},
 keywords = {Quantum Algebra, Mathematical Physics},
     year = 2004,
   adsurl = {http://adsabs.harvard.edu/abs/2004math......2433T},
  adsnote = {Provided by the SAO/NASA Astrophysics Data System}
}

@article{Borsato:2016ose,
      author         = "Borsato, Riccardo and Wulff, Linus",
      title          = "{Target space supergeometry of $\eta$ and
                        $\lambda$-deformed strings}",
      journal        = "JHEP",
      volume         = "10",
      year           = "2016",
      pages          = "045",
      doi            = "10.1007/JHEP10(2016)045",
      eprint         = "1608.03570",
      archivePrefix  = "arXiv",
      primaryClass   = "hep-th",
      reportNumber   = "IMPERIAL-TP-LW-2016-03",
      SLACcitation   = "
}

@inproceedings{drinfeld1983constant,
  title={Constant quasiclassical solutions of the Yang--Baxter quantum equation},
  author={Drinfeld, Vladimir Gershonovich},
  booktitle={Doklady Akademii Nauk},
  volume={273},
  number={3},
  pages={531--535},
  year={1983},
  organization={Russian Academy of Sciences}
}

@article{vanTongeren:2019dlq,
    author = "van Tongeren, Stijn J.",
    title = "{Unimodular jordanian deformations of integrable superstrings}",
    eprint = "1904.08892",
    archivePrefix = "arXiv",
    primaryClass = "hep-th",
    doi = "10.21468/SciPostPhys.7.1.011",
    journal = "SciPost Phys.",
    volume = "7",
    pages = "011",
    year = "2019"
}

@article{Borsato:2022ubq,
    author = "Borsato, Riccardo and Driezen, Sibylle",
    title = "{All Jordanian deformations of the $AdS_5 \times S^5$ superstring}",
    eprint = "2212.11269",
    archivePrefix = "arXiv",
    primaryClass = "hep-th",
    doi = "10.21468/SciPostPhys.14.6.160",
    journal = "SciPost Phys.",
    volume = "14",
    number = "6",
    pages = "160",
    year = "2023"
}

@article{Osten:2016dvf,
      author         = "Osten, David and van Tongeren, Stijn J.",
      title          = "{Abelian Yang–Baxter deformations and TsT
                        transformations}",
      journal        = "Nucl. Phys.",
      volume         = "B915",
      year           = "2017",
      pages          = "184-205",
      doi            = "10.1016/j.nuclphysb.2016.12.007",
      eprint         = "1608.08504",
      archivePrefix  = "arXiv",
      primaryClass   = "hep-th",
      reportNumber   = "HU-EP-16-29",
      SLACcitation   = "
}

@article{Lunin:2005jy,
      author         = "Lunin, Oleg and Maldacena, Juan Martin",
      title          = "{Deforming field theories with $U(1) \times U(1)$ global
                        symmetry and their gravity duals}",
      journal        = "JHEP",
      volume         = "0505",
      pages          = "033",
      doi            = "10.1088/1126-6708/2005/05/033",
      year           = "2005",
      eprint         = "hep-th/0502086",
      archivePrefix  = "arXiv",
      primaryClass   = "hep-th",
      SLACcitation   = "
}

@article{Leigh:1995ep,
    author = "Leigh, Robert G. and Strassler, Matthew J.",
    title = "{Exactly marginal operators and duality in four-dimensional N=1 supersymmetric gauge theory}",
    eprint = "hep-th/9503121",
    archivePrefix = "arXiv",
    reportNumber = "RU-95-2",
    doi = "10.1016/0550-3213(95)00261-P",
    journal = "Nucl. Phys. B",
    volume = "447",
    pages = "95--136",
    year = "1995"
}

@article{Hashimoto:1999ut,
      author         = "Hashimoto, Akikazu and Itzhaki, N.",
      title          = "{Noncommutative Yang-Mills and the AdS / CFT
                        correspondence}",
      journal        = "Phys. Lett.",
      volume         = "B465",
      year           = "1999",
      pages          = "142-147",
      doi            = "10.1016/S0370-2693(99)01037-0",
      eprint         = "hep-th/9907166",
      archivePrefix  = "arXiv",
      primaryClass   = "hep-th",
      reportNumber   = "NSF-ITP-99-085",
      SLACcitation   = "
}

@article{Maldacena:1999mh,
      author         = "Maldacena, Juan Martin and Russo, Jorge G.",
      title          = "{Large N limit of noncommutative gauge theories}",
      journal        = "JHEP",
      volume         = "9909",
      pages          = "025",
      doi            = "10.1088/1126-6708/1999/09/025",
      year           = "1999",
      eprint         = "hep-th/9908134",
      archivePrefix  = "arXiv",
      primaryClass   = "hep-th",
      reportNumber   = "HUTP-99-A046",
      SLACcitation   = "
}

@article{Seiberg:1999vs,
      author         = "Seiberg, Nathan and Witten, Edward",
      title          = "{String theory and noncommutative geometry}",
      journal        = "JHEP",
      volume         = "09",
      year           = "1999",
      pages          = "032",
      doi            = "10.1088/1126-6708/1999/09/032",
      eprint         = "hep-th/9908142",
      archivePrefix  = "arXiv",
      primaryClass   = "hep-th",
      reportNumber   = "IASSNS-HEP-99-74",
      SLACcitation   = "
}

@article{Meier:2023kzt,
    author = "Meier, Tim and van Tongeren, Stijn J.",
    title = "{Quadratic Twist-Noncommutative Gauge Theory}",
    eprint = "2301.08757",
    archivePrefix = "arXiv",
    primaryClass = "hep-th",
    reportNumber = "HU-EP-23/03-RTG",
    doi = "10.1103/PhysRevLett.131.121603",
    journal = "Phys. Rev. Lett.",
    volume = "131",
    number = "12",
    pages = "121603",
    year = "2023"
}

@article{Meier:2023lku,
    author = "Meier, Tim and van Tongeren, Stijn J.",
    title = "{Gauge theory on twist-noncommutative spaces}",
    eprint = "2305.15470",
    archivePrefix = "arXiv",
    primaryClass = "hep-th",
    reportNumber = "HU-EP-23/11-RTG",
    doi = "10.1007/JHEP12(2023)045",
    journal = "JHEP",
    volume = "12",
    pages = "045",
    year = "2023"
}

@article{Bergman:2000cw,
    author = "Bergman, Aaron and Ganor, Ori J.",
    title = "{Dipoles, twists and noncommutative gauge theory}",
    eprint = "hep-th/0008030",
    archivePrefix = "arXiv",
    reportNumber = "PUPT-1946",
    doi = "10.1088/1126-6708/2000/10/018",
    journal = "JHEP",
    volume = "10",
    pages = "018",
    year = "2000"
}

@article{Reshetikhin:1990ep,
    author = "Reshetikhin, N.",
    title = "{Multiparameter quantum groups and twisted quasitriangular Hopf algebras}",
    doi = "10.1007/BF00626530",
    journal = "Lett. Math. Phys.",
    volume = "20",
    pages = "331--335",
    year = "1990"
}

@article{Maldacena:2008wh,
    author = "Maldacena, Juan and Martelli, Dario and Tachikawa, Yuji",
    title = "{Comments on string theory backgrounds with non-relativistic conformal symmetry}",
    eprint = "0807.1100",
    archivePrefix = "arXiv",
    primaryClass = "hep-th",
    doi = "10.1088/1126-6708/2008/10/072",
    journal = "JHEP",
    volume = "10",
    pages = "072",
    year = "2008"
}
@article{Herzog:2008wg,
    author = "Herzog, Christopher P. and Rangamani, Mukund and Ross, Simon F.",
    title = "{Heating up Galilean holography}",
    eprint = "0807.1099",
    archivePrefix = "arXiv",
    primaryClass = "hep-th",
    reportNumber = "PUPT-2274, DCPT-08-39",
    doi = "10.1088/1126-6708/2008/11/080",
    journal = "JHEP",
    volume = "11",
    pages = "080",
    year = "2008"
}

@article{Adams:2008wt,
    author = "Adams, Allan and Balasubramanian, Koushik and McGreevy, John",
    title = "{Hot Spacetimes for Cold Atoms}",
    eprint = "0807.1111",
    archivePrefix = "arXiv",
    primaryClass = "hep-th",
    reportNumber = "MIT-CTP-3962",
    doi = "10.1088/1126-6708/2008/11/059",
    journal = "JHEP",
    volume = "11",
    pages = "059",
    year = "2008"
}

@article{Alishahiha:2003ru,
    author = "Alishahiha, Mohsen and Ganor, Ori J.",
    title = "{Twisted backgrounds, PP waves and nonlocal field theories}",
    eprint = "hep-th/0301080",
    archivePrefix = "arXiv",
    reportNumber = "IPM-P-2003-002, UCB-PTH-02-60, LBL-51922",
    doi = "10.1088/1126-6708/2003/03/006",
    journal = "JHEP",
    volume = "03",
    pages = "006",
    year = "2003"
}

@article{Beisert:2005if,
      author         = "Beisert, N. and Roiban, R.",
      title          = "{Beauty and the twist: The Bethe ansatz for twisted N=4
                        SYM}",
      journal        = "JHEP",
      volume         = "08",
      year           = "2005",
      pages          = "039",
      doi            = "10.1088/1126-6708/2005/08/039",
      eprint         = "hep-th/0505187",
      archivePrefix  = "arXiv",
      primaryClass   = "hep-th",
      reportNumber   = "PUTP-2162",
      SLACcitation   = "
}

@article{Frolov:2005dj,
      author         = "Frolov, Sergey",
      title          = "{Lax pair for strings in Lunin-Maldacena background}",
      journal        = "JHEP",
      volume         = "0505",
      pages          = "069",
      doi            = "10.1088/1126-6708/2005/05/069",
      year           = "2005",
      eprint         = "hep-th/0503201",
      archivePrefix  = "arXiv",
      primaryClass   = "hep-th",
      SLACcitation   = "
}

@article{Driezen:2024mcn,
    author = "Driezen, Sibylle and Kamath, Niranjan",
    title = "{Regularising spectral curves for homogeneous Yang-Baxter strings}",
    eprint = "2406.09811",
    archivePrefix = "arXiv",
    primaryClass = "hep-th",
    doi = "10.1016/j.physletb.2024.138971",
    journal = "Phys. Lett. B",
    volume = "857",
    pages = "138971",
    year = "2024"
}

@article{Borsato:2024sru,
    author = "Borsato, Riccardo and Driezen, Sibylle",
    title = "{Particle production in a light-cone gauge fixed Jordanian deformation of AdS5\texttimes{}S5}",
    eprint = "2412.08411",
    archivePrefix = "arXiv",
    primaryClass = "hep-th",
    doi = "10.1103/PhysRevD.111.086010",
    journal = "Phys. Rev. D",
    volume = "111",
    number = "8",
    pages = "086010",
    year = "2025"
}

@article{Nishida:2007pj,
    author = "Nishida, Yusuke and Son, Dam T.",
    title = "{Nonrelativistic conformal field theories}",
    eprint = "0706.3746",
    archivePrefix = "arXiv",
    primaryClass = "hep-th",
    reportNumber = "INT-PUB-07-16",
    doi = "10.1103/PhysRevD.76.086004",
    journal = "Phys. Rev. D",
    volume = "76",
    pages = "086004",
    year = "2007"
}

@article{Son:2008ye,
    author = "Son, D. T.",
    title = "{Toward an AdS/cold atoms correspondence: A Geometric realization of the Schrodinger symmetry}",
    eprint = "0804.3972",
    archivePrefix = "arXiv",
    primaryClass = "hep-th",
    reportNumber = "INT-PUB-08-08",
    doi = "10.1103/PhysRevD.78.046003",
    journal = "Phys. Rev. D",
    volume = "78",
    pages = "046003",
    year = "2008"
}

@article{Matsumoto:2014ubv,
    author = "Matsumoto, Takuya and Yoshida, Kentaroh",
    title = "{Yang-Baxter deformations and string dualities}",
    eprint = "1412.3658",
    archivePrefix = "arXiv",
    primaryClass = "hep-th",
    reportNumber = "KUNS-2506",
    doi = "10.1007/JHEP03(2015)137",
    journal = "JHEP",
    volume = "03",
    pages = "137",
    year = "2015"
}

@article{Kawaguchi:2014fca,
    author = "Kawaguchi, Io and Matsumoto, Takuya and Yoshida, Kentaroh",
    title = "{A Jordanian deformation of AdS space in type IIB supergravity}",
    eprint = "1402.6147",
    archivePrefix = "arXiv",
    primaryClass = "hep-th",
    reportNumber = "KUNS-2484, ITP-UU-14-07, SPIN-14-07",
    doi = "10.1007/JHEP06(2014)146",
    journal = "JHEP",
    volume = "06",
    pages = "146",
    year = "2014"
}

@article{vanTongeren:2015uha,
      author         = "van Tongeren, Stijn J.",
      title          = "{Yang–Baxter deformations, AdS/CFT, and
                        twist-noncommutative gauge theory}",
      journal        = "Nucl. Phys.",
      volume         = "B904",
      year           = "2016",
      pages          = "148-175",
      doi            = "10.1016/j.nuclphysb.2016.01.012",
      eprint         = "1506.01023",
      archivePrefix  = "arXiv",
      primaryClass   = "hep-th",
      reportNumber   = "HU-EP-15-27, HU-MATH-15-08",
      SLACcitation   = "
}

@article{Vicedo:2015pna,
    author = "Vicedo, Benoit",
    title = "{Deformed integrable \ensuremath{\sigma}-models, classical R-matrices and classical exchange algebra on Drinfel\textquoteright{}d doubles}",
    eprint = "1504.06303",
    archivePrefix = "arXiv",
    primaryClass = "hep-th",
    doi = "10.1088/1751-8113/48/35/355203",
    journal = "J. Phys. A",
    volume = "48",
    number = "35",
    pages = "355203",
    year = "2015"
}

@article{Blau:2009gd,
    author = "Blau, Matthias and Hartong, Jelle and Rollier, Blaise",
    title = "{Geometry of Schrodinger Space-Times, Global Coordinates, and Harmonic Trapping}",
    eprint = "0904.3304",
    archivePrefix = "arXiv",
    primaryClass = "hep-th",
    doi = "10.1088/1126-6708/2009/07/027",
    journal = "JHEP",
    volume = "07",
    pages = "027",
    year = "2009"
}

@article{Giaquinto:1994jx,
    author = "Giaquinto, Anthony and Zhang, James J.",
    title = "{Bialgebra actions, twists, and universal deformation formulas}",
    eprint = "hep-th/9411140",
    archivePrefix = "arXiv",
    doi = "10.1016/S0022-4049(97)00041-8",
    journal = "J. Pure Appl. Algebra",
    volume = "128",
    pages = "133--151",
    year = "1998"
}

@incollection{Gerstenhaber:1992,
  author        = "Gerstenhaber, M. and Giaquinto, A. and Schack, S. D.",
  title         = "{Quantum symmetry}",
  booktitle     = "{Quantum Groups}",
  editor        = "Kulish, P. P.",
  publisher     = "Springer Berlin Heidelberg",
  year          = "1992",
  pages         = "9--46",
  doi           = "10.1007/BFb0101176"
}

@inproceedings{Ogievetsky:1992ph,
  author       = "Ogievetsky, O.",
  title        = "{Hopf structures on the Borel subalgebra of sl(2)}",
  booktitle    = "{Proceedings of the Winter School Geometry and Physics}",
  volume       = "37",
  pages        = "185",
  year         = "1993",
  url          = "https://dml.cz/handle/10338.dmlcz/701555?show=full"
}

@article{Kulish:1998be,
  author        = "Kulish, P. P. and Lyakhovsky, V. D. and Mudrov, A. I.",
  title         = "{Extended Jordanian twists for Lie algebras}",
  journal       = "J. Math. Phys.",
  volume        = "40",
  pages         = "4569",
  year          = "1999",
  doi           = "10.1063/1.532987",
  eprint        = "math/9806014",
  archivePrefix = "arXiv",
  primaryClass  = "math"
}

@article{Tolstoy:2008zz,
    author = "Tolstoy, V. N.",
    editor = "Doebner, H. D. and Dobrev, V. K.",
    title = "{Twisted quantum deformations of Lorentz and Poincare algebras}",
    journal = "Bulg. J. Phys.",
    volume = "35",
    pages = "441--459",
    year = "2008",
    eprint={0712.3962},
    archivePrefix={arXiv},
    primaryClass={math.QA}
}

@inproceedings{Faddeev:1996iy,
    author = "Faddeev, L. D.",
    title = "{How algebraic Bethe ansatz works for integrable model}",
    booktitle = "{Les Houches School of Physics: Astrophysical Sources of Gravitational Radiation}",
    eprint = "hep-th/9605187",
    archivePrefix = "arXiv",
    pages = "pp. 149--219",
    month = "5",
    year = "1996"
}

@article{Barut:1970qf,
    author = "Barut, A. O. and Girardello, L.",
    title = "{New 'coherent' states associated with noncompact groups}",
    doi = "10.1007/BF01646483",
    journal = "Commun. Math. Phys.",
    volume = "21",
    pages = "41--55",
    year = "1971"
}

@article{Levkovich-Maslyuk:2019awk,
    author = "Levkovich-Maslyuk, Fedor",
    title = "{A review of the AdS/CFT Quantum Spectral Curve}",
    eprint = "1911.13065",
    archivePrefix = "arXiv",
    primaryClass = "hep-th",
    doi = "10.1088/1751-8121/ab7137",
    journal = "J. Phys. A",
    volume = "53",
    number = "28",
    pages = "283004",
    year = "2020"
}

@article{Meljanac:2016njp,
    author = "Meljanac, Stjepan and Meljanac, Daniel and Pacho\l{}, Anna and Pikuti\'c, Danijel",
    title = "{Remarks on simple interpolation between Jordanian twists}",
    eprint = "1612.07984",
    archivePrefix = "arXiv",
    primaryClass = "math-ph",
    doi = "10.1088/1751-8121/aa72d7",
    journal = "J. Phys. A",
    volume = "50",
    number = "26",
    pages = "265201",
    year = "2017"
}

@article{Kulish:1981gi,
    author = "Kulish, P. P. and Reshetikhin, N. Yu. and Sklyanin, E. K.",
    title = "{Yang-Baxter Equation and Representation Theory. 1.}",
    doi = "10.1007/BF02285311",
    journal = "Lett. Math. Phys.",
    volume = "5",
    pages = "393--403",
    year = "1981"
}

@article{Tarasov:1983cj,
    author = "Tarasov, V. O. and Takhtajan, L. A. and Faddeev, L. D.",
    title = "{Local Hamiltonians for integrable quantum models on a lattice}",
    doi = "10.1007/BF01018648",
    journal = "Theor. Math. Phys.",
    volume = "57",
    pages = "1059--1073",
    year = "1983"
}

@article{DEVEGA1990229,
title = {Yang-Baxter algebras, integrable theories and quantum groups},
journal = {Nuclear Physics B - Proceedings Supplements},
volume = {18},
number = {1},
pages = {229-282},
year = {1990},
issn = {0920-5632},
doi = {https://doi.org/10.1016/0920-5632(90)90651-A},
author = {H.J. {De Vega}}
}

@article{Belliard:2018pvg,
    author = "Belliard, Samuel and Slavnov, Nikita A. and Vallet, Benoit",
    title = "{Modified Algebraic Bethe Ansatz: Twisted XXX Case}",
    eprint = "1804.00597",
    archivePrefix = "arXiv",
    primaryClass = "math-ph",
    doi = "10.3842/SIGMA.2018.054",
    journal = "SIGMA",
    volume = "14",
    pages = "054",
    year = "2018"
}

@article{Kazakov:2018ugh,
    author = "Kazakov, Vladimir",
    editor = "Ge, Mo-Lin and Niemi, Antti J. and Phua, Kok Khoo and Takhtajan, Leon A.",
    title = "{Quantum Spectral Curve of $\gamma$-twisted ${\cal N}=4$ SYM theory and fishnet CFT}",
    booktitle = {Ludwig Faddeev Memorial Volume},
    chapter = {},
    eprint = "1802.02160",
    archivePrefix = "arXiv",
    primaryClass = "hep-th",
    reportNumber = "LPTENS-18-02, LPTENS-18/02",
    doi = "10.1142/9789813233867_0016",
    pages = "293--342",
    year = "2018"
}

@article{Kazakov:2015efa,
    author = "Kazakov, Vladimir and Leurent, Sebastien and Volin, Dmytro",
    title = "{T-system on T-hook: Grassmannian Solution and Twisted Quantum Spectral Curve}",
    eprint = "1510.02100",
    archivePrefix = "arXiv",
    primaryClass = "hep-th",
    doi = "10.1007/JHEP12(2016)044",
    journal = "JHEP",
    volume = "12",
    pages = "044",
    year = "2016"
}

@article{Faddeev:1994zg,
    author = "Faddeev, L. D. and Korchemsky, G. P.",
    title = "{High energy QCD as a completely integrable model}",
    eprint = "hep-th/9404173",
    archivePrefix = "arXiv",
    reportNumber = "ITP-SB-94-14",
    doi = "10.1016/0370-2693(94)01363-H",
    journal = "Phys. Lett. B",
    volume = "342",
    pages = "311--322",
    year = "1995"
}

@article{Bazhanov:2010ts,
    author = "Bazhanov, Vladimir V. and Lukowski, Tomasz and Meneghelli, Carlo and Staudacher, Matthias",
    title = "{A Shortcut to the Q-Operator}",
    eprint = "1005.3261",
    archivePrefix = "arXiv",
    primaryClass = "hep-th",
    reportNumber = "AEI-2010-023, HU-EP-10-24, HU-MATHEMATIK-2010-7",
    doi = "10.1088/1742-5468/2010/11/P11002",
    journal = "J. Stat. Mech.",
    volume = "1011",
    pages = "P11002",
    year = "2010"
}

@article{Baxter:1971sapm,
  author         = "Baxter, R. J.",
  title          = "{Generalized Ferroelectric Model on a Square Lattice}",
  journal        = "Stud. Appl. Math.",
  volume         = "50",
  number         = "1",
  pages          = "51--69",
  year           = "1971",
  doi            = "10.1002/sapm197150151",
  eprint         = "https://onlinelibrary.wiley.com/doi/pdf/10.1002/sapm197150151",
  SLACcitation   = "
}

@article{Baxter:1972aop,
  author         = "Baxter, Rodney J.",
  title          = "{Partition function of the Eight-Vertex lattice model}",
  journal        = "Annals Phys.",
  volume         = "70",
  number         = "1",
  pages          = "193--228",
  year           = "1972",
  doi            = "10.1016/0003-4916(72)90335-1",
  url            = "https://www.sciencedirect.com/science/article/pii/0003491672903351",
  SLACcitation   = "
}

@article{Sklyanin:1991ss,
    author = "Sklyanin, E. K.",
    title = "{Quantum inverse scattering method. Selected topics}",
    eprint = "hep-th/9211111",
    archivePrefix = "arXiv",
    reportNumber = "HU-TFT-91-51",
    month = "10",
    year = "1991"
}

@article{Bazhanov:2010jq,
    author = "Bazhanov, Vladimir V. and Frassek, Rouven and Lukowski, Tomasz and Meneghelli, Carlo and Staudacher, Matthias",
    title = "{Baxter Q-Operators and Representations of Yangians}",
    eprint = "1010.3699",
    archivePrefix = "arXiv",
    primaryClass = "math-ph",
    reportNumber = "HU-EP-10-28, AEI-2010-116",
    doi = "10.1016/j.nuclphysb.2011.04.006",
    journal = "Nucl. Phys. B",
    volume = "850",
    pages = "148--174",
    year = "2011"
}
@article{Frassek:2011aa,
    author = "Frassek, Rouven and Lukowski, Tomasz and Meneghelli, Carlo and Staudacher, Matthias",
    title = "{Baxter Operators and Hamiltonians for 'nearly all' Integrable Closed $\mathfrak{gl}(n)$ Spin Chains}",
    eprint = "1112.3600",
    archivePrefix = "arXiv",
    primaryClass = "math-ph",
    reportNumber = "HU-MATHEMATIK-2011-22, HU-EP-11-52, AEI-2011-081",
    doi = "10.1016/j.nuclphysb.2013.06.006",
    journal = "Nucl. Phys. B",
    volume = "874",
    pages = "620--646",
    year = "2013"
}
@article{Kirch:2004mk,
    author = "Kirch, M. and Manashov, A. N.",
    title = "{Noncompact SL(2,R) spin chain}",
    eprint = "hep-th/0405030",
    archivePrefix = "arXiv",
    reportNumber = "RUB-TP2-03-04",
    doi = "10.1088/1126-6708/2004/06/035",
    journal = "JHEP",
    volume = "06",
    pages = "035",
    year = "2004"
}

@article{Frenkel:2013uda,
    author = "Frenkel, Edward and Hernandez, David",
    title = "{Baxter{\textquoteright}s relations and spectra of quantum integrable models}",
    eprint = "1308.3444",
    archivePrefix = "arXiv",
    primaryClass = "math.QA",
    doi = "10.1215/00127094-3146282",
    journal = "Duke Math. J.",
    volume = "164",
    number = "12",
    pages = "2407--2460",
    year = "2015"
}
@article{Felder_2017,
   title={Baxter operators and asymptotic representations},
   volume={23},
   ISSN={1420-9020},
   url={http://dx.doi.org/10.1007/s00029-017-0320-z},
   DOI={10.1007/s00029-017-0320-z},
   number={4},
   journal={Selecta Mathematica},
   publisher={Springer Science and Business Media LLC},
   author={Felder, Giovanni and Zhang, Huafeng},
   year={2017},
   month=mar, pages={2947–2975} 
   }

   @article{Zhang:2018sej,
    author = "Zhang, Huafeng",
    title = "{Yangians and Baxter{\textquoteright}s relations}",
    eprint = "1808.02294",
    archivePrefix = "arXiv",
    primaryClass = "math.QA",
    doi = "10.1007/s11005-020-01285-x",
    journal = "Lett. Math. Phys.",
    volume = "110",
    number = "8",
    pages = "2113--2141",
    year = "2020"
}
   
@article{Loebbert:2016cdm,
    author = "Loebbert, Florian",
    title = "{Lectures on Yangian Symmetry}",
    eprint = "1606.02947",
    archivePrefix = "arXiv",
    primaryClass = "hep-th",
    reportNumber = "HU-EP-16-12",
    doi = "10.1088/1751-8113/49/32/323002",
    journal = "J. Phys. A",
    volume = "49",
    number = "32",
    pages = "323002",
    year = "2016"
}

@article{Gaudin:1992ci,
    author = "Gaudin, M. and Pasquier, V.",
    title = "{The periodic Toda chain and a matrix generalization of the bessel function's recursion relations}",
    reportNumber = "SACLAY-SPH-T-92-014",
    journal = "J. Phys. A",
    volume = "25",
    pages = "5243--5252",
    year = "1992"
}

@article{Beisert:2003yb,
    author = "Beisert, Niklas and Staudacher, Matthias",
    title = "{The N=4 SYM integrable super spin chain}",
    eprint = "hep-th/0307042",
    archivePrefix = "arXiv",
    reportNumber = "AEI-2003-057",
    doi = "10.1016/j.nuclphysb.2003.08.015",
    journal = "Nucl. Phys. B",
    volume = "670",
    pages = "439--463",
    year = "2003"
}

@article{Beisert:2010jr,
      author         = "Beisert, Niklas and others",
      title          = "{Review of AdS/CFT Integrability: An Overview}",
      journal        = "Lett. Math. Phys.",
      volume         = "99",
      year           = "2012",
      pages          = "3-32",
      doi            = "10.1007/s11005-011-0529-2",
      eprint         = "1012.3982",
      archivePrefix  = "arXiv",
      primaryClass   = "hep-th",
      reportNumber   = "AEI-2010-175, CERN-PH-TH-2010-306, HU-EP-10-87,
                        HU-MATH-2010-22, KCL-MTH-10-10, UMTG-270, UUITP-41-10",
      SLACcitation   = "
}

@article{Bombardelli:2016rwb,
    author = "Bombardelli, Diego and Cagnazzo, Alessandra and Frassek, Rouven and Levkovich-Maslyuk, Fedor and Loebbert, Florian and Negro, Stefano and Sz{\'e}cs{\'e}nyi, Istvan M. and Sfondrini, Alessandro and van Tongeren, Stijn J. and Torrielli, Alessandro",
    title = "{An integrability primer for the gauge-gravity correspondence: An introduction}",
    eprint = "1606.02945",
    archivePrefix = "arXiv",
    primaryClass = "hep-th",
    reportNumber = "CNRS-16-03, DCPT-16-19, DESY-16-083, DMUS-MP-16-09, HU-EP-16-13, NORDITA-2016-33, HU-MATH-16-08",
    doi = "10.1088/1751-8113/49/32/320301",
    journal = "J. Phys. A",
    volume = "49",
    number = "32",
    pages = "320301",
    year = "2016"
}

@article{Kulish:1970ut,
    author = "Kulish, P. P. and Faddeev, L. D.",
    title = "{Asymptotic conditions and infrared divergences in quantum electrodynamics}",
    reportNumber = "D70-07927",
    doi = "10.1007/BF01066485",
    journal = "Theor. Math. Phys.",
    volume = "4",
    pages = "745",
    year = "1970"
}

@article{vanTongeren:2016eeb,
    author = "van Tongeren, Stijn J.",
    title = "{Almost abelian twists and AdS/CFT}",
    eprint = "1610.05677",
    archivePrefix = "arXiv",
    primaryClass = "hep-th",
    reportNumber = "HU-EP-16-35, HU-MATH-16-19",
    doi = "10.1016/j.physletb.2016.12.002",
    journal = "Phys. Lett. B",
    volume = "765",
    pages = "344--351",
    year = "2017"
}

@article{Araujo:2017jkb,
    author = "Araujo, T. and Bakhmatov, I. and Colg{\'a}in, E. {\'O} and Sakamoto, J. and Sheikh-Jabbari, M. M. and Yoshida, K.",
    title = "{Yang-Baxter $\sigma$-models, conformal twists, and noncommutative Yang-Mills theory}",
    eprint = "1702.02861",
    archivePrefix = "arXiv",
    primaryClass = "hep-th",
    reportNumber = "DMUS-MP-17-03, KUNS-2663, IPM-P-2017-004, APCTP-PRE2017-003",
    doi = "10.1103/PhysRevD.95.105006",
    journal = "Phys. Rev. D",
    volume = "95",
    number = "10",
    pages = "105006",
    year = "2017"
}

@article{Kruczenski:2003gt,
    author = "Kruczenski, Martin",
    title = "{Spin chains and string theory}",
    eprint = "hep-th/0311203",
    archivePrefix = "arXiv",
    reportNumber = "BRX-TH-539",
    doi = "10.1103/PhysRevLett.93.161602",
    journal = "Phys. Rev. Lett.",
    volume = "93",
    pages = "161602",
    year = "2004"
}

@article{Ahn:2010ws,
    author = "Ahn, Changrim and Bajnok, Zoltan and Bombardelli, Diego and Nepomechie, Rafael I.",
    title = "{Twisted Bethe equations from a twisted S-matrix}",
    eprint = "1010.3229",
    archivePrefix = "arXiv",
    primaryClass = "hep-th",
    reportNumber = "UMTG-268",
    doi = "10.1007/JHEP02(2011)027",
    journal = "JHEP",
    volume = "02",
    pages = "027",
    year = "2011"
}

@article{deLeeuw:2012hp,
    author = "de Leeuw, Marius and van Tongeren, Stijn J.",
    title = "{The spectral problem for strings on twisted AdS$_5 \times$ S$^5$}",
    eprint = "1201.1451",
    archivePrefix = "arXiv",
    primaryClass = "hep-th",
    doi = "10.1016/j.nuclphysb.2012.03.004",
    journal = "Nucl. Phys. B",
    volume = "860",
    pages = "339--376",
    year = "2012"
}

@article{McLoughlin:2006cg,
    author = "McLoughlin, Tristan and Swanson, Ian",
    title = "{Integrable twists in AdS/CFT}",
    eprint = "hep-th/0605018",
    archivePrefix = "arXiv",
    doi = "10.1088/1126-6708/2006/08/084",
    journal = "JHEP",
    volume = "08",
    pages = "084",
    year = "2006"
}

@article{Beisert:2024wqq,
    author = {Beisert, Niklas and K{\"o}nig, Benedikt},
    title = "{Yangian form-alism for planar gauge theories}",
    eprint = "2411.16176",
    archivePrefix = "arXiv",
    primaryClass = "hep-th",
    doi = "10.1063/5.0253127",
    journal = "J. Math. Phys.",
    volume = "66",
    number = "6",
    pages = "062301",
    year = "2025"
}

@article{Borsato:2023oru,
    author = "Borsato, Riccardo and Driezen, Sibylle and Hoare, Ben and Retore, Ana L. and Seibold, Fiona K.",
    title = "{Inequivalent light-cone gauge-fixings of strings on AdSn{\texttimes}Sn backgrounds}",
    eprint = "2312.17056",
    archivePrefix = "arXiv",
    primaryClass = "hep-th",
    doi = "10.1103/PhysRevD.109.106023",
    journal = "Phys. Rev. D",
    volume = "109",
    number = "10",
    pages = "106023",
    year = "2024"
}

@article{Meier:2025tjq,
    author = "Meier, Tim and van Tongeren, Stijn J.",
    title = "{Integrable spin chains in twisted maximally supersymmetric Yang-Mills theory}",
    eprint = "2507.18626",
    archivePrefix = "arXiv",
    primaryClass = "hep-th",
    reportNumber = "HU-EP-25/27",
    month = "7",
    year = "2025"
}

@article{Matsumoto:2015jja,
    author = "Matsumoto, Takuya and Yoshida, Kentaroh",
    title = "{Yang{\textendash}Baxter sigma models based on the CYBE}",
    eprint = "1501.03665",
    archivePrefix = "arXiv",
    primaryClass = "hep-th",
    reportNumber = "KUNS-2534",
    doi = "10.1016/j.nuclphysb.2015.02.009",
    journal = "Nucl. Phys. B",
    volume = "893",
    pages = "287--304",
    year = "2015"
}

@article{Kawaguchi:2013lba,
    author = "Kawaguchi, Io and Matsumoto, Takuya and Yoshida, Kentaroh",
    title = "{Schroedinger sigma models and Jordanian twists}",
    eprint = "1305.6556",
    archivePrefix = "arXiv",
    primaryClass = "hep-th",
    reportNumber = "KUNS-2451",
    doi = "10.1007/JHEP08(2013)013",
    journal = "JHEP",
    volume = "08",
    pages = "013",
    year = "2013"
}

@article{Driezen:2025izd,
    author = "Driezen, Sibylle and Levkovich-Maslyuk, Fedor and Molines, Adrien",
    title = "{Integrability for the spectrum of Jordanian AdS/CFT}",
    eprint = "2511.11521",
    archivePrefix = "arXiv",
    primaryClass = "hep-th",
    month = "11",
    year = "2025"
}

\end{bibtex}

\bibliographystyle{nb}
\bibliography{\jobname}

\end{document}